\documentclass[aps,prd,floatfix,showkeys,showpacs,10pt]{revtex4}

\usepackage[overload]{textcase}
\usepackage[dvips]{graphics}
\usepackage{hyperref}
\usepackage{amssymb}
\usepackage{amsmath}
\usepackage{graphicx,multirow}
\usepackage{pstricks}
\usepackage{color}
\usepackage{placeins}
\makeatletter
\newcounter{subsubsubsection}[subsubsection]
\def\subsubsubsectionmark#1{}

\def\subsubsubsection{\@startsection
     {subsubsubsection}{4}{\z@} {-3.25ex plus -1
     ex minus -.2ex}{1.5ex plus .2ex}{\normalsize\it}}
\def\l@subsubsubsection{\@dottedtocline{4}{4.8em}
     {4.2em}}
\newcounter{subsubsubsubsection}[subsubsubsection]
\def\subsubsubsubsectionmark#1{}

\def\subsubsubsubsection{\@startsection
     {subsubsubsubsection}{5} {\z@} {-3.25ex plus -1
     ex minus -.2ex}{1.5ex plus .2ex}{\normalsize\bf}}
\def\l@subsubsubsubsection{\@dottedtocline{5}
     {5.8em}{5.2em}}

\def\etal{{\it et al.}}

\def\half{{\textstyle{1\over2}}}
\def\thalf{{\textstyle{3\over2}}}
\def\fhalf{{\textstyle{5\over2}}}

\def\hlf{{{1\over2}}}

\def\>{\rangle}
\def\<{\langle}

\def\E#1{10^{#1}}

\def\A{{\cal A}}
\def\B{{\cal B}}

\def\D{{\cal D}}
\def\F{{\cal F}}

\def\I{{\cal I}}
\def\J{{\cal J}}

\def\M{{\cal M}}

\def\O{{\cal O}}
\def\P{{\cal P}}

\def\R{{\cal R}}

\def\X{{\cal X}}

\def\alf{\alpha}

\def\betpm{\beta_{+-}}
\def\betmp{\beta_{-+}}

\def\gm{\gamma}
\def\Lb{\Lambda_b}
\def\Lc{\Lambda_c}
\def\Ls{\Lambda}
\def\Lstar{\Lambda^{(*)}}
\def\elp{\ell^+}
\def\elm{\ell^-}
\def\Ppm{{\cal P}^\pm}
\def\epm{\hat{e}^\pm}
\def\spm{\vec{\xi}^\pm}

\def\PNm{{\cal P}_N^-}
\def\PNp{{\cal P}_N^+}
\def\sh{\hat{s}}
\def\zh{\hat{z}}
\def\ym{y^{-}}
\def\yp{y^{+}}

\def\mlh{\hat{m}_\ell}
\def\mbh{\hat{m}_b}
\def\mlb{m_{\Lambda_b}}
\def\mls{m_{\Lambda}}
\def\plb{p_{\Lambda_b}}
\def\pls{p_{\Lambda}}
\def\slb{s_{\Lambda_b}}
\def\sls{s_{\Lambda}}
\def\Re{\text{Re}}
\def\Im{\text{Im}}

\def\vvp{v\cdot v'}
\def\vq{\mlb-\mls\vvp}
\def\vpq{\mlb\vvp-\mls}

\def\detvvpm{\eta^{VV(\betpm)}}
\def\detaapm{\eta^{AA(\betpm)}}
\def\detvvmp{\eta^{VV(\betmp)}}
\def\detaamp{\eta^{AA(\betmp)}}

\def\={&=&}
\def\<{\langle}
\def\>{\rangle}
\def\jp{{J^P}}
\def\slash#1{#1 \hskip -0.5em / }
\def\beq{\begin{equation}}
\def\eeq{\end{equation}}
\def\nn{\nonumber}
\def\beqy{\begin{eqnarray}}
\def\eeqy{\end{eqnarray}}
\def\beqynn{\begin{eqnarray*}}
\def\eeqynn{\end{eqnarray*}}

\def\xiv{\xi^V}
\def\xia{\xi^A}
\def\xit{\xi^T}
\def\xz{\xi^{(0)}}
\def\xo{\xi^{(1)}}
\def\xt{\xi^{(2)}}
\def\zz{\zeta^{(0)}}
\def\zo{\zeta^{(1)}}

\def\lqcd{\Lambda_\text{QCD}}
\def\GeV{\text{GeV}}
\def\sbar{\overline{s}}
\def\ubar{\overline{u}}

\begin{document}

\markboth{L. Mott, W. Roberts}
{Lepton polarization asymmetries for FCNC decays of the $\Lambda_b$ baryon}

\title{\bf Lepton polarization asymmetries for FCNC decays of the $\Lambda_b$ baryon}

\author{L. Mott}
\affiliation{Department of Physics, Florida State University, Tallahassee, FL 32306\\ Department of Physics, Florida A \& M University, Tallahassee, FL 32307}

\author{W. Roberts}
\affiliation{Department of Physics, Florida State University, Tallahassee, FL 32306}

\begin{abstract}
Branching ratios, lepton forward-backward asymmetries, and lepton polarization asymmetries for the flavor-changing neutral current (FCNC) dileptonic decays of the $\Lb$ baryon to the ground state and a number of excited state $\Ls$ baryons are calculated using form factors extracted using wave functions from a constituent quark model. The SM branching ratios for the transition to the ground state calculated using these quark model form factors are consistent with the recent measurement reported by the LHCb collaboration. It is shown that the lepton polarization asymmetries are largely insensitive to the transition form factors and, therefore, to the effects of QCD in the nonperturbative regime. These observables can therefore  provide somewhat model independent ways of extracting various combinations of the Wilson coefficients.
\end{abstract}


\maketitle

\section{Introduction}

Flavor-changing neutral current (FCNC) processes involving heavy hadrons are of significant interest. These Glashow-Iliopoulos-Maiani (GIM)
suppressed decays are important sources of information about the loop structure of the Standard Model (SM) and are useful for constraining new
physics beyond the standard model (SM). The main problem one has in investigating processes involving these transitions is in the evaluation of the hadronic matrix elements of the weak currents. These hadronic matrix elements are usually written in terms of several unknown form factors that parametrize the uncalculable nonperturbative QCD dynamics. Heavy quark effective theory (HQET) \cite{hqet}, quark models \cite{isgw,barik}, QCD sum rules (QCDSR) \cite{qcdsr}, lattice QCD (LQCD) \cite{lattice}, etc., have been employed in the approximation or modeling of these form factors.

When investigating the FCNC decays of heavy hadrons, several experimentally measurable quantities contain valuable information about the Wilson
coefficients that enter into the effective Hamiltonian that describes these decays. Among these are observables involving the final state leptons. These lepton asymmetries are important because, being defined as ratios of decay rates, they are expected to be less sensitive to the nonperturbative QCD dynamics. Thus, they should offer somewhat model independent ways of determining the values of the Wilson coefficients, which is crucial to the determination of any new physics.

Theoretical investigations into FCNC processes involving $B$ mesons have been carried out extensively \cite{buras,buchalla,falk,greub,cho,robertsledroit,roberts,hewett,aliev1,aliev2,aliev3,aliev4,aliev5,aliev6,melikhov1,melikhov2,melikhov3,huangli,burdman,goto,huang1,huang2,yan,huang3,huang4,colangelo1,colangelo2,chua,ali,chang,itlan,beneke,kruger,safir,chen3,chen4,chen5,chen6,frank,arda,choudhury,bpsw,ghinculov,bashiry,defazio,ferrandes}. Relativistic quark models \cite{melikhov1}, perturbative QCD (pQCD) \cite{chen3}, light-cone sum rules (LCSR) \cite{aliev1,ali,arda,yan}, along with other techniques, have been employed to estimate hadronic matrix elements and model the necessary form factors. These form factors have been used to calculate observables for both radiative and dileptonic decays both within the SM and for various beyond-the-SM scenarios, such as supersymmetric (SUSY) models \cite{yan,chua,chen4}, models with universal extra dimensions (UEDs) \cite{ferrandes}, and other new physics (NP) scenarios \cite{chen6,arda}.

Similarly, there have been theoretical efforts to estimate these matrix elements for FCNC processes involving $\Lb$ baryons \cite{morob,aslam,wang,chen,chen1,chen2,he,zolf,mannel,huang,cheng,manwang,feldyip,detmold}. The majority of what has been put forth involves employing HQET to reduce the number of independent form factors to two universal form factors, valid for all currents, and then a model of some kind is used to extract these two form factors. LCSR \cite{aslam,wang,huang,manwang,feldyip}, QCDSR \cite{chen,chen1,chen2,cheng,zolf}, multipole model parametrizations (PM) \cite{chen,chen2,mannel}, pQCD \cite{he}, bag models \cite{cheng}, and LQCD \cite{detmold} have all been employed to calculate form factors for transitions to ground state $\Ls$s.

In a previous work \cite{morob}, a nonrelativistic constituent quark model was employed in calculating the transition form factors for $\Lb\to\Lstar$. Two approximation schemes were used to compute the form factors. The first method involved the use of single component wave functions obtained from a variational diagonalization of a quark model Hamiltonian. To compute the matrix elements, the quark operators were reduced to their Pauli form with the form factors being extracted analytically. Form factors obtained from this approximation are called SCA form factors. This method had been used previously \cite{pervin,pervin1} to obtain reasonable estimates of the decay rates for the semileptonic decays $\Lb\to\Lc\ell\overline{\nu}_\ell$ and $\Lc\to\Ls\ell\overline{\nu}_\ell$.

In the second approximation scheme, the full relativistic form of the quark current and the full quark model wave function were kept in a numerical extraction of the form factors. These form factors are called MCN form factors. Both the SCA and MCN form factors were shown to obey the expectations of leading order HQET. Both sets of form factors were used to compute the differential decay rates, branching ratios (BRs), and lepton forward-backward asymmetries (FBAs) for transitions to ground state $\Ls$ and a number of its excited states. The BRs obtained using both sets of form factors with SM Wilson coefficients were found to lie just below the lower limit of the CDF collaboration's measurement of $\Lb\to\Ls(1115)\mu^+\mu^-$, while the BRs obtained with SUSY Wilson coefficients were consistent with the CDF observation.

In this paper, we examine the weak baryonic FCNC process $\Lb\to\Lstar\elp\elm\,\,\,\,(\ell=\mu,\tau)$. We will reexamine the BRs and FBAs of our previous work \cite{morob} with updated MCN form factors. We will also examine the longitudinal, transverse, and normal components of the lepton polarization asymmetries (LPAs). It has been shown that for $B$ meson decays that the LPAs are largely independent of the form factors in certain limits \cite{buras,robertsledroit,roberts,aliev1,aliev2,aliev3,aliev4}. We explore whether LPAs for the decays of the $\Lb$ are also independent of the form factors in any kinematic regimes.

The rest of this paper is organized as follows: in Section \ref{sec:merfba}, we present the decay amplitude, as well as expressions for the decay rates, and forward-backward asymmetries. Additionally, we present the form of the lepton polarization asymmetries in this section. In Section \ref{sec:hqet}, we present the relationships among the form factors expected from HQET, as well as the HQET forms for the BRs, FBAs and LPAs. Numerical results for the form factors, branching ratios, lepton forward-backward asymmetries, and lepton polarization asymmetries are presented in Section \ref{sec:results}. We present our conclusions and outlook in Section \ref{sec:concl}. Some details of the calculation are shown in the Appendices.

\section{Decay Rates and Lepton Asymmetries\label{sec:merfba}}

\subsection{Decay Amplitude\label{sec:bme}}

The amplitude for the dileptonic decay of the $\Lb$ baryon is
\begin{equation}
i{\M}(\Lb\to\Ls\ell^+\ell^-)=\frac{G_{F}}{\sqrt{2}}\frac{\alpha_{em}}{2\pi}V_{tb}V^{*}_{ts}(H_{1}^{\mu}L_{\mu}^{(V)}+H_{2}^{\mu}L_{\mu}^{(A)}),
\end{equation}
where, for transitions to states with $J=1/2$,
\beqy
H_1^\mu&=&\bar{u}(\pls,\sls)\bigg[\gm^\mu\bigg(A_1+B_1\gm_5\bigg)+v^\mu\bigg(A_2+B_2\gm_5\bigg)+v'^\mu\bigg(A_3+B_3\gm_5\bigg)\bigg]u(\plb,\slb),\nn\\
\\
H_2^\mu&=&\bar{u}(\pls,\sls)\bigg[\gm^\mu\bigg(D_1+E_1\gm_5\bigg)+v^\mu\bigg(D_2+E_2\gm_5\bigg)+v'^\mu\bigg(D_3+E_3\gm_5\bigg)\bigg]u(\plb,\slb).\nn\\
\eeqy
For transitions to states with $J=3/2$, 
\beqy
H_1^\mu&=&\bar{u}_\alpha(\pls,\sls)\bigg[v^\alpha\bigg(\gm^\mu\bigg(A_1+B_1\gm_5\bigg)+v^\mu\bigg(A_2+B_2\gm_5\bigg)+
v'^\mu\bigg(A_3+B_3\gm_5\bigg)\bigg)\nn\\&& +g^{\alpha\mu}\bigg(A_4+B_4\gm_5\bigg)\bigg]u(\plb,\slb), \\
H_2^\mu&=&\bar{u}_\alpha(\pls,\sls)\bigg[v^\alpha\bigg(\gm^\mu\bigg(D_1+E_1\gm_5\bigg)+v^\mu\bigg(D_2+E_2\gm_5\bigg)+
v'^\mu\bigg(D_3+E_3\gm_5\bigg)\bigg)\nn\\&& +g^{\alpha\mu}\bigg(D_4+E_4\gm_5\bigg)\bigg]u(\plb,\slb),
\eeqy
while for those with $J=5/2$,
\beqy
H_1^\mu&=&\bar{u}_{\alpha\beta}(\pls,\sls)v^\alpha\bigg[v^\beta\bigg(\gm^\mu\bigg(A_1+B_1\gm_5\bigg)+v^\mu\bigg(A_2+B_2\gm_5\bigg)+
v'^\mu\bigg(A_3+B_3\gm_5\bigg)\bigg)\nn\\&& +g^{\beta\mu}\bigg(A_4+B_4\gm_5\bigg)\bigg]u(\plb,\slb), \\
H_2^\mu&=&\bar{u}_{\alpha\beta}(\pls,\sls)v^\alpha\bigg[v^\beta\bigg(\gm^\mu\bigg(D_1+E_1\gm_5\bigg)+v^\mu\bigg(D_2+E_2\gm_5\bigg)+
v'^\mu\bigg(D_3+E_3\gm_5\bigg)\bigg)\nn\\&& +g^{\beta\mu}\bigg(D_4+E_4\gm_5\bigg)\bigg]u(\plb,\slb).
\eeqy
For transitions to states with natural parity spinors,
\beqy
A_i&=&-\frac{2m_b}{q^2}C_7F^{T}_i+C_9F_i,\nn\\
B_i&=&-\frac{2m_b}{q^2}C_7G^{T}_i-C_9G_i,\nn\\
D_i&=&C_{10}F_i, \,\,\,\,\ E_i=-C_{10}G_i,
\eeqy
and for transitions to states with unnatural parity,
\beqy
A_i&=&-\frac{2m_b}{q^2}C_7G^{T}_i-C_9G_i,\nn\\
B_i&=&-\frac{2m_b}{q^2}C_7F^{T}_i+C_9F_i,\nn\\
D_i&=&-C_{10}G_i, \,\,\,\,\ E_i=C_{10}F_i.
\eeqy
The $C_i$ are the Wilson coefficients and the expressions for the $F_i^{(T)}$ and $G_i^{(T)}$ etc., are given in Appendix \ref{sec:hme}.

In our analysis of the dileptonic decays, we include the long distance contributions coming from the charmonium resonances $J/\psi$, $\psi'$, $\ldots$ 
etc. To include these resonant contributions, we replace the Wilson coefficient $C_9$ in Eq. \ref{eq:h1mu} with the effective coefficient
\begin{equation}
C_{9}^{eff}=C_9+Y_{SD}(z,s')+Y_{LD}(s'),
\end{equation}
where $z=m_c/m_b$ and $s'=q^2/m_b^2$. $Y_{SD}$ contains the short distance (SD) contributions from the four-quark operators far from the charmonium 
resonance regions and $Y_{LD}$ are the long distance (LD) contributions from the four-quark operators near the resonances. The SD term can be calculated 
reliably in the perturbative theory; the same cannot be done for the LD contributions. The LD contributions are usually parametrized using a Breit-Wigner 
formalism by making use of vector meson dominance (VMD) and the factorization approximation (FA). The explicit expressions for $Y_{SD}$ and $Y_{LD}$ are 
\cite{buras,aslam,wang,chen,chen1,chen2}
\begin{eqnarray}
Y_{SD}(z,s')&=&(3C_1+C_2+3C_3+C_4+3C_5+C_6)h(z,s')-\nonumber \\&&\frac{1}{2}(4C_3+4C_4+3C_5+C_6)h(1,s')-\frac{1}{2}(C_3+3C_4)h(0,s')+
\nonumber \\&&\frac{2}{9}(3C_3+C_4+3C_5+C_6), \\
Y_{LD}(s')&=&\frac{3\pi}{\alpha_{em}^2}(3C_1+C_2+3C_3+C_4+3C_5+C_6)\nonumber \\&&\times
\sum_{j=J/\psi,\psi',\ldots}\kappa_j\frac{m_j\Gamma(j\rightarrow\ell^+\ell^-)}{q^2-m_j^2+im_j\Gamma_j},
\end{eqnarray}
where $\kappa_j$ are phenomonological parameters introduced to compensate for VMD and FA, $m_j$, $\Gamma_j$, and $\Gamma(j\rightarrow\ell^+\ell^-)$ are the 
masses, total widths, and partial widths of the resonances, respectively, and
\begin{eqnarray}
h(z,s')&=&-\frac{8}{9}\ln z+\frac{8}{27}+\frac{16z^2}{9s'}-\frac{2}{9}\left(2+\frac{4z^2}{s'}\right)\left| 1-\frac{4z^2}{s'}\right|^{1/2}\nonumber 
\\&&\times\bigg\{\Theta\left(1-\frac{4z^2}{s'}\right) \bigg[\ln\left(\frac{1+\sqrt{1-\frac{4z^2}{s'}}}{1-\sqrt{1-\frac{4z^2}{s'}}}\right)-i\pi\bigg] 
+2\Theta\left(\frac{4z^2}{s'}-1\right)\nonumber \\&&\times\arctan\left(\frac{1}{\sqrt{\frac{4z^2}{s'}-1}}\right)
\bigg\}, \nonumber \\
h(0,s')&=&\frac{8}{27}-\frac{4}{9}\ln s'+\frac{4}{9}i\pi.
\end{eqnarray}

\subsection{Decays Rates and Forward-Backward Asymmetries\label{sec:drfba}}

\begin{figure}
\centerline{\includegraphics[width=3.0in]{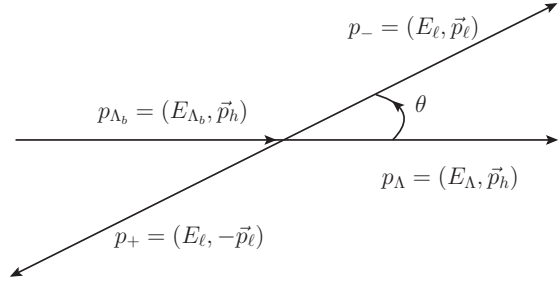}}
\caption{Kinematics of the dilepton rest frame.}\label{fig:dilep}
\end{figure}

The decay rate is
\begin{equation}
d\Gamma=\frac{1}{2m_{\Lambda_b}}\left(\prod_{f}{\frac{d^{3}p_{f}}{(2\pi)^{3}}\frac{1}{2E_{f}}}\right)(2\pi)^{4}
\delta^{(4)}\left(p_{\Lambda_b}-\sum_{f}{p_{f}}\right)|\overline{\cal M}|^2.
\end{equation}
For unpolarized baryons, $|\overline{\cal M}|^2$ is the squared amplitude averaged over the initial polarization and summed over the final 
polarizations,
\beq
|\overline{\cal M}|^2= \frac{G_F^2\alpha_{em}^2}{2^4\pi^2}\left|V_{tb}V_{ts}^{*}\right|^2\left(H_a^{\mu\nu}L_{a\mu\nu}+H_b^{\mu\nu}L_{b\mu\nu}+
H_c^{\mu\nu}L_{c\mu\nu}+H_d^{\mu\nu}L_{d\mu\nu}\right).
\eeq
For unpolarized leptons, the leptonic tensors $L_{f\mu\nu}$ are
\begin{eqnarray}
L_{a\mu\nu}&=&\sum_{spin} L_\mu^{(V)\dag} L_\nu^{(V)}=4\left[p_{+\mu}p_{-\nu}+p_{+\nu}p_{-\mu}-(p_{-}\cdot p_{+}+m_\ell^2)g_{\mu\nu}\right], \\
L_{b\mu\nu}&=&\sum_{spin} L_\mu^{(A)\dag} L_\nu^{(A)}=4\left[p_{+\mu}p_{-\nu}+p_{+\nu}p_{-\mu}-(p_{-}\cdot p_{+}-m_\ell^2)g_{\mu\nu}\right], \\
L_{c\mu\nu}&=&\sum_{spin} L_\mu^{(V)\dag} L_\nu^{(A)}=4i\varepsilon_{\mu\nu\alpha\beta} p_{-}^{\alpha} p_{+}^{\beta}, \\
L_{d\mu\nu}&=&\sum_{spin} L_\mu^{(A)\dag} L_\nu^{(V)}=4i\varepsilon_{\mu\nu\alpha\beta} p_{-}^{\alpha} p_{+}^{\beta}.
\end{eqnarray}
The hadronic tensors $H_f^{\mu\nu}$ are 
\begin{eqnarray}
H_a^{\mu\nu}&=&\sum_{pol} H_1^{\mu\dag} H_1^{\nu},\,\,\,\,\,
H_b^{\mu\nu}=\sum_{pol} H_2^{\mu\dag} H_2^{\nu}, \\
H_c^{\mu\nu}&=&\sum_{pol} H_1^{\mu\dag} H_2^{\nu},\,\,\,\,\,
H_d^{\mu\nu}=\sum_{pol} H_2^{\mu\dag} H_1^{\nu}.
\end{eqnarray}
The most general Lorentz structure for each of these hadronic tensors is
\beq
H_f^{\mu\nu}=-\alpha^f g^{\mu\nu}+\beta^f_{++} Q^\mu Q^\nu+\beta^f_{+-} Q^\mu q^\nu+\beta^f_{-+} q^\mu Q^\nu+\beta^f_{--} q^\mu q^\nu+
i\gamma^f\varepsilon^{\mu\nu\alpha\beta}Q_\alpha q_\beta,
\label{eq:hmunu}
\eeq
where $Q=\plb+\pls$ is the total baryon 4-momentum and $q=\plb-\pls$ is the 4-momentum transfer.

We carry out our calculations in the dilepton rest frame (see Fig. \ref{fig:dilep}). In this frame
\beqy
E_{\Lb}&=&\frac{\mlb}{2\sqrt{\sh}}(1-r+\sh),\,\,\,\, E_{\Ls}=\frac{\mlb}{2\sqrt{\sh}}(1-r-\sh),\nn\\
p_h&=&\frac{\mlb}{2}\sqrt{\frac{\phi(\sh)}{\sh}},\,\,\,\,E_\ell=\frac{\mlb}{2}\sqrt{\sh},\,\,\,\,p_\ell=\frac{\mlb}{2}\sqrt{\sh\psi(\sh)},\nn
\eeqy
where $p_h$ is the magnitude of the 3-momentum of either baryon in this frame. In addition, we have defined
\beqy
\sh&\equiv&q^2/\mlb^2, \,\,\,\, r\equiv m_\Lambda^2/\mlb^2, \,\,\,\, \hat m_\ell\equiv m_\ell/m_{\Lambda_b}, \nonumber\\
\phi(\sh)&=&(1-r)^2-2(1+r)\sh+\sh^2,\,\,\,\,\psi(\sh)=1-4 \hat m_\ell^2/\sh.\nn
\eeqy
Performing the contractions, the differential decay rate becomes
\begin{equation}
\frac{d^2\Gamma}{d\sh d\zh}= \frac{m_{\Lambda_b}G_F^2\alpha_{em}^2}{2^{13}\pi^5}\left|V_{tb}V_{ts}^{*}\right|^2 \sqrt{\phi(\sh)\psi(\sh)} 
{\cal F}_0 (\sh,\zh),
\label{eq:d2gam0}
\end{equation}
where $\zh=\cos\theta$. In these decays, $4\hat m_\ell^2\leq\sh\leq (1-\sqrt{r})^2$ and $-1\leq\zh\leq 1$. The normalized rate ${\cal F}_0 (\sh,\zh)$ has 
the form
\begin{equation}
{\cal F}_0 (\sh,\zh)={\cal I}_0 (\sh)+\zh {\cal I}_1 (\sh)+\zh^2 {\cal I}_2 (\sh),
\label{eq:r0}
\end{equation}
where
\begin{equation}
{\cal I}_0(\sh)=\alpha^a A_\alpha+\beta_{++}^a A_{++}+\alpha^b B_\alpha+\beta_{++}^b B_{++}+\beta_{+-}^b B_{+-}+\beta_{-+}^b B_{-+}+\beta_{--}^b B_{--},
\label{eq:i0}
\end{equation}
and
\begin{eqnarray}
A_\alpha&=&4m_{\Lambda_b}^2 (2\hat m_\ell^2+\sh), \,\,\,\,\
A_{++}=2m_{\Lambda_b}^4 \phi(\sh), \nonumber \\
B_\alpha&=&4m_{\Lambda_b}^2 (\sh-6\hat m_\ell^2), \,\,\,\,\
B_{++}=2m_{\Lambda_b}^4 \left(\phi(\sh)+4\hat m_\ell^2 \left(2(1+r)-\sh\right)\right), \nonumber \\
B_{+-}&=&B_{-+}=8m_{\Lambda_b}^4 \hat m_\ell^2 (1-r), \,\,\,\,\
B_{--}=8m_{\Lambda_b}^4 \hat m_\ell^2 \sh.
\end{eqnarray}
The terms proportional to $\zh$ and $\zh^2$ are
\begin{eqnarray}
{\cal I}_1(\sh)&=&4\mlb^4\sh\sqrt{\phi(\sh)\psi(\sh)}(\gm^c+\gm^d),
\label{eq:i1} \\
{\cal I}_2(\sh)&=&-2\mlb^4\phi(\sh)\psi(\sh)(\beta_{++}^a +\beta_{++}^b),
\label{eq:i2}
\end{eqnarray}
respectively. Integrating over $\zh$ in Eq. \ref{eq:d2gam0} yields
\begin{equation}
\frac{d\Gamma}{d\sh}=\frac{m_{\Lambda_b}G_F^2\alpha_{em}^2}{2^{12}\pi^5}\left|V_{tb}V_{ts}^{*}\right|^2 \sqrt{\phi(\sh)\psi(\sh)}{\cal R}_0(s),
\label{eq:dgam}
\end{equation}
where
\beq
{\cal R}_0(\sh)={\cal I}_0(\sh)+\frac{1}{3}{\cal I}_2(\sh).
\label{eq:r0s}
\eeq
The functions ${\cal I}_0$ and ${\cal I}_2$ are given in Eqs. \ref{eq:i0} and \ref{eq:i2}. The explicit forms of the coefficients $\alpha$, 
$\beta_{\pm\pm}$, and $\gamma$ are given in \cite{morob}.

Another observable of interest is the forward-backward asymmetry (FBA) which is defined as
\begin{equation}
{\cal A}_{FB}(\sh)=\frac{1}{d\Gamma/d\sh}\left[\int_{0}^{1} d\zh\frac{d^2\Gamma}{d\sh d\zh}-\int_{-1}^{0} d\zh\frac{d^2\Gamma}{d\sh d\zh}\right].
\label{eq:afb1}
\end{equation}
Using Eqs. \ref{eq:d2gam0} and \ref{eq:r0}, we find that
\begin{equation}
{\cal A}_{FB}(\sh)=\frac{{\cal I}_1(\sh)}{2\left[{\cal I}_0(\sh)+\frac{1}{3}{\cal I}_2(\sh)\right]}=\frac{\R_{FB}(\sh)}{\R_0(\sh)},
\label{eq:afb2}
\end{equation}
where $\R_{FB}(\sh)={\cal I}_1(\sh)/2$ and $\I_1$ is given in Eq. \ref{eq:i1}.

\subsection{Polarized Leptons\label{sec:pollep}}


Polarization observables such as lepton polarization asymmetries (LPAs) are also sensitive to physics beyond the Standard Model and like FBAs may be less sensitive to the model used to extract the form factors. The spin four-vectors for the leptons are given as
\beq
S^\pm=\bigg(\frac{\vec{p}_\pm\cdot\vec{\xi}^\pm}{m_\ell},\vec{\xi}^\pm+\frac{\vec{p}_\pm\cdot\vec{\xi}^\pm}{m_\ell(E_\ell+m_\ell)}\vec{p}_\pm\bigg),
\eeq
where $\vec{\xi}^\pm$ are the leptons' spins in their rest frames. The unit vectors along the longitudinal, transverse and normal components of polarization are defined to be
\beq
\hat{e}^\pm_L=\frac{\vec{p}_\pm}{|\vec{p}_\pm|}, \,\,\,\,\,
\hat{e}^\pm_T=\frac{\vec{\pls}\times\vec{p}_\pm}{|\vec{\pls}\times\vec{p}_\pm|}, \,\,\,\,\,
\hat{e}^\pm_N=\hat{e}^\pm_L\times\hat{e}^\pm_T.
\eeq
Using the spin projectors $\frac{1}{2}(1+\gm_5\slash{S}^\pm)$, the polarization asymmetries can now be calculated.

When a single lepton is polarized, the differential decay rate becomes
\beq
\frac{d^2\Gamma}{d\sh d\zh}= \frac{m_{\Lambda_b}G_F^2\alpha_{em}^2}{2^{14}\pi^5}\left|V_{tb}V_{ts}^{*}\right|^2 \sqrt{\phi(\sh)\psi(\sh)} 
{\cal F}_1(\sh,\zh),
\label{eq:d2gam1}
\eeq
where ${\cal F}_1(\sh,\zh)$ has the form
\beq
\F_1(\sh,\zh)=\F_0(\sh,\zh)+\vec{\F}^\pm(\sh,\zh)\cdot\spm.
\eeq
$\F_0(\sh,\zh)$ is given in Eq. \ref{eq:r0} and
\beq
\vec{\F}^\pm(\sh,\zh)=\F^\pm_L(\sh,\zh)\epm_L+\F^\pm_T(\sh,\zh)\epm_T+\F^\pm_N(\sh,\zh)\epm_N,
\label{eq:vecr}
\eeq
with
\beq
{\cal F}^\pm_i (\sh,\zh)={\cal J}^\pm_{0i}(\sh)+\sqrt{1-\zh^2}{\cal J}^{\prime\pm}_{0i}(\sh)+\zh {\cal J}^{\pm}_{1i}(\sh)+
\zh\sqrt{1-\zh^2}{\cal J}^{\prime\pm}_{1i}(\sh)+\zh^2{\cal J}^{\pm}_{2i}(\sh).
\label{eq:r1}
\eeq
Integrating Eq. \ref{eq:d2gam1} over $d\zh$, we find that the polarized rate is related to the unpolarized rate by
\beq
\frac{d\Gamma(\vec{\xi}^\pm)}{d\sh}=\frac{1}{2}\left(\frac{d\Gamma}{d\sh}\right)_{0}\left(1+\vec{{\cal P}}^\pm\cdot\vec{\xi}^\pm\right),
\label{eq:dgpol}
\eeq
where $(d\Gamma/ds)_0$ is the unpolarized rate given in Eq. \ref{eq:dgam},
\beq
\vec{{\cal P}}^\pm={\cal P}^\pm_L\hat{e}^\pm_L+{\cal P}^\pm_T\hat{e}^\pm_T+{\cal P}^\pm_N\hat{e}^\pm_N,
\eeq
and the ${\cal P}_x$ are the single lepton polarization asymmetries (SLPAs). The SLPAs can be extracted as
\beq
\Ppm_x(\sh)=\frac{d\Gamma(\xi^\pm_x=+1)-d\Gamma(\xi^\pm_x=-1)}{d\Gamma(\xi^\pm_x=+1)+d\Gamma(\xi^\pm_x=-1)},
\eeq
where $\xi^\pm_x=\epm_x\cdot\spm$ and $x=L,T,N$ for longitudinal, transverse, and normal components, respectively. From Eq. \ref{eq:dgpol}, we find that 
the SLPAs can be written as
\beqy
\Ppm_x(\sh)=\frac{{\cal R}_{x}^{\pm}(\sh)}{{\cal R}_0 (\sh)},
\eeqy
with ${\cal R}_0$ given in Eq. \ref{eq:r0s} and
\beq
{\cal R}_{x}^{\pm}(\sh)={\cal J}_{0 x}^{\pm}(\sh)+\frac{\pi}{4}{\cal J}_{0 x}^{\prime\pm}(\sh)+\frac{1}{3}{\cal J}_{2 x}^{\pm}(\sh).\nn\\
\label{eq:ri}
\eeq
The nonzero $\J$-functions for the longitudinal component are
\beqy
{\cal J}_{0 L}^{\pm}(\sh)&=&\alpha^c C_{\alpha L}^\pm+\beta_{++}^c C_{++ L}^\pm+\alpha^d D_{\alpha L}^\pm+\beta_{++}^d D_{++ L}^\pm,\nn\\
{\cal J}_{2 L}^{\pm}(\sh)&=&\pm 2\mlb^4\phi(\sh)\sqrt{\psi(\sh)}(\beta_{++}^c+\beta_{++}^d),
\label{eq:jl}
\eeqy
where
\beqy
C_{\alpha L}^\pm&=&D_{\alpha L}^\pm=\mp 4\mlb^2\sh\sqrt{\psi(\sh)},\nn\\
C_{++ L}^\pm&=&D_{++ L}^\pm=\mp2\mlb^4\phi(\sh)\sqrt{\psi(\sh)}.
\label{eq:jlcoeff}
\eeqy
Thus, for the longitudinal component, we have
\beq
\R_L^\pm(\sh)=\J_{0L}^\pm(\sh)+\frac{1}{3}\J_{2L}^\pm(\sh).
\eeq
From Eqs. \ref{eq:jl} and \ref{eq:jlcoeff}, we see that $\P_L^-=-\P_L^+$.

The nonzero $\J$-functions for the transverse component are
\beq
{\cal J}_{0 T}^{\prime\pm}(\sh)=\mp 4i\mlh\mlb^4\sqrt{\sh\phi(\sh)\psi(\sh)}(\beta_{+-}^b-\beta_{-+}^b\mp\gm^c\pm\gm^d).
\eeq
It was shown in \cite{morob} that
\beqy
\betpm^b&=&\sum_{j k}\left[\detvvpm_{j k}D_j^{*} D_k+\detaapm_{j k}E_j^{*} E_k\right],\nn\\
\betmp^b&=&\sum_{j k}\left[\detvvmp_{j k}D_j^{*} D_k+\detaamp_{j k}E_j^{*} E_k\right],\nn
\eeqy
where $D_j$ and $E_j$ are given in Eq. \ref{eq:natpar_dilep} for states with natural parity and in Eq. \ref{eq:unnatpar_dilep} states with 
unnatural parity. We note that $\detvvmp_{j k}=\detvvpm_{k j}$ and $\detaamp_{j k}=\detaapm_{k j}$. Since $D_j^{*}D_k$ and $E_j^{*}E_k$ are proportional to $|C_{10}|^2$, it follows that $\betmp^b=\betpm^b$ and
\beq
{\cal J}_{0 T}^{\prime\pm}(\sh)=4i\mlh\mlb^4\sqrt{\sh\phi(\sh)\psi(\sh)}(\gm^c-\gm^d).
\label{eq:j0pt}
\eeq
Thus,
\beq
\R_T^\pm(\sh)=\frac{\pi}{4}\J_{0T}^{\prime\pm}(\sh).
\eeq
From Eq. \ref{eq:j0pt}, we see that $\P_T^+=\P_T^-$.

The nonzero $\J$-functions for the normal component are
\beqy
{\cal J}_{0 N}^{\prime\pm}(\sh)&=&\gamma^a A_{\gamma N}^\pm+\beta_{++}^c C_{++ N}^\pm+\beta_{+-}^c C_{+- N}^\pm+\beta_{++}^d D_{++ N}^\pm+
\beta_{-+}^d D_{-+ N}^\pm,\nn\\
\eeqy
with
\beqy
A_{\gamma N}^\pm&=&\mp 8\mlh\mlb^4\sqrt{\sh\phi(\sh)},\nn\\
C_{++ N}^\pm&=&D_{++ N}^\pm=-4\mlh\mlb^4(1-r)\sqrt{\frac{\phi(\sh)}{\sh}}\nn\\
C_{+- N}^\pm&=&D_{-+ N}^\pm=-4\mlh\mlb^4\sqrt{\sh\phi(\sh)}.
\eeqy
Therefore,
\beq
\R_N^\pm(\sh)=\frac{\pi}{4}\J_{0N}^{\prime\pm}(\sh).
\eeq

It should be noted that the longitudinal component of the lepton polarization is a parity odd and CP-even observable just like the FBAs; however, the transverse component is T-odd. The explicit forms of the various $\alpha$, $\beta_{\pm\pm}$, and $\gamma$ are given in \cite{morob}.

\section{Heavy Quark Effective Theory\label{sec:hqet}}

It is well known that for heavy-to-light transitions, to leading order in heavy quark effective theory (HQET), only two independent form factors are required to parametrize hadronic matrix elements for baryon transitions. In this section, we present the matrix elements of the hadronic currents in terms of the HQET form factors and give the relationships among these form factors and those of the full theory as presented in Appendix \ref{sec:hme}. We also present the expressions for the normalized rates and lepton asymmetries in the heavy quark limit.

\subsection{Form Factor Relations}

\subsubsection{$J=1/2$}

For transitions to any state with $\jp=1/2^+$, the hadronic matrix elements for any current operator $\Gamma$ have the form
\beq
\<\Lstar(\pls)\mid\sbar\Gamma b\mid\Lb(v)\>=\ubar(\pls)(\xz_1+\slash{v}\xz_2)\Gamma u(v),
\eeq
and for transitions to $\jp=1/2^-$,
\beq
\<\Lstar(\pls)\mid\sbar\Gamma b\mid\Lb(v)\>=\ubar(\pls)(\zz_1+\slash{v}\zz_2)\gamma_5\Gamma u(v).
\eeq
Here, the $\xi_i\,\,(\zeta_i)$ are the form factors for transitions to the state with natural (unnatural) parity. Thus, in the limit where the $b$ quark is infinitely heavy, the form factors defined in Appendix \ref{sec:hme} satisfy
\begin{eqnarray}
F_3&=&G_3=H_3=H_4=0, \,\,\,\, F_2=G_2=-H_2=2\xi^{(0)}_2, \nonumber \\ F_1&=&\xi^{(0)}_1-\xi^{(0)}_2, \,\,\,\, G_1=H_1=\xi^{(0)}_1+\xi^{(0)}_2,
\label{eq:hqet12p}
\end{eqnarray}
for transitions to $\jp=1/2^+$, while for transitions to $\jp=1/2^-$, we have
\begin{eqnarray}
F_3&=&G_3=H_3=H_4=0, \,\,\,\, F_2=G_2=H_2=-2\zeta^{(0)}_2, \nonumber \\ F_1&=&-\left[\zeta^{(0)}_1+\zeta^{(0)}_2\right], \,\,\,\,
G_1=-H_1=-\left[\zeta^{(0)}_1-\zeta^{(0)}_2\right].
\label{eq:hqet12m}
\end{eqnarray}

\subsubsection{$J=3/2$}

For transitions to any state with $\jp=3/2^-$, the hadronic matrix elements have the form
\beq
\<\Lstar(\pls)\mid\sbar\Gamma b\mid\Lb(v)\>=\ubar_\alpha(\pls)v^\alpha(\xo_1+\slash{v}\xo_2)\Gamma u(v),
\eeq
and for transitions to $\jp=3/2^+$,
\beq
\<\Lstar(\pls)\mid\sbar\Gamma b\mid\Lb(v)\>=\ubar_\alpha(\pls)v^\alpha(\zo_1+\slash{v}\zo_2)\gamma_5\Gamma u(v).
\eeq
In the heavy quark limit, for transitions to states with $J^P=3/2^{-}$, we find that
\begin{eqnarray}
F_3&=&G_3=H_3=F_4=G_4=H_4=H_5=H_6=0, \,\,\,\, F_2=G_2=-H_2=2\xi^{(1)}_2, \nonumber \\ F_1&=&\xi^{(1)}_1-\xi^{(1)}_2, \,\,\,\,
G_1=H_1=\xi^{(1)}_1+\xi^{(1)}_2,
\label{eq:hqet32m}
\end{eqnarray}
while for transitions to states with $J^P=3/2^{+}$, we get
\begin{eqnarray}
F_3&=&G_3=H_3=F_4=G_4=H_4=H_5=H_6=0, \,\,\,\, F_2=G_2=H_2=-2\zeta^{(1)}_2, \nonumber \\ F_1&=&-\left[\zeta^{(1)}_1+\zeta^{(1)}_2\right], \,\,\,\,
G_1=-H_1=-\left[\zeta^{(1)}_1-\zeta^{(1)}_2\right].
\label{eq:hqet32p}
\end{eqnarray}

\subsubsection{$J=5/2$}

For transitions to any state with $\jp=5/2^+$, the hadronic matrix elements have the form
\beq
\<\Lstar(\pls)\mid\sbar\Gamma b\mid\Lb(v)\>=\ubar_{\alpha\beta}(\pls)v^\alpha v^\beta(\xt_1+\slash{v}\xt_2)\Gamma u(v),
\eeq
In the heavy quark limit, we find that
\begin{eqnarray}
F_3&=&G_3=H_3=F_4=G_4=H_4=H_5=H_6=0, \,\,\,\, F_2=G_2=-H_2=2\xi^{(2)}_2, \nonumber \\ F_1&=&\xi^{(2)}_1-\xi^{(2)}_2, \,\,\,\,
G_1=H_1=\xi^{(2)}_1+\xi^{(2)}_2.
\label{eq:hqet52p}
\end{eqnarray}

\subsection{Decay Rates and Asymmetries}

\label{sec:hqetasym}

For convenience, we denote the HQET form factors $\xi_i^{(n)}$ and $\zeta_i^{(n)}$ generically as $\aleph_i$. In terms of these form factors, the normalized rate (Eq. \ref{eq:r0s}) for transitions to any of the states we consider can be written 
\beq
\R_0(\sh)=(\aleph_1)^2\rho(\sh,r)\D_0(\sh),\label{eq:r12p}
\eeq
where
\beq
\D_0(\sh)=\D_0^{(0)}+\D_0^{(1)}\frac{\aleph_2}{\aleph_1}+\D_0^{(2)}\bigg(\frac{\aleph_2}{\aleph_1}\bigg)^2,\label{eq:d12p}
\eeq
and
\beqy
\D^{(0)}_0(\sh)&=&64\mbh\mlb^4(1-r-\sh)\bigg(1+\frac{2\mlh^2}{\sh}\bigg)\Re(C_9^*C_7)+\frac{64}{3\sh}\mbh^2\mlb^4[2\phi(\sh)+3\sh(1+r-\sh)]\bigg(1+
\frac{2\mlh^2}{\sh}\bigg)|C_7|^2 \nn\\&& +\frac{16}{3}\mlb^4\bigg[\phi(\sh)\bigg(1+\frac{2\mlh^2}{\sh}\bigg)+3\sh(1+r-\sh)\bigg](|C_9|^2+|C_{10}|^2)+
32\mlb^4\mlh^2(1+r-\sh)(|C_9|^2-|C_{10}|^2),\nn\\
\D^{(1)}_0(\sh)&=&128\mbh\mlb^4\sqrt{r}(1-r+\sh)\bigg(1+\frac{2\mlh^2}{\sh}\bigg)\Re(C_9^*C_7)+\frac{256}{3\sh}\mbh^2\mlb^4\sqrt{r}[\phi(\sh)+3\sh]
\bigg(1+\frac{2\mlh^2}{\sh}\bigg)|C_7|^2 \nn\\&& +\frac{32}{3}\mlb^4\sqrt{r}\bigg[\phi(\sh)\bigg(1+\frac{2\mlh^2}{\sh}\bigg)+6\sh\bigg]
(|C_9|^2+|C_{10}|^2)+128\mlb^4\mlh^2\sqrt{r}(|C_9|^2-|C_{10}|^2),\nn\\
\D^{(2)}_0(\sh)&=&-64\mbh\mlb^4\bigg[\phi(\sh)-(1-r-\sh)\bigg]\bigg(1+\frac{2\mlh^2}{\sh}\bigg)\Re(C_9^*C_7)+\frac{64}{3\sh}\mbh^2\mlb^4\bigg[
2(r-\sh)\phi(\sh)+3\sh(1+r-\sh)\bigg] \nn\\&& \times\bigg(1+\frac{2\mlh^2}{\sh}\bigg)|C_7|^2
+\frac{16}{3}\mlb^4\bigg[(r-\sh)\phi(\sh)\bigg(1+\frac{2\mlh^2}{\sh}\bigg)+3\sh(1+r-\sh)\bigg](|C_9|^2+|C_{10}|^2)
\nn\\&& +32\mlb^4\mlh^2(1+r-\sh)(|C_9|^2-|C_ { 10 } |^2).\label{eq:d0}
\eeqy
The funcion $\rho(\sh,r)$ takes the form
\beq
\rho(\sh,r)=\begin{cases} 1,& J^P=\half^\pm;\\[+5pt]
             \frac{\phi(\sh)}{6r},&J^P=\thalf^\pm;\\[+5pt]
              \frac{\phi(\sh)^2}{40r^2},&J^P=\fhalf^+.           
            \end{cases}
\eeq

In the heavy quark limit, the numerators of all of the asymmetries can now be written 
\beq
\R_\X(\sh)=(\aleph_1)^2\rho(\sh,r)\D_\X(\sh) \,\,\,\,\, ({\cal X}=FB,L,T,N),\label{eq:rx}
\eeq
with the asymmetries becoming
\beq
\A_\X(\sh)=\frac{\R_\X}{\R_0}=\frac{\D_\X}{\D_0}.\label{eq:ax}
\eeq
The $\D_\X$ all have the form
\beq
\D_{\cal X}^{\pm}(\sh)=\D^{(0)\pm}_{\cal X}(\sh)+\D^{(1)\pm}_{\cal X}(\sh)\frac{\aleph_2}{\aleph_1}+\D^{(2)\pm}_{\cal X}(\sh)\bigg(\frac{\aleph_2}{\aleph_1}\bigg)^2\label{eq:dx}.
\eeq

For the FBAs,
\beqy
\D^{(0)}_{FB}(\sh)&=&32\mbh\mlb^4\sqrt{\phi(\sh)\psi(\sh)}\Re(C_7^*C_{10})+16\mlb^4\sh\sqrt{\phi(\sh)\psi(\sh)}\Re(C_9^*C_{10}), \nn\\
\D^{(1)}_{FB}(\sh)&=&64\mbh\mlb^4\sqrt{r\phi(\sh)\psi(\sh)}\Re(C_7^*C_{10}), \nn\\
\D^{(2)}_{FB}(\sh)&=&32\mbh\mlb^4\sqrt{\phi(\sh)\psi(\sh)}(r-\sh)\Re(C_7^*C_{10})-16\mlb^4\sh\sqrt{\phi(\sh)\psi(\sh)}\Re(C_9^*C_{10}).\label{eq:dfb}
\eeqy

For the SLPAs, 
\beqy
\D^{(0)\pm}_L(\sh)&=&\mp64\mbh\mlb^4\sqrt{\psi(\sh)}(1-r-\sh)\Re(C_7^*C_{10})\mp\frac{32}{3}\mlb^4\sqrt{\psi(\sh)}
\bigg[\phi(\sh)+3\sh(1+r-\sh)\bigg]\Re(C_9^*C_{10}),\nn\\
\D^{(1)\pm}_L(\sh)&=&\mp128\mbh\mlb^4\sqrt{r\psi(\sh)}(1-r+\sh)\Re(C_7^*C_{10})\mp\frac{64}{3}\mlb^4\sqrt{r\psi(\sh)}
\bigg[\phi(\sh)+6\sh\bigg]\Re(C_9^*C_{10}),\nn\\
\D^{(2)\pm}_L(\sh)&=&\pm64\mbh\mlb^4\sqrt{\psi(\sh)}[\phi(\sh)-(1-r-\sh)]\Re(C_7^*C_{10})\mp\frac{32}{3}\mlb^4\sqrt{\psi(\sh)}\bigg[
(r-\sh)\phi(\sh) \nn\\&& +3\sh(1+r-\sh)\bigg]\Re(C_9^*C_ {10}),\label{eq:dl}
\eeqy
\beqy
\D^{(0)\pm}_T(\sh)&=&-16\pi\mlh\mbh\mlb^4\sqrt{\frac{\phi(\sh)\psi(\sh)}{\sh}}\Im(C_7^*C_{10})-
8\pi\mlh\mlb^4\sqrt{\sh\phi(\sh)\psi(\sh)}\Im(C_9^*C_{10}),\nn\\
\D^{(1)\pm}_T(\sh)&=&-32\pi\mlh\mbh\mlb^4\sqrt{\frac{r\phi(\sh)\psi(\sh)}{\sh}}\Im(C_7^*C_{10}),\nn\\
\D^{(2)\pm}_T(\sh)&=&-16\pi\mlh\mbh\mlb^4(r-\sh)\sqrt{\frac{\phi(\sh)\psi(\sh)}{\sh}}\Im(C_7^*C_{10})+
8\pi\mlh\mlb^4\sqrt{\sh\phi(\sh)\psi(\sh)}\Im(C_9^*C_{10}),\label{eq:dt}
\eeqy
\beqy
\D_N^{(0)\pm}(\sh)&=&\mp32\pi\mbh\mlh\mlb^4\sqrt{\frac{\phi(\sh)}{\sh}}\Re(C_9^*C_7)-16\pi\mbh\mlh\mlb^4\sqrt{\frac{\phi(\sh)}{\sh}}\Re(C_7^*C_{10}) 
\nn\\&& -8\pi\mlh\mlb^4(1-r)\sqrt{\frac{\phi(\sh)}{\sh}}\Re(C_9^*C_{10}) \mp\frac{32\pi}{\sh}\mlh\mbh^2\mlb^4(1-r)\sqrt{\frac{\phi(\sh)}{\sh}}|C_7|^2
\mp8\pi\mlh\mlb^4\sqrt{\sh\phi(\sh)}|C_9|^2, \nn\\
\D_N^{(1)\pm}(\sh)&=&\mp64\pi\mbh\mlh\mlb^4\sqrt{\frac{r\phi(\sh)}{\sh}}\Re(C_9^*C_7)-32\pi\mbh\mlh\mlb^4\sqrt{\frac{r\phi(\sh)}{\sh}}\Re(C_7^*C_{10}) 
\nn\\&& -16\pi\mlh\mlb^4(1-r+\sh)\sqrt{\frac{r\phi(\sh)}{\sh}}\Re(C_9^*C_{10})
\mp\frac{64\pi}{\sh}\mlh\mbh^2\mlb^4(1-r+\sh)\sqrt{\frac{r\phi(\sh)}{\sh}}|C_7|^2, \nn\\
\D_N^{(2)\pm}(\sh)&=&\mp32\pi\mbh\mlh\mlb^4(r-\sh)\sqrt{\frac{\phi(\sh)}{\sh}}\Re(C_9^*C_7)-
16\pi\mbh\mlh\mlb^4(r-\sh)\sqrt{\frac{\phi(\sh)}{\sh}}\Re(C_7^*C_{10}) \nn\\&& 
-8\pi\mlh\mlb^4[r-(r-\sh)^2]\sqrt{\frac{\phi(\sh)}{\sh}}\Re(C_9^*C_{10})\mp
\frac{32\pi}{\sh}\mlh\mbh^2\mlb^4[r-(r-\sh)^2]\sqrt{\frac{\phi(\sh)}{\sh}}|C_7|^2 \nn\\&& \pm8\pi\mlh\mlb^4\sqrt{\sh\phi(\sh)}|C_9|^2.\label{eq:dn}
\eeqy

It is worth emphasizing that the ${\cal D}^{(i)}_{\cal X}\,\, ({\cal X}=0, FB, L, T, N)$ appearing in the differential decay rates and asymmetries are completely independent of the spin and parity of the daughter baryon. Furthermore, the ${\cal D}^{(0)}_{\cal X}$ for each of the observables we discuss is identical with the analogous expression that arises for the free-quark decay, $b\to s\ell^+\ell^-$, in the limits $m_{\Lambda_b} \to m_b$ and $m_\Lambda\to m_s$. These results indicate that the shapes of the FBAs and single lepton polarization asymmetries should be determined predominantly by the Wilson coefficients and kinematics, independent of the quantum numbers of the $\Lambda^*$ produced in the decay. 

\section{Results\label{sec:results}}

\subsection{Form Factors}


For the numerical calculations, we use form factors that result from the SCA and MCN approaches outlined in \cite{morob}. The parameters for the quark model wave functions used in the extraction of the form factors are taken from \cite{roper} and are given in Tables \ref{hampar} and \ref{qmtab}. 

The form factors calculated using the SCA model have the form
\beq
F(\sh)=A_0\exp\left(-\frac{3m_q^2}{2\tilde m_\Lambda^2}\frac{p_\Lambda^2}{\alpha_{\lambda\lambda'}^2}\right),
\label{eq:scaff}
\eeq
where $A_0$ is the value of the form factor at the nonrecoil point, $m_q$ is the mass of each light quark,
$\alpha_{\lambda\lambda'}=\sqrt{(\alpha_{\lambda}^{2}+\alpha_{\lambda'}^{2})/2}$, $\tilde{m}_\Lambda=m_s+2m_q$, and $\pls=\mlb\sqrt{\phi(\sh)}/2$ is the daughter baryon momentum in the $\Lb$ rest frame. The explicit expressions for these form factors are given in \cite{morob}. 

The form factors calculated using the MCN model are parametrized to have the form
\beq
F(\sh)=(a_0+a_2\pls^2+a_4\pls^4)\exp\left(-\frac{3m_q^2}{2\tilde m_\Lambda^2}\frac{p_\Lambda^2}{\alpha_{\lambda\lambda'}^2}\right).
\label{eq:mcnff}
\eeq
The parameters $a_0$, $a_2$, and $a_4$ for the vector and axial vector form factors are given in Table \ref{vaff}. The parameters for the tensor form factors are given in Table \ref{tff}.

\begin{table}[h!]
\caption{Hamiltonian parameters obtained from a fit to a selection of known baryons.}
\label{hampar}
\begin{tabular}{cccccccccc}
\hline\small $m_q$ &\small $m_s$ &\small $m_c$ &\small $m_b$ &\small $b$ &\small $\alf_{coul}$ &\small $\alf_{con}$ &\small $\alf_{SO}$ &\small $\alf_{tens}$ &\small $C_{qqq}$ \\
\small (${\text{GeV}}$) &\small (${\text{GeV}}$) &\small (${\text{GeV}}$) &\small (${\text{GeV}}$) &\small (${\text{GeV}^2}$) & & &\small (${\text{GeV}}$) & &\small ($\text{GeV}$) \\ \hline
\small $0.2848$ &\small $0.5553$ &\small $1.8182$ &\small $5.2019$ &\small $0.1540$ &\small $\approx0.0$ &\small $1.0844$ &\small $0.9321$ &\small $-0.2230$ &\small $-1.4204$ \\ \hline
\end{tabular}
\end{table}

\FloatBarrier

The SCA form factors have all of their kinematic dependence in the form of a Gaussian in the momentum of the daughter baryon (Eq. \ref{eq:scaff}), calculated in the rest frame of the parent, whereas the MCN form factors have an additional multiplicative polynomial dependence in the daughter baryon's momentum (Eq. \ref{eq:mcnff}). Depending on the relative sizes of $a_0$, $a_2$, and $a_4$ in Eq. \ref{eq:mcnff}, the SCA and MCN form factors can have quite different shapes \cite{morob}.

In Section \ref{sec:hqet}, relationships among the form factors of the full theory and those of leading order HQET are presented. It is expected that for transitions to all states that
\beq
F_3=F_4=G_3=G_4=H_3=H_4=H_5=H_6=0.\nn
\eeq
For transitions to states with natural parity it is expected that
\beq
F_2=G_2=-H_2, \,\,\,\, G_1=H_1,\nn
\eeq
while for transitions to states with unnatural parity, 
\beq
F_2=G_2=H_2, \,\,\,\, G_1=-H_1.\nn
\eeq
From the explicit forms of the SCA form factors given in \cite{morob} and the MCN parameters given in Tables \ref{vaff} and \ref{tff}, we see that both sets of form factors satisfy the above relations.

\begin{table}[h!]
\caption{Baryon masses and wave function size parameters, $\alf_\rho$ and $\alf_\lambda$, for states considered in this work. All values are in GeV.}
\label{qmtab}
\begin{tabular}{ccccc}
\hline\small State, $J^{P}$ &\small Experiment &\small Model &\small $\alf_\lambda$  &\small $\alf_\rho$ \\ \hline
\small$\Lambda_b(5620)\,1/2^{+}$ &\small $5.62$ &\small $5.61$ &\small $0.443$ &\small $0.385$  \\ \hline
\small$\Lambda(1115)\,1/2^{+}$   &\small $1.12$ &\small $1.10$ &\small $0.387$ &\small $0.372$  \\ \hline
\small$\Lambda(1600)\,1/2^{+}$   &\small $1.60$ &\small $1.71$ &\small $0.387$ &\small $0.372$  \\ \hline
\small$\Lambda(1405)\,1/2^{-}$   &\small$1.41$  &\small $1.48$ &\small $0.333$ &\small $0.320$  \\ \hline
\small$\Lambda(1520)\,3/2^{-}$   &\small $1.52$ &\small $1.53$ &\small $0.333$ &\small $0.308$  \\ \hline
\small$\Lambda(1890)\,3/2^{+}$   &\small $1.89$ &\small $1.81$ &\small $0.325$ &\small $0.303$  \\ \hline
\small$\Lambda(1820)\,5/2^{+}$   &\small $1.82$ &\small $1.81$ &\small $0.325$ &\small $0.303$  \\ \hline
\end{tabular}
\end{table}

\begin{table}
\caption{Coefficients in the parametrization of the vector and axial-vector form factors obtained in the MCN approach.}\label{vaff}
\begin{tabular}{cccccccccc}
\hline                      & $a_n(\text{GeV}^{-n})$ & $F_1$ & $F_2$ & $F_3$ & $F_4$ & $G_1$ & $G_2$ & $G_3$ & $G_4$ \\ \hline
                            &  $a_0$ & $1.14$       & $-0.137$     & $-0.0614$    & $-$ & $0.923$     & $-0.171$    & $0.0779$    & $-$ \\
$\Lambda_b\to\Lambda(1115)$ &  $a_2$ & $0.0434$     & $-0.00408$   & $0.00249$    & $-$ & $0.0466$    & $0.000477$  & $-0.00162$  & $-$ \\
                            &  $a_4$ & $-0.000879$  & $0.00164$    & $-0.00113$   & $-$ & $-0.000293$ & $0.00121$   & $0.000386$  & $-$ \\ \hline
                            &  $a_0$ & $0.420$      & $-0.328$     & $0.0470$     & $-$ & $0.116$     & $-0.352$    & $-0.0304$   & $-$ \\
$\Lambda_b\to\Lambda(1600)$ &  $a_2$ & $0.433$      & $-0.122$     & $-0.0297$    & $-$ & $0.286$     & $-0.140$    & $0.0424$    & $-$ \\
                            &  $a_4$ & $0.0102$     & $0.00362$    & $-0.000788$  & $-$ & $0.0139$    & $-0.00434$  & $0.000227$  & $-$ \\ \hline
                            &  $a_0$ & $0.245$      & $-0.907$     & $0.112$      & $-$ & $1.09$      & $-0.806$    & $0.00493$   & $-$ \\
$\Lambda_b\to\Lambda(1405)$ &  $a_2$ & $0.130$      & $-0.0980$    & $0.0178$     & $-$ & $0.212$     & $-0.0901$   & $-0.0141$   & $-$ \\
                            &  $a_4$ & $0.00123$    & $0.00674$    & $-0.000442$  & $-$ & $-0.00439$  & $0.00610$   & $0.000269$  & $-$ \\ \hline
                            &  $a_0$ & $-1.30$      & $0.259$      & $0.120$      & $-0.0323$    & $-0.907$     & $0.310$      & $-0.153$     &
$0.0532$ \\
$\Lambda_b\to\Lambda(1520)$ &  $a_2$ & $-0.0828$    & $0.0149$     & $0.00232$    & $-0.00895$   & $-0.0747$    & $0.0130$     & $-0.00269$   &
$-0.00154$ \\
                            &  $a_4$ & $0.00500$    & $-0.00308$   & $0.000161$   & $0.00225$    & $0.00278$    & $-0.00313$   & $0.000130$   &
$-0.000350$ \\ \hline
                            &  $a_0$ & $-0.369$     & $1.31$       & $-0.218$     & $0.0457$     & $-1.60$      & $1.13$       & $0.0212$     &
$-0.0142$ \\
$\Lambda_b\to\Lambda(1890)$ &  $a_2$ & $-0.150$     & $0.0819$     & $-0.0235$    & $0.0128$     & $-0.222$     & $0.0761$     & $0.0195$     &
$-0.00540$ \\
                            &  $a_4$ & $-0.000684$  & $0.000189$   & $-0.0000847$ & $-0.00217$   & $-0.000644$  & $-0.000109$  & $-0.000261$  &
$0.000566$ \\ \hline
                            &  $a_0$ & $1.84$       & $-0.426$     & $-0.190$     & $0.0768$     & $1.18$       & $-0.520$       & $0.270$      &
$-0.114$ \\
$\Lambda_b\to\Lambda(1820)$ &  $a_2$ & $0.0899$     & $-0.00558$   & $-0.00640$   & $0.00385$    & $0.0909$     & $-0.00116$     & $0.00270$     &
$-0.00259$ \\
                            &  $a_4$ & $-0.000260$  & $0.000242$   & $-0.000159$  & $-0.0000581$ & $-0.000923$  & $-0.000167$    & $0.000501$   &
$0.0000289$ \\ \hline
\end{tabular}
\end{table}

\begin{table}
\caption{Coefficients in the parametrization of the tensor form factors obtained in the MCN approach.}\label{tff}
\begin{tabular}{cccccccc}
\hline                      & $a_n(\text{GeV}^{-n})$ & $H_1$ & $H_2$ & $H_3$ & $H_4$ & $H_5$ & $H_6$ \\ \hline
                            &  $a_0$ & $0.930$      & $0.165$      & $-0.0771$    & $-0.0135$    & $-$ & $-$ \\
$\Lambda_b\to\Lambda(1115)$ &  $a_2$ & $0.0193$     & $0.0269$     & $-0.000141$  & $-0.00309$   & $-$ & $-$ \\
                            &  $a_4$ & $-0.00123$   & $-0.00274$   & $0.0000809$  & $0.00110$    & $-$ & $-$ \\ \hline
                            &  $a_0$ & $0.124$      & $0.346$      & $0.0294$     & $-0.00441$   & $-$ & $-$ \\
$\Lambda_b\to\Lambda(1600)$ &  $a_2$ & $0.290$      & $0.136$      & $-0.0425$    & $-0.0191$    & $-$ & $-$ \\
                            &  $a_4$ & $0.00846$    & $0.00109$    & $-0.000222$  & $0.00102$    & $-$ & $-$ \\ \hline
                            &  $a_0$ & $-1.07$      & $-0.805$     & $0.00262$    & $-0.101$     & $-$ & $-$ \\
$\Lambda_b\to\Lambda(1405)$ &  $a_2$ & $-0.213$     & $-0.0920$    & $-0.0141$    & $-0.00522$   & $-$ & $-$ \\
                            &  $a_4$ & $0.00457$    & $0.00629$    & $0.000257$   & $0.000386$   & $-$ & $-$ \\ \hline
                            &  $a_0$ & $-1.02$      & $-0.200$     & $0.153$      & $0.0516$     & $-0.0524$    & $0.0156$ \\
$\Lambda_b\to\Lambda(1520)$ &  $a_2$ & $-0.0360$    & $-0.0516$    & $0.00271$    & $-0.00354$   & $0.00158$    & $0.00218$ \\
                            &  $a_4$ & $0.000522$   & $0.00538$    & $-0.000132$  & $0.000186$   & $-0.000349$  & $-0.00109$ \\ \hline
                            &  $a_0$ & $1.57$       & $1.12$       & $0.0233$     & $0.186$      & $-0.0136$    & $0.0316$ \\
$\Lambda_b\to\Lambda(1890)$ &  $a_2$ & $0.222$      & $0.0768$     & $0.0197$     & $0.00454$    & $-0.00587$   & $0.00775$  \\
                            &  $a_4$ & $0.000636$   & $-0.000109$  & $-0.000271$  & $0.000307$   & $0.000576$   & $-0.00166$ \\ \hline
                            &  $a_0$ & $1.41$       & $0.286$      & $-0.270$     & $-0.0855$    & $0.114$      & $-0.0365$ \\
$\Lambda_b\to\Lambda(1820)$ &  $a_2$ & $0.0484$     & $0.0436$     & $-0.00270$   & $0.00432$    & $0.00261$    & $0.00114$ \\
                            &  $a_4$ & $0.000542$   & $-0.00130$   & $-0.000501$  & $-0.000387$  & $-0.0000296$ & $-0.00000613$ \\ \hline
\end{tabular}
\end{table}

\vskip 0.2in
\begin{figure}
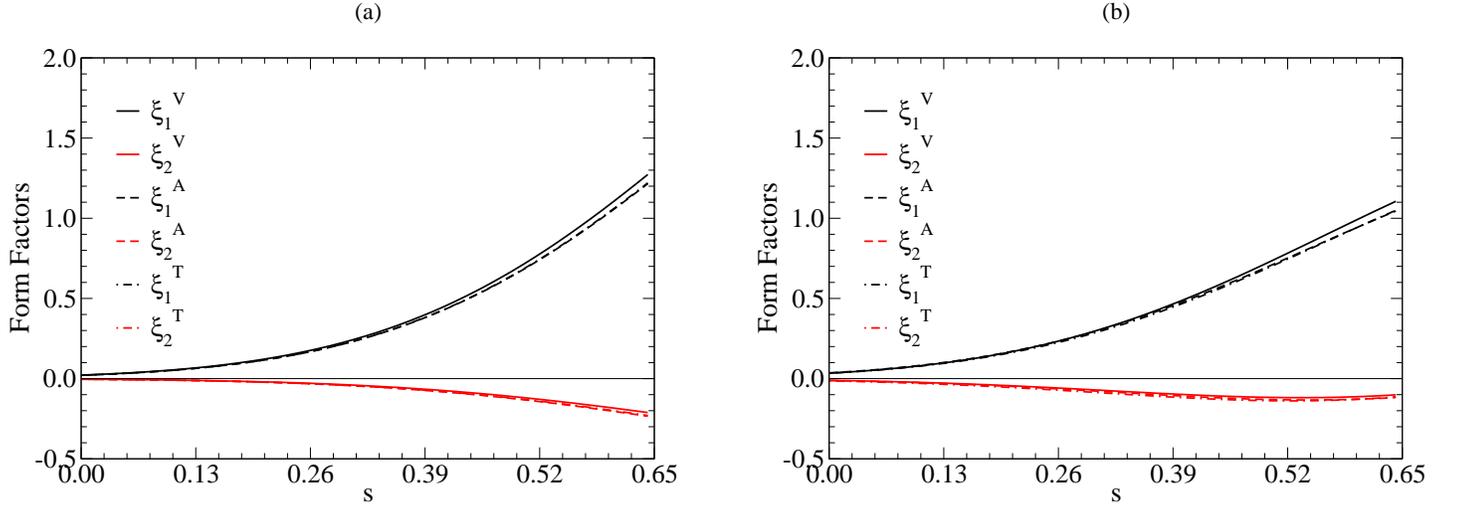

\centerline{\includegraphics[width=3.5in]{hqet_ff_12p_sca.eps}\hskip30pt
\includegraphics[width=3.5in]{hqet_ff_12p_mcn.eps}}
\caption{HQET form factors for $\Lambda_{b}\rightarrow\Lambda(1115),\,\, J^P=1/2^{+}$ as a function of $\hat s=q^{2}/m_{\Lambda_{b}}^{2}$. The graphs
show the HQET form factors calculated using the vector, axial vector, and tensor form factors using (a) SCA and (b) MCN models.}\label{fig:hqet12p}
\end{figure}

\begin{figure}
\centerline{\includegraphics[width=3.5in]{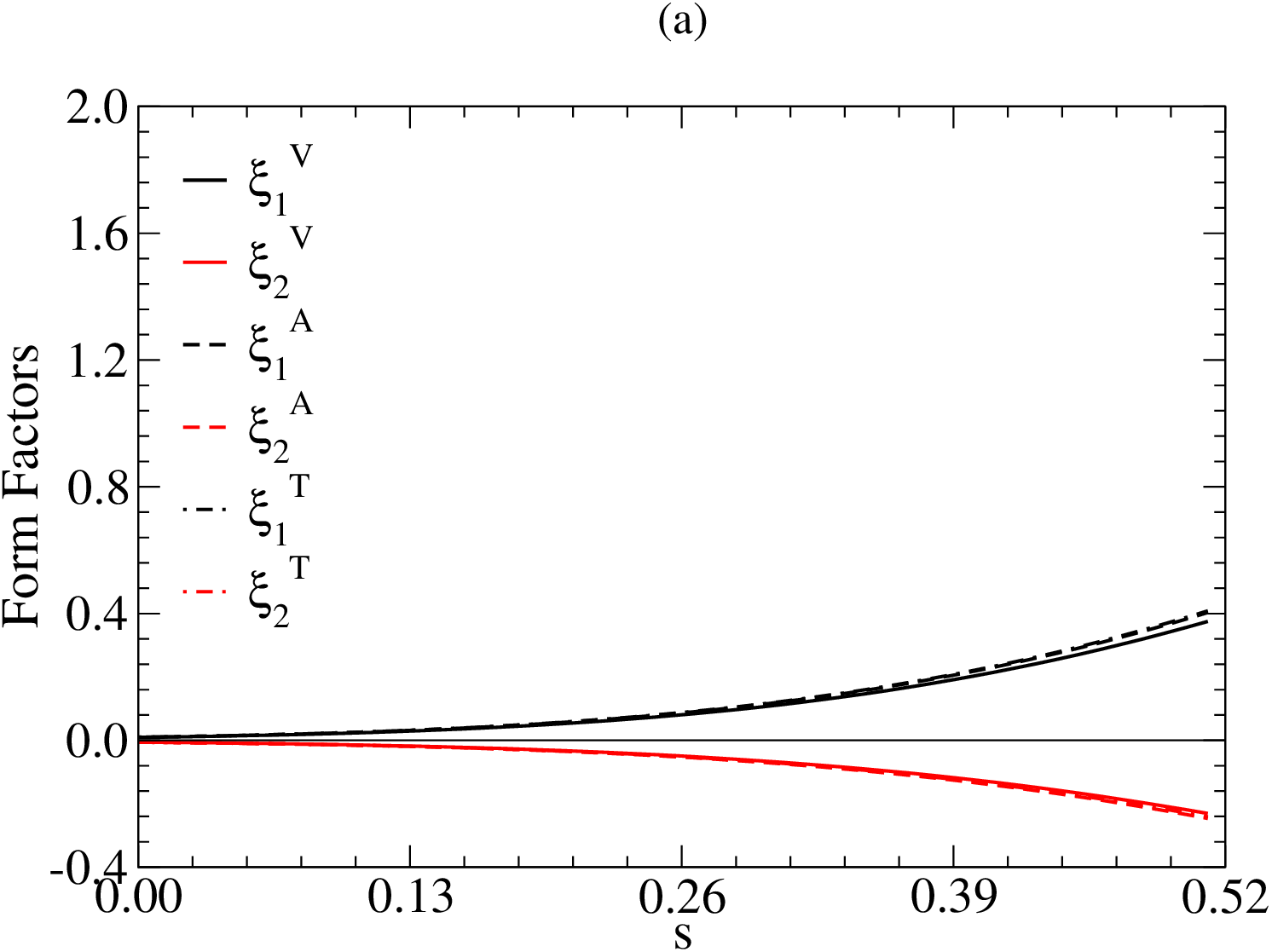}\hskip30pt
\includegraphics[width=3.5in]{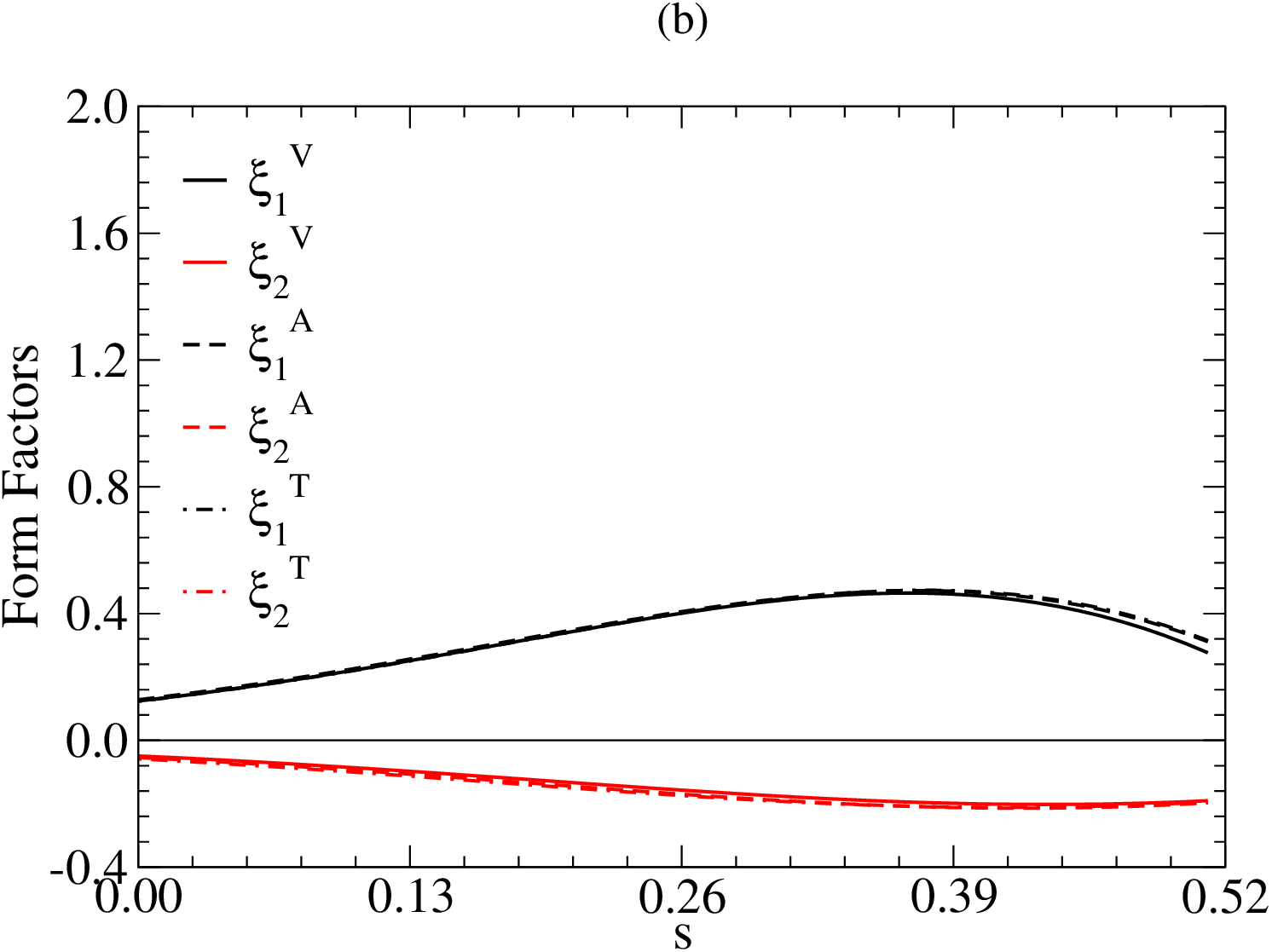}}
\caption{Same as Fig. \ref{fig:hqet12p}, but for $\Lambda_{b}\rightarrow\Lambda(1600),\,\, J^P=1/2^{+}$.}\label{fig:hqet12p1r}
\end{figure}

\begin{figure}[t]
\centerline{\includegraphics[width=3.5in]{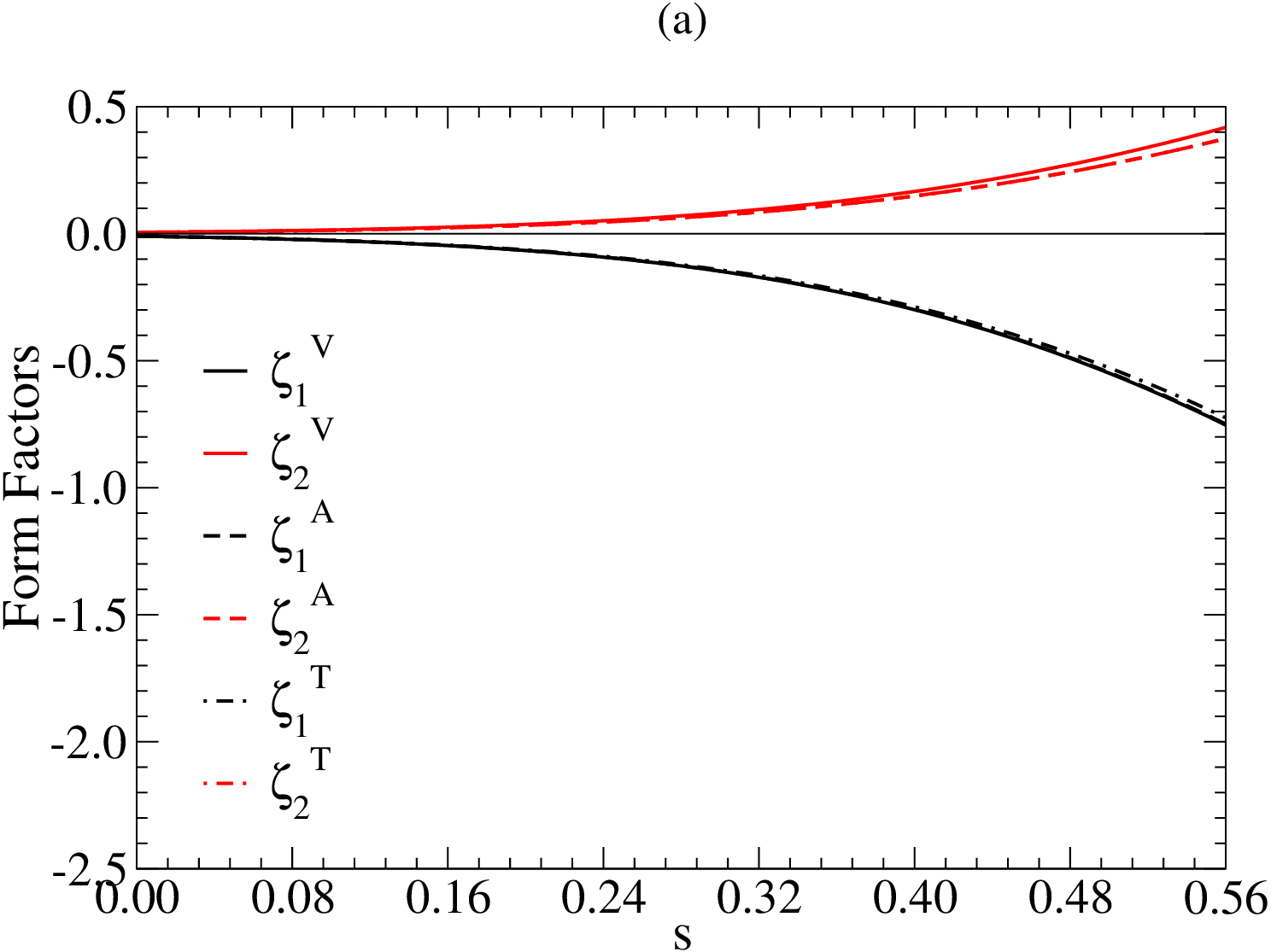}\hskip30pt
\includegraphics[width=3.5in]{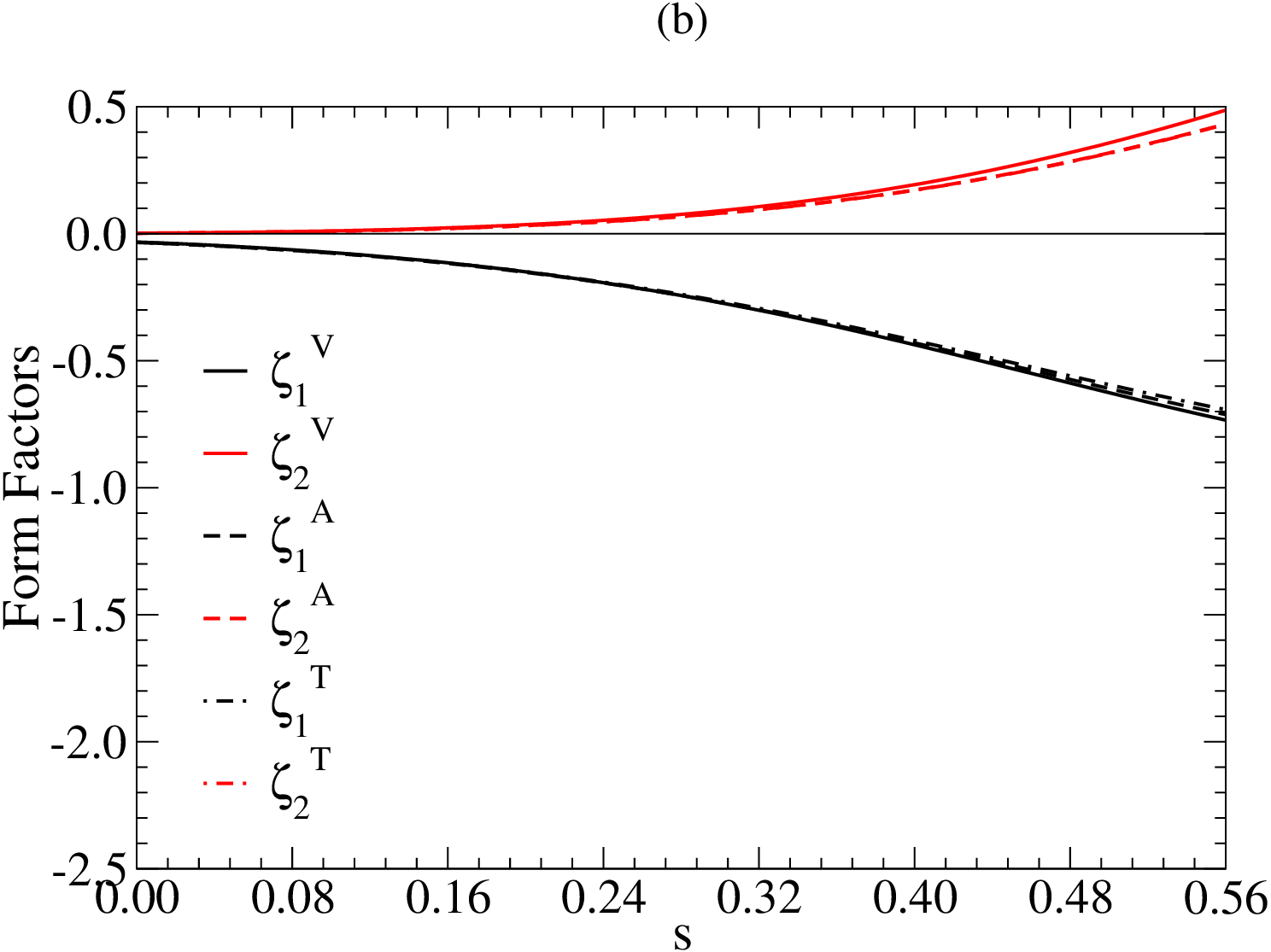}}
\caption{Same as Fig. \ref{fig:hqet12p}, but for $\Lambda_{b}\rightarrow\Lambda(1405),\,\, J^P=1/2^{-}$.}\label{fig:hqet12m}
\end{figure}

\begin{figure}
\centerline{\includegraphics[width=3.5in]{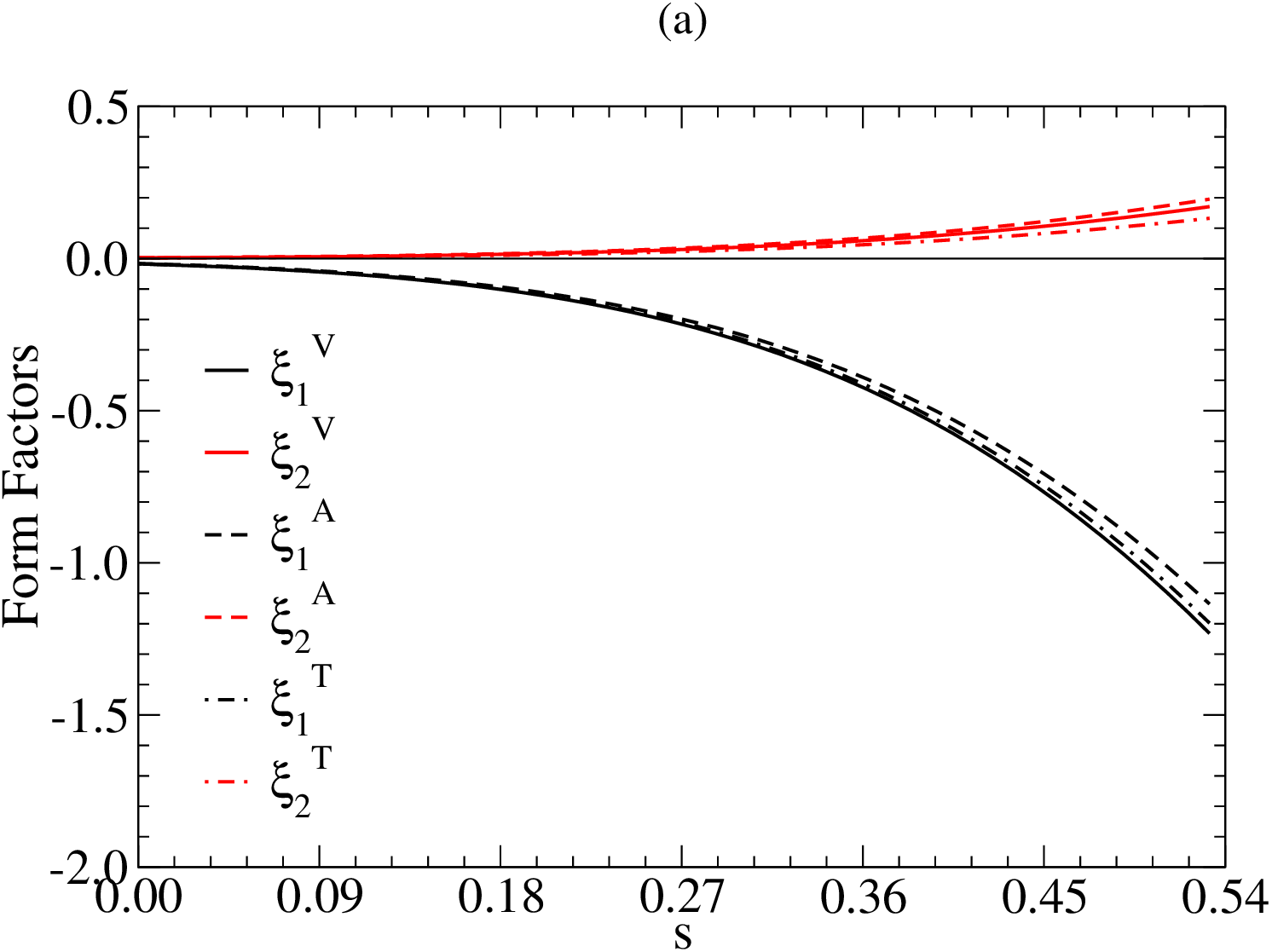}\hskip30pt
\includegraphics[width=3.5in]{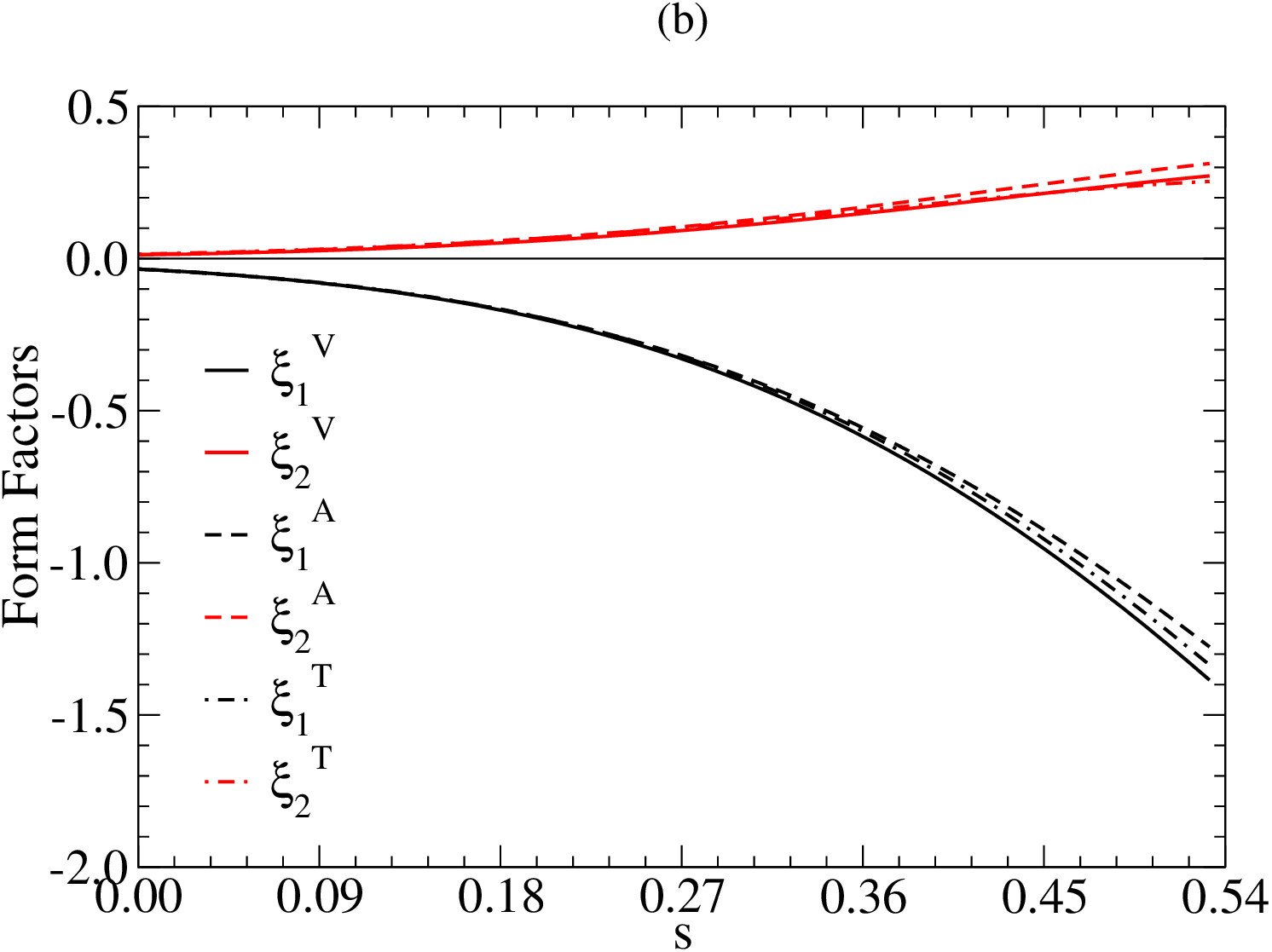}}
\caption{Same as Fig. \ref{fig:hqet12p}, but for $\Lambda_{b}\rightarrow\Lambda(1520),\,\, J^P=3/2^{-}$.}\label{fig:hqet32m}
\end{figure}

\begin{figure}
\centerline{\includegraphics[width=3.5in]{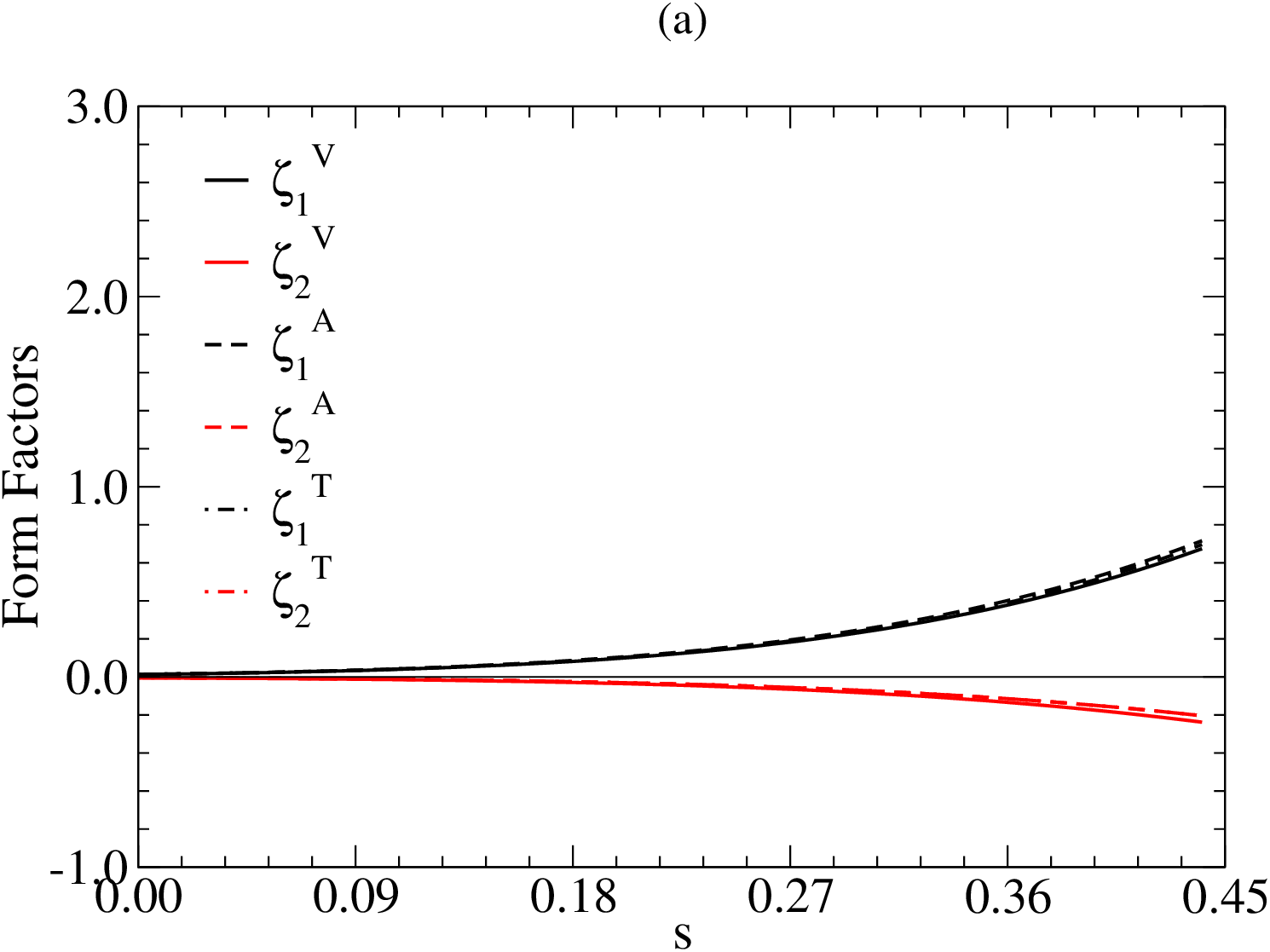}\hskip30pt
\includegraphics[width=3.5in]{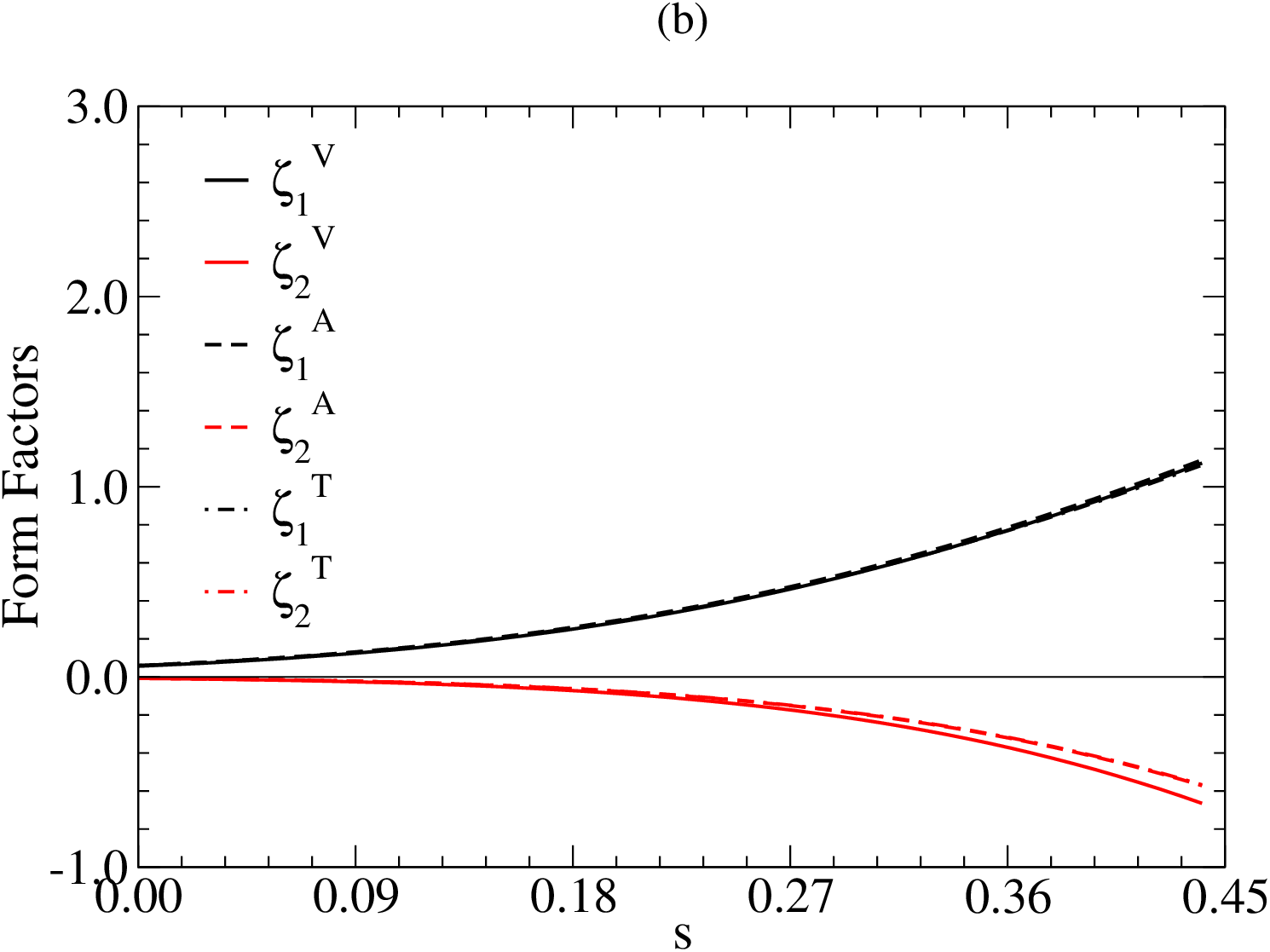}}
\caption{Same as Fig. \ref{fig:hqet12p}, but for $\Lambda_{b}\rightarrow\Lambda(1890),\,\, J^P=3/2^{+}$.}\label{fig:hqet32p}
\end{figure}

\begin{figure}
\centerline{\includegraphics[width=3.5in]{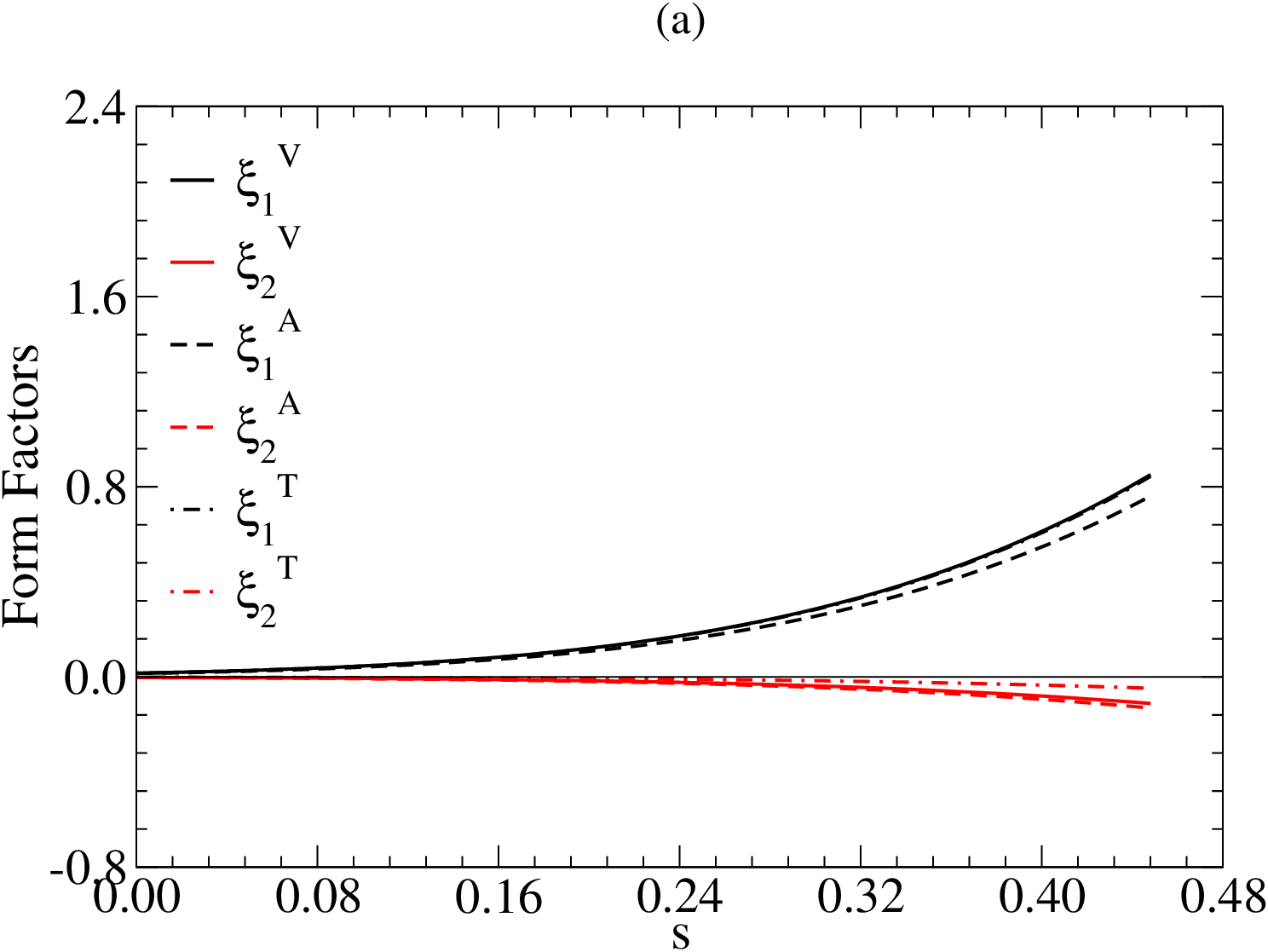}\hskip30pt
\includegraphics[width=3.5in]{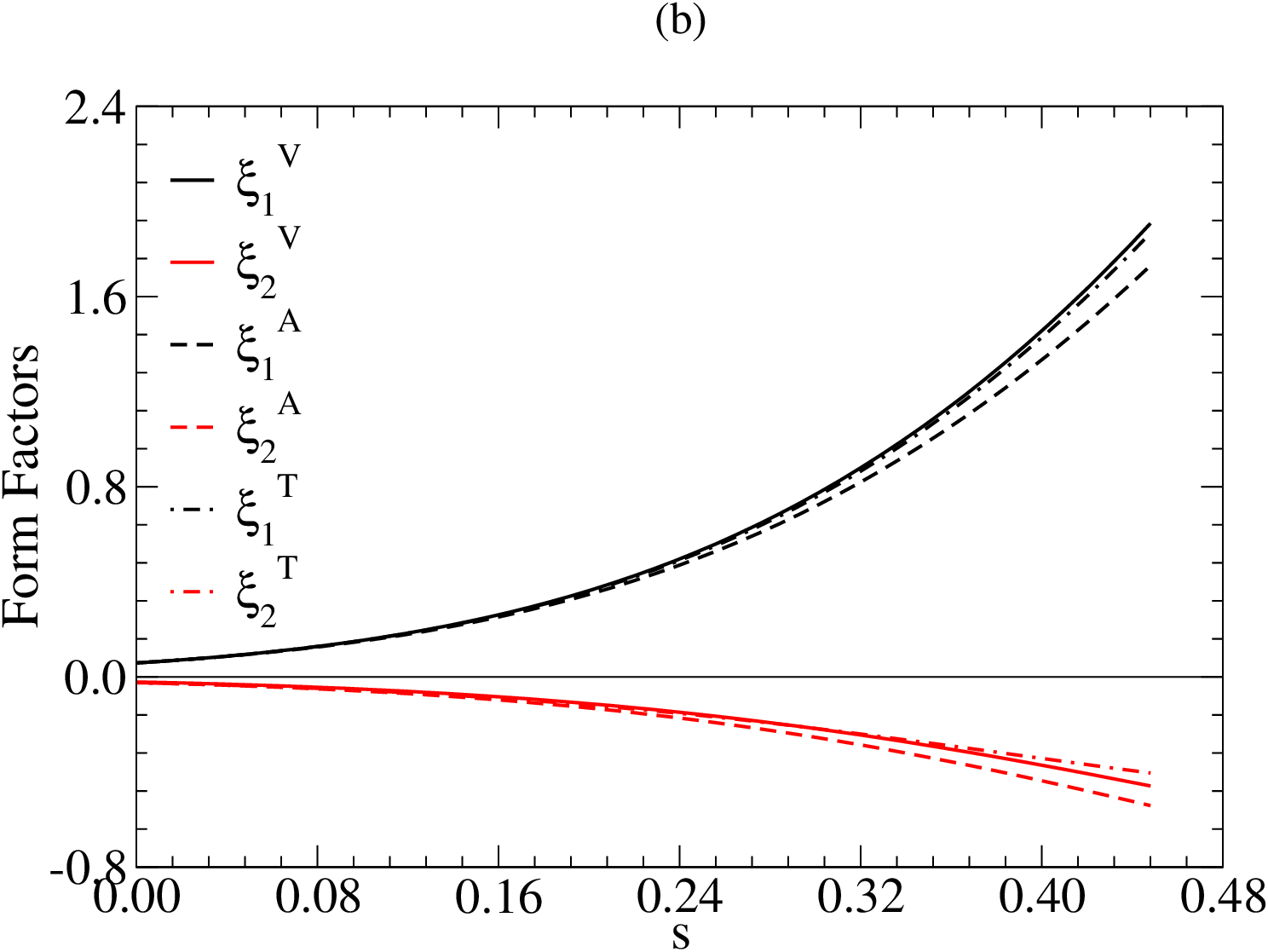}}
\caption{Same as Fig. \ref{fig:hqet12p}, but for $\Lambda_{b}\rightarrow\Lambda(1820),\,\, J^P=5/2^{+}$.}\label{fig:hqet52p}
\end{figure}

The expressions for the leading order HQET predictions given in Section \ref{sec:hqet} can be inverted to give
\beqynn
\xi_1&=&F_1+F_2/2=G_1-G_2/2 =H_1+H_2/2,\\
\xi_2&=&F_2/2=G_2/2=-H_2/2, 
\eeqynn
for transitions to states with natural parity, and
\beqynn
\zeta_1&=&-(F_1-F_2/2)=-(G_1+G_2/2)=H_1-H_2/2,\\
\zeta_2&=&-F_2/2=-G_2/2=-H_2/2,
\eeqynn
for transitions to states with unnatural parity. It is useful to extract the $\xi_i(\zeta_i)$ independently from the vector, axial vector and tensor form factors, and compare the three different forms obtained in this way. Figs. \ref{fig:hqet12p}-\ref{fig:hqet52p} show the three extractions for these form factors from both the SCA and MCN models for all transitions we consider here. As can be seen, the form factors obtained using the axial vector and tensor form factors are virtually identical in both models. The set obtained from the vector form factors is also in very good agreement with the other sets as well. This shows clearly that the form factors obtained from both the SCA and MCN models satisfy the leading order expectations of HQET. However, the curves from the three extractions should not be expected to be identical, since the expressions for the $F_i$, $G_i$ and $H_i$ in terms of universal HQET `Isgur-Wise' type functions will receive corrections due to the finite mass of the $b$ quark.

Additionally, we see that the $\xi_2(\zeta_2)$ fall off much faster than $\xi_1(\zeta_1)$ as $q^2\to0$, and is approximately zero for small $q^2$. This implies that $\xi_1(\zeta_1)$ is dominant in this region and the matrix elements can be described by a single form factor. This is consistent with what is expected from soft collinear effective theory (SCET) \cite{manwang,feldyip}.

\begin{table}
\caption{Ratio of HQET form factors, $\xi_2/\xi_1$ (natural parity) or $\zeta_2/\zeta_1$ (unnatural parity), at the nonrecoil point (maximum $q^2$). The HQET form factors are extracted independently from the vector, axial vector, and tensor form factors of the full theory using both the SCA and MCN models.}\label{hqet_ratio}
\begin{tabular}{c|ccc|ccc|ccc}
\hline                    & & $\xiv_2/\xiv_1$ & & & $\xia_2/\xia_1$ & & & $\xit_2/\xit_1$ &  \\
                            & \small SCA & & \small MCN & \small SCA & & \small MCN & \small SCA & & \small MCN \\ \hline
$\Lambda_b\to\Lambda(1115)$ & $-0.166$ & & $-0.064$ & $-0.193$ & & $-0.085$ & $-0.188$ & & $-0.081$ \\ \hline
$\Lambda_b\to\Lambda(1600)$ & $-0.615$ & & $-0.641$ & $-0.611$ & & $-0.603$ & $-0.590$ & & $-0.582$  \\ \hline
$\Lambda_b\to\Lambda(1405)$ & $-0.555$ & & $-0.623$ & $-0.499$ & & $-0.587$ & $-0.517$ & & $-0.603$  \\ \hline
$\Lambda_b\to\Lambda(1520)$ & $-0.139$ & & $-0.111$ & $-0.172$ & & $-0.146$ & $-0.111$ & & $-0.089$  \\ \hline
$\Lambda_b\to\Lambda(1890)$ & $-0.353$ & & $-0.640$ & $-0.285$ & & $-0.546$ & $-0.293$ & & $-0.554$  \\ \hline
$\Lambda_b\to\Lambda(1820)$ & $-0.130$ & & $-0.131$ & $-0.172$ & & $-0.181$ & $-0.056$ & & $-0.092$  \\ \hline
\end{tabular}
\end{table}

Another expectation of HQET is that the form factors for $\Lb\to\Ls$ should be the same as those for $\Lambda_c\to\Lambda$, up to terms of order $1/m_Q$, where $m_Q$ is the mass of the heavy quark. The CLEO collaboration has extracted the ratio $\xi_2/\xi_1=-0.25\pm0.14\pm0.08$ at the nonrecoil point for the transition $\Lambda_c\to\Ls(1115)$ \cite{cleo3}. Table \ref{hqet_ratio} shows our predictions for this ratio with all three independent extractions from both the SCA and MCN form factors. For the transition to the ground state, our values from both the SCA and MCN models for this ratio are consistent with the value reported by CLEO.

\begin{table}
\caption{Parameters for LCSR, QCDSR, and PM form factors.}
\begin{tabular}{ccc|ccc|ccc}
\hline
 & LCSR & & & QCDSR & & & PM & \\ \hline
$f_2(0)$                   & & $0.15$  & $\xi_1(0)$       & & $0.462$     & $\lqcd(\GeV)$ & & $0.20$ \\
$a_1$                      & & $2.94$  & $a_1(\GeV^{-2})$ & & $-0.0182$   & $N_1$         & & $52.32$ \\
$a_2$                      & & $2.31$  & $a_2(\GeV^{-4})$ & & $-0.000176$ & $N_2$         & & $-13.08$ \\ \hline
$g_2(0)(\GeV^{-1})$        & & $0.013$ & $\xi_2(0)$       & & $-0.077$    & & \\
$a_1$                      & & $2.91$  & $a_1(\GeV^{-2})$ & & $-0.0685$   & & \\
$a_2$                      & & $2.24$  & $a_2(\GeV^{-4})$ & & $0.00146$   & & \\ \hline
\end{tabular}\label{srpmff}
\end{table}

In addition to the SCA and MCN form factors, we also use form factors from light-cone sum rules (LCSR), QCD sum rules (QCDSR), and a multipole model (PM) for transitions to the ground state. In these approaches, only the two universal HQET form factors are used. The LCSR form factors are parametrized to have the form
\beq
f=\frac{f(0)}{1-a_1\sh+a_2\sh^2},
\eeq
where $f(0)$ is the value of the form factor at $q^2=0$. The parameters for the LCSR form factors are taken from \cite{aslam} and are presented in Table \ref{srpmff}. We note here that the form factors $f_2$ and $g_2$ in Table \ref{srpmff} are not the HQET form factors but are related to them by
\beq
\xi_1=f_2-\sqrt{r}\mlb g_2,\,\,\,\,\,\, \xi_2=\mlb g_2.
\eeq
The QCDSR form factors have the form
\beq
\xi=\frac{\xi(0)}{1+a_1q^2+a_2q^4},
\eeq
where $\xi(0)$ is the value of the form factor at $q^2=0$. The parameters for the QCDSR form factors are taken from \cite{chen} and are also presented in Table \ref{srpmff}. The PM form factors are given by
\beq
\xi_i=N_i\bigg(\frac{\lqcd}{\lqcd+z}\bigg)^2,
\eeq
where $z=\pls\cdot\plb/\mlb=\mlb(1+r-\sh)/2$. The parameters for these PM form factors are also taken form \cite{chen} and are given in Table \ref{srpmff} as well.

\subsection{Branching Ratios}


\begin{table}
\caption{SM and SUSY values for the Wilson coefficients. In the SUSY model we use, only $C_7$, $C_9$, and $C_{10}$ get modified from their SM values.}
{\begin{tabular}{cccccccccc}
\hline &\small $C_1$ &\small $C_2$ &\small $C_3$ &\small $C_4$ &\small $C_5$ &\small $C_6$ &\small $C_7$ &\small $C_9$ &\small $C_{10}$ \\ \hline
\small SM &\small $-0.243$ &\small $1.105$ &\small $0.011$ &\small $-0.025$ &\small $0.007$ &\small $-0.031$ &\small $-0.312$ &\small $4.193$ &\small
$-4.578$ \\ \hline
\small SUSY & & & & & & &\small $0.376$ &\small $4.767$ &\small $-3.735$ \\ \hline
\end{tabular}\label{wilco}}
\end{table}

In this section, we present branching ratios (BRs) for dileptonic decays using both sets of form factors. The results we present are obtained using Wilson coefficients that have been calculated in the standard model (SM), with both sets of form factors. We also examine one scenario that arises beyond the SM, namely a supersymmetric (SUSY) extension to the SM, but there we use the MCN form factors exclusively. In our numerical calculations, the SM values of the Wilson coefficients are taken from \cite{huang} and the SUSY values are taken from \cite{aslam}. These values are presented in Table \ref{wilco}. In \cite{huang}, the Wilson coefficients are evaluated using a naive dimensional regularization scheme at the scale $\mu=5.0\text{ GeV}$. The top  quark mass is taken to be $m_t=174\text{ GeV}$ and the cut-off $\Lambda_{\overline{MS}}=225\text{ MeV}$, where $\overline{MS}$ denotes the modified minimal subtraction scheme. The SUSY model used here is referred to as SUSYI in Ref. \cite{aslam}. SUSYI corresponds 
to the regions of parameter space where supersymmetry can destructively contribute and can change the sign of $C_7$, but contributions from neutral Higgs bosons are neglected.

Since the BRs for the $e$ and $\mu$ channels are essentially the same, in what follows, we present the results for the $\mu$ and $\tau$ channels only.

\begin{table}
\caption{Branching ratios for $\Lambda_b\rightarrow \Lambda^{(*)}\mu^{+}\mu^{-}$ in units of $10^{-6}$. The numbers in the column labeled SM1 are obtained using the SCA form factors with standard model Wilson Coefficients. The numbers in the column labeled SM2 are also obtained with SM Wilson coefficients, but using the MCN form factors. The numbers in the column labeled SUSY are obtained using the MCN form factors with Wilson coefficients from a supersymmetric scenario. The column labeled LD refers to the long distance contributions of the charmonium resonances, with `a' indicating that these contributions have been neglected, and `b' indicating that they have been included. In this table, it is assumed that the $\Lambda(1600)$ is the first radial excitation. The lifetime of the $\Lambda_b$ is taken from the Particle Data Listings \cite{pdg}.}
{\begin{tabular}{cccccccccc}
\hline\small State, $J^{P}$       & LD & SM1 &\small SM2 &\small SUSY &\small LCSR  &\small QCDSR &\small PM & CDF \cite{cdf2} & LHCb \cite{lhcb2} \\ \hline
\small$\Lambda(1115)\,1/2^{+}$    & a &\small $0.66$  &\small $0.59$  &\small $0.86$  &\small $6.4$ &\small $2.0$ &\small $1.2$ & $1.73\pm0.42\pm0.55$ & $0.96\pm0.16\pm0.13\pm0.21$ \\
                                  & b &\small $21$    &\small $20$    &\small $20$    &\small $154$  &\small $108$  &\small $60$ & $-$ & $-$\\ \hline
\small$\Lambda(1600)\,1/2^{+}$    & a &\small $0.030$ &\small $0.23$  &\small $0.38$  & $-$ & $-$ & $-$& $-$ & $-$ \\
                                  & b &\small $2.6$   &\small $24$    &\small $24$    & $-$ & $-$ & $-$& $-$ & $-$ \\ \hline
\small$\Lambda(1405)\,1/2^{-}$    & a &\small $0.10$  &\small $0.15$  &\small $0.23$  & $-$ & $-$ & $-$& $-$ & $-$ \\
                                  & b &\small $5.9$   &\small $11$    &\small $11$    & $-$ & $-$ & $-$ & $-$ & $-$\\ \hline
\small$\Lambda(1520)\,3/2^{-}$    & a &\small $0.14$  &\small $0.16$  &\small $0.25$  & $-$ & $-$ & $-$ & $-$ & $-$\\
                                  & b &\small $14$    &\small $16$    &\small $16$    & $-$ & $-$ & $-$& $-$ & $-$ \\ \hline
\small$\Lambda(1890)\,3/2^{+}$    & a &\small $0.020$ &\small $0.057$ &\small $0.099$ & $-$ & $-$ & $-$ & $-$ & $-$\\
                                  & b &\small $2.6$   &\small $6.8$   &\small $6.8$   & $-$ & $-$ & $-$& $-$ & $-$ \\ \hline
\small$\Lambda(1820)\,5/2^{+}$    & a &\small $0.014$ &\small $0.056$ &\small $0.10$  & $-$ & $-$ & $-$ & $-$ & $-$\\
                                  & b &\small $5.6$   &\small $6.3$   &\small $6.4$   & $-$ & $-$ & $-$ & $-$ & $-$\\ \hline
\end{tabular}\label{br1}}
\end{table}

\begin{table}
\caption{Branching ratios for $\Lambda_b\rightarrow \Lambda^{(*)}\tau^{+}\tau^{-}$ in units of $10^{-6}$. The columns are labeled as in Table
\ref{br1}.}
{\begin{tabular}{cccccccc}
\hline\small State, $J^{P}$     & LD &\small SM1 &\small SM2 &\small SUSY &\small LCSR  &\small QCDSR  &\small PM  \\ \hline
\small$\Lambda(1115)\,1/2^{+}$  & a &\small $0.24$  &\small $0.21$  &\small $0.36$  &\small $2.5$ &\small $0.17$ &\small $0.24$ \\
                                & b &\small $0.59$  &\small $0.52$  &\small $0.67$  &\small $5.0$ &\small $0.74$ &\small $0.75$ \\ \hline
\small$\Lambda(1600)\,1/2^{+}$  & a &\small $<0.01$ &\small $<0.01$ &\small $0.015$ & $-$ & $-$ & $-$ \\
                                & b &\small $0.033$ &\small $0.096$ &\small $0.10$  & $-$ & $-$ & $-$ \\ \hline
\small$\Lambda(1405)\,1/2^{-}$  & a &\small $0.026$ &\small $0.025$ &\small $0.046$ & $-$ & $-$ & $-$ \\
                                & b &\small $0.12$  &\small $0.15$  &\small $0.17$  & $-$ & $-$ & $-$ \\ \hline
\small$\Lambda(1520)\,3/2^{-}$  & a &\small $0.015$ &\small $0.014$ &\small $0.028$ & $-$ & $-$ & $-$ \\
                                & b &\small $0.14$  &\small $0.14$  &\small $0.15$  & $-$ & $-$ & $-$ \\ \hline
\small$\Lambda(1890)\,3/2^{+}$  & a &\small $<0.01$ &\small $<0.01$ &\small $<0.01$ & $-$ & $-$ & $-$ \\
                                & b &\small $<0.01$ &\small $<0.01$ &\small $<0.01$ & $-$ & $-$ & $-$ \\ \hline
\small$\Lambda(1820)\,5/2^{+}$  & a &\small $<0.01$ &\small $<0.01$ &\small $<0.01$ & $-$ & $-$ & $-$ \\
                                & b &\small $0.018$ &\small $<0.01$ &\small $<0.01$ & $-$ & $-$ & $-$ \\ \hline
\end{tabular}\label{br2}}
\end{table}

The branching ratios predicted for the $\mu$ and $\tau$ channels are presented in Tables \ref{br1} and \ref{br2}, respectively. Each table displays the results for the SM calculations using two models for the form factors, as well as one SUSY scenario with the MCN form factors. In addition, results obtained omitting and including the long distance (LD) contributions are presented. For ease of discussion, we will refer to the results obtained in the SM as SM1 for the SCA form factors, and SM2 for the MCN form factors. SM1a and SM2a will refer to results with LD contributions omitted, while SM1b and SM2b will refer to results with LD contributions included. Finally, SUSYa and SUSYb will refer to the results obtained using Wilson coefficients from the supersymmetric extension to the standard model discussed above, with the SUSYa (SUSYb) results obtained when LD contributions are omitted (included). For transitions to the ground state, we also compare our model predictions with those made by LCSR, QCDSR, and PM 
form factors using SM Wilson coefficients and Eqs. \ref{eq:r12p}-\ref{eq:d0}, as well as with the recent experimental results from the CDF collaboration \cite{cdf2} and the more recent LHCb collaboration \cite{lhcb2}.

As can be seen in Tables \ref{br1} and \ref{br2}, SM1, SM2, and SUSY predictions for the transitions to $\Ls(1890)$ and $\Ls(1820)$ are very small, and are therefore unlikely to be observed: these decay modes will not be discussed any further in this section.

\subsubsection*{$\it\jp=1/2^+$}

\begin{figure}[t]
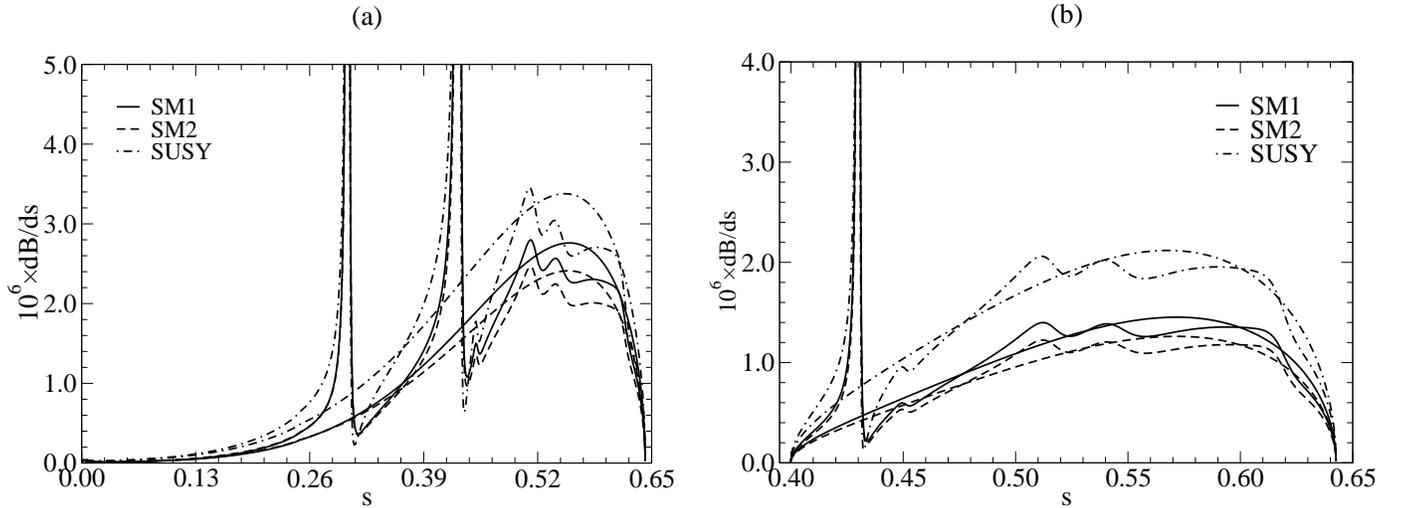

\centerline{\includegraphics[width=3.5in]{mu_dbr_12p.eps}\,\,\,\,\,\,
\includegraphics[width=3.5in]{tau_dbr_12p.eps}}
\caption{$d\B/d\sh$ for (a) $\Lambda_{b}\rightarrow\Lambda(1115)\mu^{+}\mu^{-}$ and (b) $\Lambda_{b}\rightarrow\Lambda(1115)\tau^{+}\tau^{-}$
without and with long distance (LD) contributions from charmonium resonances. The solid curves represent rates obtained from SM1, the dashed curves are from SM2, and the dot-dashed curves are from SUSY.}\label{fig:br12p}
\end{figure}

The differential BRs for the decays to the ground state are shown in Fig. \ref{fig:br12p}. In the graphs, the solid curves are SM1 predictions, the dashed curves are SM2 results, and the dot-dashed curves are from SUSY. The BRs are enhanced by the resonance contributions in both the $\mu$ and $\tau$ channels, but the enhancement is significantly less in the $\tau$ channel, as seen in Tables \ref{br1} and \ref{br2}. In the $\mu$ channel, the BRs we obtain in both SM scenarios and SUSY are smaller than those obtained by the LCSR, QCDSR and PM predictions. However, our SMa results lie just below the margin of error of the CDF measurement but within that of the LHCb measurement. Our SUSYa prediction is within the margin of error of both the CDF and LHCb measurements. It can also be seen in Table \ref{br1} that the LCSR prediction is much larger than both reported experimental values. In the $\tau$ channel, our results in models SM and SUSY are smaller than the LCSR prediction for the ground state, but are 
comparable to the results obtained using QCDSR and PM.

\begin{figure}
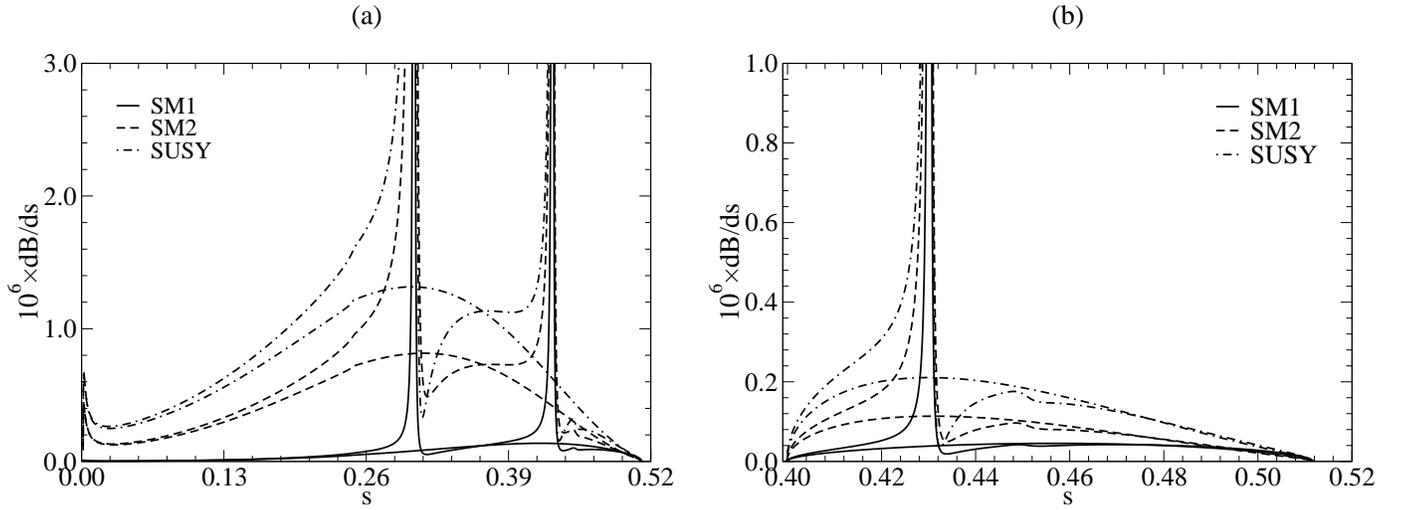

\centerline{\includegraphics[width=3.5in]{mu_dbr_12p1r.eps}\,\,\,\,\,\, \includegraphics[width=3.5in]{tau_dbr_12p1r.eps}}
\caption{Same as Fig. \ref{fig:br12p} but for $\Lambda_{b}\rightarrow\Lambda(1600)$.}\label{fig:br12p1r}
\end{figure}

Fig. \ref{fig:br12p1r} shows the differential BRs for decays to $\Ls(1600)$, the first radial excitation. The SM1 predictions shown in Tables \ref{br1} and \ref{br2} are much smaller than the SM2 predictions for both decay channels. Furthermore, SM2b predicts that decays to this state are the dominant rare decay mode of the $\Lambda_b$.  The truncations of the quark currents and the form factors in SM1 have lead to significant
underestimates of the BRs for decays to this state in both the $\mu$ and $\tau$ channels.

\subsubsection*{$\it\jp=1/2^-$}

\begin{figure}[t]
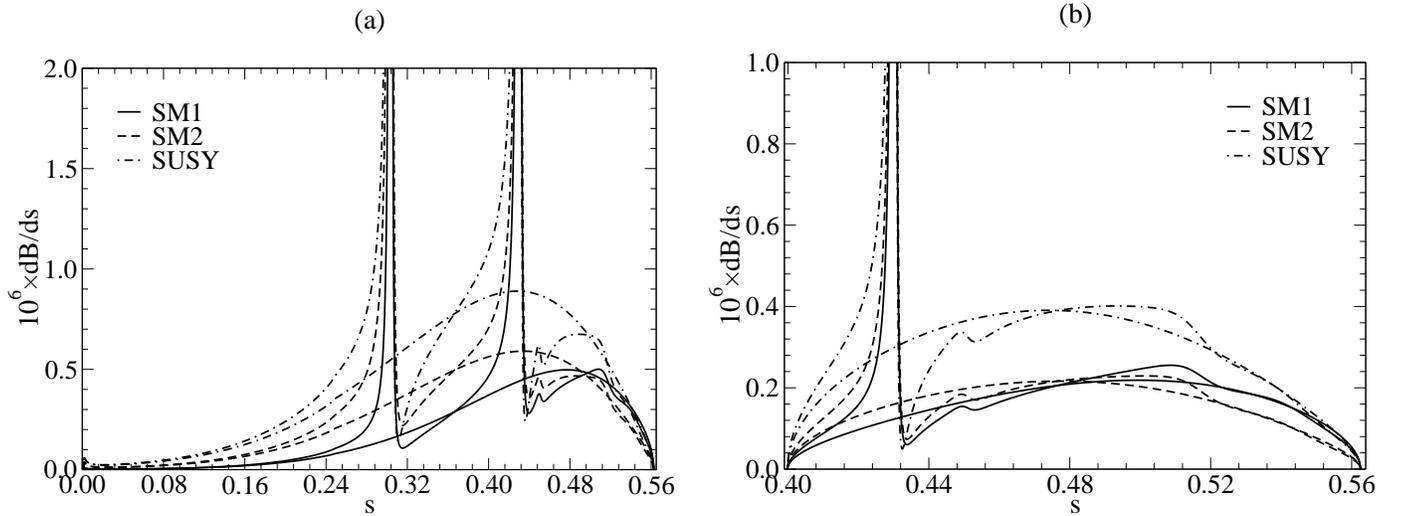

\centerline{\includegraphics[width=3.5in]{mu_dbr_12m.eps}\,\,\,\,\,\, 
\includegraphics[width=3.5in]{tau_dbr_12m.eps}}
\caption{Same as Fig. \ref{fig:br12p} but for $\Lambda_{b}\rightarrow\Lambda(1405)$.}\label{fig:br12m}
\end{figure}

Figs. \ref{fig:br12m}(a) and (b) show the differential BRs to the lowest-lying $1/2^-$ state, the $\Lambda(1405)$, assuming that it is a
three-quark state (there is at least one other suggestion for the structure of this state in the literature \cite{l1405}). From the graphs, it can be seen that SM2 predicts a larger BR than SM1, and this is seen in Tables \ref{br1} and \ref{br2}. The SUSY BRs are significantly larger in both decay channels over the entire kinematic range.

\subsubsection*{$\it\jp=3/2^-$}

\begin{figure}[t]
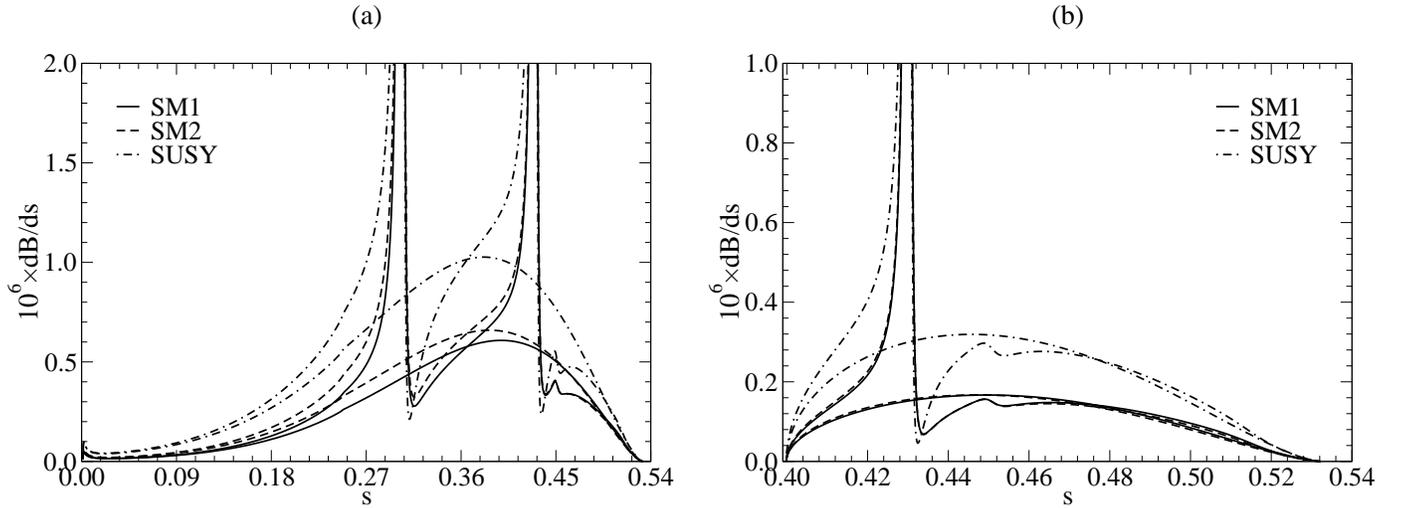

\centerline{\includegraphics[width=3.5in]{mu_dbr_32m.eps}\,\,\,\,\,\,  \includegraphics[width=3.5in]{tau_dbr_32m.eps}}
\caption{Same as Fig. \ref{fig:br12p} but for $\Lambda_{b}\rightarrow\Lambda(1520)$.}\label{fig:br32m}
\end{figure}

Figs. \ref{fig:br32m}(a) and (b) show the differential BRs for decays to $\Lambda(1520)$. The curves indicate that SM1 and SM2 make similar predictions for this channel, and this is borne out by the numbers in Tables \ref{br1} and \ref{br2}. The SUSY rates are larger in both the $\mu$ and $\tau$ channels. For this state, SM2b predicts that the BR into this state is very similar to the BR into the ground state.

We conclude this section by noting that our SM1, SM2, and SUSY predictions agree with the more recent and more precise measurement of the LHCb collaboration. SM1 predicts that decays to the ground state dominate the rare decays of the $\Lambda_b$ in the $\mu$ channel, but SM2b indicates that decays to the first radial excitation are the dominant mode, with the BR for $\Lambda(1520)$ being similar in magnitude to the BR for the ground state. In addition, the BR for the $\Lambda(1405)$ mode is only slightly smaller than the BR for decays the ground state. These results imply that searches for rare decays of the $\Lambda_b$, such as those being carried out by the LHCb collaboration, should include excited final states, as the BRs for these decay modes can be sizable. This is consistent with the current experimental status in the rare decays of the $B$ meson, where decays to the $K$ and $K^*$ account for less than half of the inclusive dileptonic decay rate.

\subsection{Lepton Asymmetries}

\subsubsection{Lepton Forward-Backward Asymmetries}

\begin{table}[t]
\caption{Integrated forward-backward asymmetry for $\Lambda_b\rightarrow \Lambda^{(*)}\mu^{+}\mu^{-}$. The columns are labeled as in Table \ref{br1}.}
{\begin{tabular}{cccccccc}
\hline State, $J^{P}$    & LD & SM1 & SM2 & SUSY & LCSR & QCDSR & PM \\ \hline
$\Lambda(1115)\,1/2^{+}$ & a & $-0.1336$ & $-0.1216$ & $-0.1626$ & $-0.0126$ & $-0.1318$ & $-0.1209$ \\
                         & b & $-0.1127$ & $-0.1026$ & $-0.1486$ & $-0.0091$ & $-0.1108$ & $-0.1011$ \\ \hline
$\Lambda(1600)\,1/2^{+}$ & a & $-0.0729$ & $-0.0888$ & $-0.1296$ & $-$ & $-$ & $-$ \\
                         & b & $-0.0575$ & $-0.0732$ & $-0.1150$ & $-$ & $-$ & $-$ \\ \hline
$\Lambda(1405)\,1/2^{-}$ & a & $-0.1005$ & $-0.1218$ & $-0.1607$ & $-$ & $-$ & $-$ \\
                         & b & $-0.0821$ & $-0.1015$ & $-0.1449$ & $-$ & $-$ & $-$ \\ \hline
$\Lambda(1520)\,3/2^{-}$ & a & $-0.0687$ & $-0.0745$ & $-0.1196$ & $-$ & $-$ & $-$ \\
                         & b & $-0.0566$ & $-0.0619$ & $-0.1063$ & $-$ & $-$ & $-$ \\ \hline
$\Lambda(1890)\,3/2^{+}$ & a & $-0.0475$ & $-0.0714$ & $-0.1118$ & $-$ & $-$ & $-$  \\
                         & b & $-0.0439$ & $-0.0633$ & $-0.0980$ & $-$ & $-$ & $-$ \\ \hline
$\Lambda(1820)\,5/2^{+}$ & a & $-0.0142$ & $-0.0313$ & $-0.0840$ & $-$ & $-$ & $-$  \\
                         & b & $-0.0260$ & $-0.0298$ & $-0.0753$ & $-$ & $-$ & $-$ \\ \hline
\end{tabular}\label{afb1}}
\end{table}

\begin{table}
\caption{Integrated forward-backward asymmetry for $\Lambda_b\rightarrow \Lambda^{(*)}\tau^{+}\tau^{-}$. The columns are labeled as in Table
\ref{br1}.}
{\begin{tabular}{cccccccc}
\hline State, $J^{P}$    & LD & SM1 & SM2 & SUSY & LCSR & QCDSR & PM  \\ \hline
$\Lambda(1115)\,1/2^{+}$ & a & $-0.0337$ & $-0.0297$ & $-0.0241$ & $-0.0066$ & $-0.0391$ & $-0.0335$ \\
                         & b & $-0.0318$ & $-0.0280$ & $-0.0238$ & $-0.0060$ & $-0.0371$ & $-0.0317$  \\ \hline
$\Lambda(1600)\,1/2^{+}$ & a & $-0.0118$ & $-0.0105$ & $-0.0086$ & $-$ & $-$ & $-$ \\
                         & b & $-0.0106$ & $-0.0094$ & $-0.0082$ & $-$ & $-$ & $-$  \\ \hline
$\Lambda(1405)\,1/2^{-}$ & a & $-0.0219$ & $-0.0240$ & $-0.0189$ & $-$ & $-$ & $-$  \\
                         & b & $-0.0203$ & $-0.0222$ & $-0.0183$ & $-$ & $-$ & $-$  \\ \hline
$\Lambda(1520)\,3/2^{-}$ & a & $-0.0072$ & $-0.0076$ & $-0.0067$ & $-$ & $-$ & $-$  \\
                         & b & $-0.0063$ & $-0.0067$ & $-0.0064$ & $-$ & $-$ & $-$  \\ \hline
$\Lambda(1890)\,3/2^{+}$ & a & $-0.0003$ & $-0.0015$ & $-0.0013$ & $-$ & $-$ & $-$  \\
                         & b & $-0.0002$ & $-0.0010$ & $-0.0010$ & $-$ & $-$ & $-$  \\ \hline
$\Lambda(1820)\,5/2^{+}$ & a & $0.0019$  & $0.0013$  & $0.0001$  & $-$ & $-$ & $-$  \\
                         & b & $0.0006$  & $0.0014$  & $0.0002$  & $-$ & $-$ & $-$  \\ \hline
\end{tabular}\label{afb2}}
\end{table}

\begin{table}[t]
\caption{Zeroes of the forward-backward asymmetry for $\Lambda_b\rightarrow \Lambda^{(*)}\mu^{+}\mu^{-}$ without LD contributions. The columns are labeled as in Table \ref{br1}.}
{\begin{tabular}{ccccccc}
\hline State, $J^{P}$    & SM1 & SM2 & SUSY & LCSR & QCDSR & PM  \\ \hline
$\Lambda(1115)\,1/2^{+}$ & $0.097$ & $0.094$ & $-$ & $0.189$ & $0.112$ & $0.112$ \\ \hline
$\Lambda(1600)\,1/2^{+}$ & $0.134$ & $0.094$ & $-$ & $-$ & $-$ & $-$ \\ \hline
$\Lambda(1405)\,1/2^{-}$ & $0.123$ & $0.098$ & $-$ & $-$ & $-$ & $-$ \\ \hline
$\Lambda(1520)\,3/2^{-}$ & $\hphantom{333}0.108\,\,\,\,\,0.522\hphantom{333}$ & $\hphantom{333}0.095\,\,\,\,\,0.524\hphantom{333}$ & $0.527$ & $-$ & $-$ & $-$ \\ \hline
$\Lambda(1890)\,3/2^{+}$ & $\hphantom{333}0.106\,\,\,\,\,0.430\hphantom{333}$ & $\hphantom{333}0.094\,\,\,\,\,0.437\hphantom{333}$ & $0.438$ & $-$ & $-$ & $-$  \\ \hline
$\Lambda(1820)\,5/2^{+}$ & $\hphantom{333}0.127\,\,\,\,\,0.417\hphantom{333}$ & $\hphantom{333}0.105\,\,\,\,\,0.429\hphantom{333}$ & $0.440$ & $-$ & $-$ & $-$  \\ \hline
\end{tabular}\label{s01}}
\end{table}

\begin{table}
\caption{Zeroes of the forward-backward asymmetry for $\Lambda_b\rightarrow \Lambda^{(*)}\tau^{+}\tau^{-}$ without LD contributions. The columns are labeled as in Table \ref{br1}.}
{\begin{tabular}{cccc}
\hline State, $J^{P}$    & SM1 & SM2 & SUSY \\ \hline
$\Lambda(1520)\,3/2^{-}$ & $0.522$ & $0.524$ & $0.527$ \\ \hline
$\Lambda(1890)\,3/2^{+}$ & $0.431$ & $0.437$ & $0.438$ \\ \hline
$\Lambda(1820)\,5/2^{+}$ & $0.416$ & $0.428$ & $0.440$ \\ \hline
\end{tabular}\label{s02}}
\end{table}

The differential forward-backward asymmetries (FBAs) ${\cal A}_{FB}(\sh)$ are shown in  Figs. \ref{fig:asym12p}-\ref{fig:asym32m}. The key to the curves is the same as the differential BRs. In addition to the differential asymmetries, the zeroes in the asymmetries also contain
information on the Wilson coefficients, and are therefore of interest. It is also useful to introduce the integrated forward-backward asymmetry $\<{\cal A}_{FB}\>$ in order to characterize the typical value of the FBA. This integrated FBA is defined as
\beq
\<{\cal A}_{FB}\>=\int_{4\mlh^2}^{(1-\sqrt{r})^2}{\cal A}_{FB}(\sh)d\sh.
\eeq

The integrated FBAs we obtain are shown in Tables \ref{afb1} and \ref{afb2}. The column labels have the same meaning as with the branching ratios shown in Tables \ref{br1} and \ref{br2}. We also compare our model predictions with LCSR, QCDSR and PM predictions for transitions to the ground state using SM Wilson coefficients and Eqs. \ref{eq:d12p}-\ref{eq:dfb}. Tables \ref{s01} and \ref{s02} show the locations of the zeroes in the FBAs without LD contributions.

\subsubsection*{$\it\jp=1/2^+$}

\begin{figure}[t]
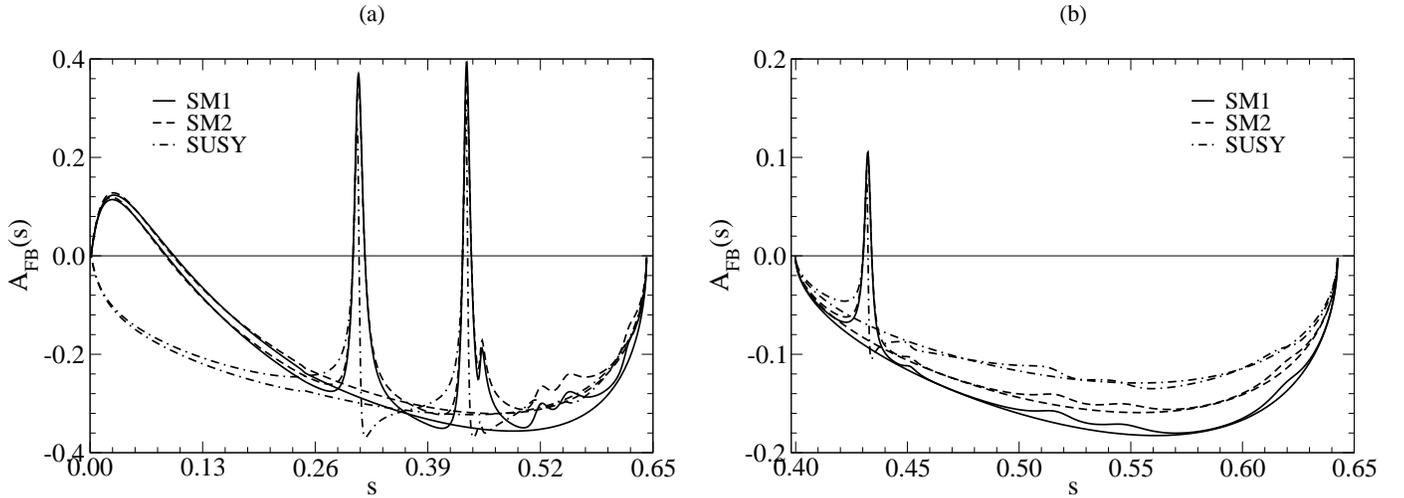

\centerline{\includegraphics[width=3.5in]{mu_fba_12p.eps}\,\,\,\,\,\,
\includegraphics[width=3.5in]{tau_fba_12p.eps}}
\caption{${\cal A}_{FB}(\hat s)$ for (a) $\Lambda_{b}\rightarrow\Lambda(1115)\mu^{+}\mu^{-}$ and (b)
$\Lambda_{b}\rightarrow\Lambda(1115)\tau^{+}\tau^{-}$. The solid curves arise from the SM1 model, the dashed curves from SM2, and the dot-dashed
curves from the SUSY scenario.}\label{fig:asym12p}
\end{figure}

\begin{figure}
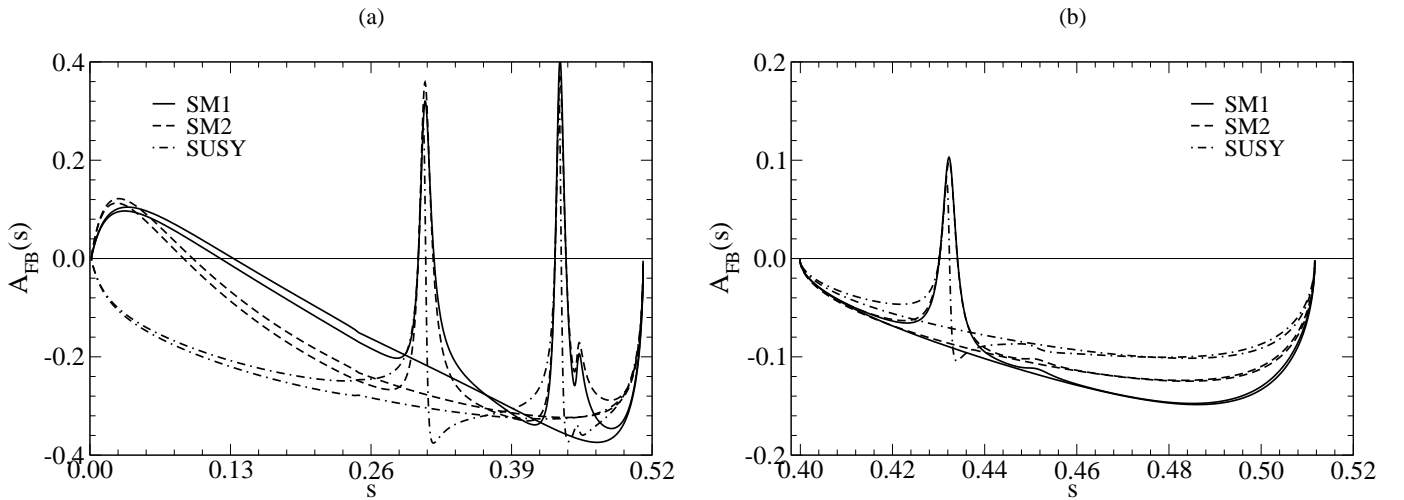

\centerline{\includegraphics[width=3.5in]{mu_fba_12p1r.eps}\,\,\,\,\,\, \includegraphics[width=3.5in]{tau_fba_12p1r.eps}}
\caption{Same as Fig. \ref{fig:asym12p} but for $\Lambda_{b}\rightarrow\Lambda(1600)$.}\label{fig:asym12p1r}
\end{figure}

Figs. \ref{fig:asym12p}(a) and (b) show the differential FBAs for decays to the ground state, in the $\mu$ and $\tau$ channels, respectively. The values of the integrated asymmetries, $\<{\cal A}_{FB}\>$, obtained in the SM1 and SM2  models, agree well with QCDSR and PM predictions for both channels, but is roughly an order of magnitude larger than LCSR predictions. It is very interesting to note that for this final state, the integrated asymmetry appears to be largely independent of the model employed for the form factors. The results of QCDSR and PM are consistent with this observation, but the results of LCSR depart significantly from this. The curves of Figs. \ref{fig:asym12p}(a) and (b) also indicate that the differential FBA for decays to this state are largely independent of the SCA or MCN form factors used.

The FBAs for decays to $\Lambda(1600)$ for the $\mu$ and $\tau$ channels are shown in Figs. \ref{fig:asym12p1r}(a) and (b), respectively. In the $\mu$ channel, the SM1 and SM2 models lead to significantly different FBAs over most of the kinematic range, and this is reflected in the values of the integrated FBAs. In addition, the locations of the zeroes are quite different in the two models. In the $\tau$ channel, the curves from the SM1 and SM2 models are closer than in the $\mu$ channel, and the integrated FBAs are very close.

The condition for the position of the zero(s) for decays to states with $\jp=1/2^+$ is \cite{morob}
\beq
\Re(C_9^{*}C_{10})=\frac{2\mbh}{\sh_0}\Re(C_7^{*}C_{10})\bigg(\frac{F^T_1G_1-G^T_1F_1}{2\mlb F_1G_1}\bigg).
\label{eq:zero12}
\eeq
This relation holds for $\jp=1/2^-$ as well. From this relation, we see that $\sh_0$ depends on the two combinations of Wilson coefficients,
$\Re\left( C_7^{*}C_{10}\right)$ and $\Re \left(C_9^{*}C_{10}\right)$, and a ratio of form factors. For this final state, there is a single possible zero when LD contributions are omitted: LD contributions introduce other zeroes, as can be clearly seen in Figs. \ref{fig:asym12p}(a) and \ref{fig:asym12p1r}(a). Using the form factors obtained in the two models we consider, the values of $\sh_0$ are shown in Table \ref{s01} for the $\mu$ channel. These values are in good agreement with QCDSR and PM predictions. However, the LCSR result is nearly twice as large as our result.

For decays to states with $J=1/2$, apart from the endpoints and resonance regions, there are no zeroes in the FBAs in either channel for the SUSY model we use. The zeroes that occur in the SM scenarios can be traced to the opposite signs of $C_7$ and $C_9$. In the SUSY model that we employ here, $C_7$ and $C_9$ have the same sign, and the condition for the zero in Eq. \ref{eq:zero12} can no longer be satisfied. In addition, in the $\tau$ channel, even in the SM1 and SM2 scenarios, no zeroes are possible, apart from those induced by the resonance effects.

\subsubsection*{$\it\jp=1/2^-$}

\begin{figure}[t]
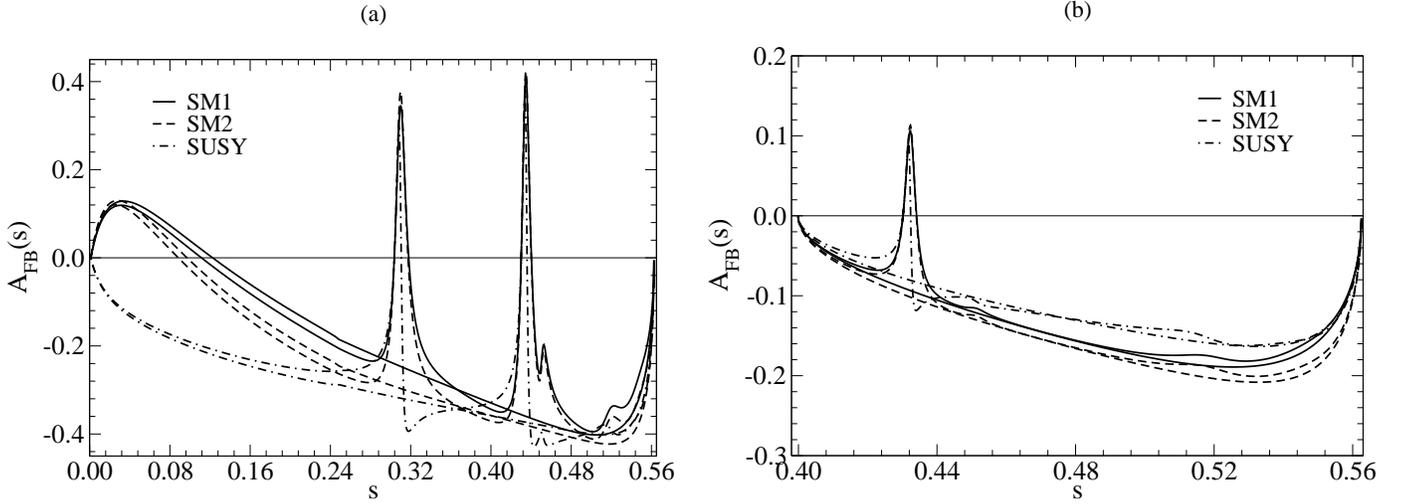

\centerline{\includegraphics[width=3.5in]{mu_fba_12m.eps}\,\,\,\,\,\, 
\includegraphics[width=3.5in]{tau_fba_12m.eps}}
\caption{Same as Fig. \ref{fig:asym12p} but for $\Lambda_{b}\rightarrow\Lambda(1405)$.}\label{fig:asym12m}
\end{figure}

The FBAs predicted for decays to the $\Lambda(1405)$ are shown in Fig. \ref{fig:asym12m}. In these figures, we see that the predictions from SM1 and SM2 are somewhat different, and the locations of the zero are also different (see Table \ref{s01}).  As with the decay to states with $J^P=1/2^+$, there are no zeroes in the FBA in the SUSY scenario we consider here. There are also no zeroes, apart from those induced by the resonance contributions, in the $\tau$ channel for any of the scenarios considered.

\subsubsection*{$\it\jp=3/2^-$}

\begin{figure}
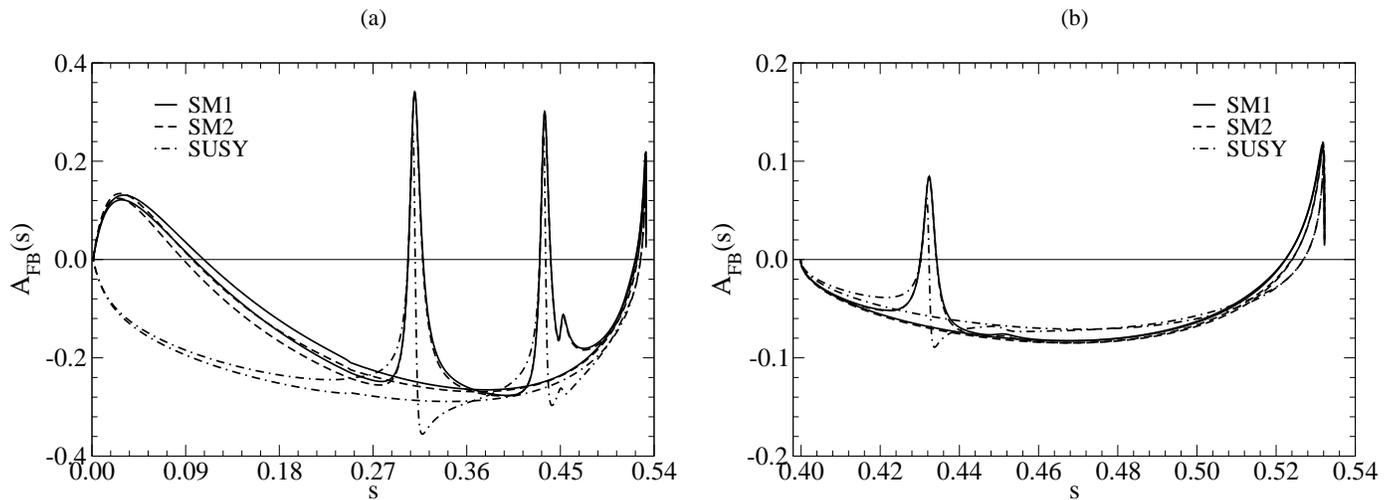

\centerline{\includegraphics[width=3.5in]{mu_fba_32m.eps}\,\,\,\,\,\,  \includegraphics[width=3.5in]{tau_fba_32m.eps}}
\caption{Same as Fig. \ref{fig:asym12p} but for $\Lambda_{b}\rightarrow\Lambda(1520)$.}\label{fig:asym32m}
\end{figure}

The predictions for the FBAs in the decays to the $\Lambda(1520)$ are shown in Fig. \ref{fig:asym32m}. The SM1 and SM2 models give slightly different asymmetries in the $\mu$ channel. Even without the LD contributions, the structure of the asymmetry arising from these decays is richer than in the decays to states with $J=1/2$.

For $J^P=3/2^-$, the condition for the positions of the zeroes is \cite{morob}
\beq
\Re(C_9^{*}C_{10})=\frac{2\mbh}{\sh_0}\Re(C_7^{*}C_{10})\bigg(\frac{X_1}{2\mlb Y_1}\bigg),
\label{eq:zero32}
\eeq
where
\beqy
X_1\=-\frac{\phi(\sh_0)}{r}\left(F^T_1G_1-G^T_1F_1\right) +\ym(\sh_0)\left(F^T_1G_4-G^T_4F_1\right) +\nn\\&&\yp(\sh_0)\left(F^T_4G_1-G^T_1F_4\right)
+2\left(F^T_4G_4-G^T_4F_4\right),\nn\\
Y_1\=-\frac{\phi(\sh_0)}{r}F_1G_1+\ym(\sh_0) F_1G_4+\yp(\sh_0) F_4G_1+2F_4G_4.\label{eq:zero32m}
\eeqy
Here $\phi(\sh)=(1-r)^2-2(1+r)\sh+\sh^2$ and $y^\pm(\sh)=\left[(1\pm\sqrt{r})^2-\sh\right]/\sqrt{r}$. Similar expressions can be found for $\jp=3/2^{+},5/2^{+}$. \cite{morob} As we can see in Fig. \ref{fig:asym32m}, apart from the resonance region, there are two zeroes for this mode in the $\mu$ channel in the SM. The zero at the larger value of $\sh$ is also present in the $\tau$ channel. This is quite different from the case with $J=1/2$ where there is only one zero in the $\mu$ channel and none for the $\tau$. The positions of the zeroes in the $\mu$ channel are shown in Table \ref{s01}. In the SUSY scenario that we explore, there is only one zero in the $\mu$ channel, and it sits at $\sh=0.527$. In the $\tau$ channel, there is a single zero in the FBA, and its position is largely independent of the model used for the form factors.

The zeroes for the FBAs in decays to states with spin 1/2, along with the first zero in decays to the states with higher spin that we have considered, all lie relatively close to each other, despite the more complicated expression for the location of the zeroes for the states with higher spin. However, in Eq. \ref{eq:zero32m} terms in $F_4$, $G_4$, $F_4^T$ and $G_4^T$ are negligible (they are exactly zero in the limit of an infinitely heavy $b$ quark), so that the expression for the location of the zeroes reduces to one that is identical to the case of spin 1/2, Eq. \ref{eq:zero12}. Furthermore
\beq
\frac{F^T_1G_1-G^T_1F_1}{2\mlb F_1G_1}\approx 1+{\cal O}\left(\frac{\xi_2}{\xi_1}\right)+{\cal O}\left(\frac{\Lambda_{\rm QCD}}{m_b}\right),
\eeq
for states with natural parity, or
\beq
\frac{F^T_1G_1-G^T_1F_1}{2\mlb F_1G_1}\approx 1+{\cal O}\left(\frac{\zeta_2}{\zeta_1}\right)+{\cal O}\left(\frac{\Lambda_{\rm QCD}}{m_b}\right),
\eeq
for states with unnatural parity, for the states we have examined. This means that the location of the zero, up to corrections ${\cal
O}\left(\frac{\Lambda_{\rm QCD}}{m_b}\right)$ and ${\cal O}\left(\frac{\xi_2}{\xi_1}\right)$ or ${\cal O}\left(\frac{\zeta_2}{\zeta_1}\right)$, is approximately given by
\beq
\sh_0\approx -2\mbh\frac{\Re(C_7^{*}C_{10})}{\Re(C_9^{*}C_{10})},\label{eq:global0}
\eeq
independent of the angular momentum of the final state (at least, up to spin $5/2$), and of the form factors. Using the SM Wilson coefficients along with the physical mass of the $\Lb$ and the accepted mass of the $b$ quark, this gives a value of 0.121. This number, obtained in this simplifying limit, is in surprisingly good agreement with the values of the lower zeroes shown in Table \ref{s01}, for all states. Of course, there must be deviations from this simple limit, as the $b$ quark is not infinitely heavy. However, it may be possible to systematically estimate the corrections to the value of 0.121.

In addition, in the limit of an infinitely heavy $b$ quark, the location of the zeroes is given by Eq. \ref{eq:global0} (up to the corrections mentioned), since $F_4$, $G_4$, $F_4^T$ and $G_4^T$ all vanish explicitly in this limit. This means that in this limit, there can only be one zero in the FBAs if the form factors are treated in the strict heavy-quark limit. This also means that the location of this second zero may be sensitively dependent on the form factors, the angular momentum of the state being considered and, of course, on the Wilson coefficients. Surprisingly, the results shown in Table \ref{s02} suggest that the most important dependence is the angular momentum of the daughter baryon.

\begin{figure}\vskip 18pt
\centerline{\includegraphics[width=3.5in]{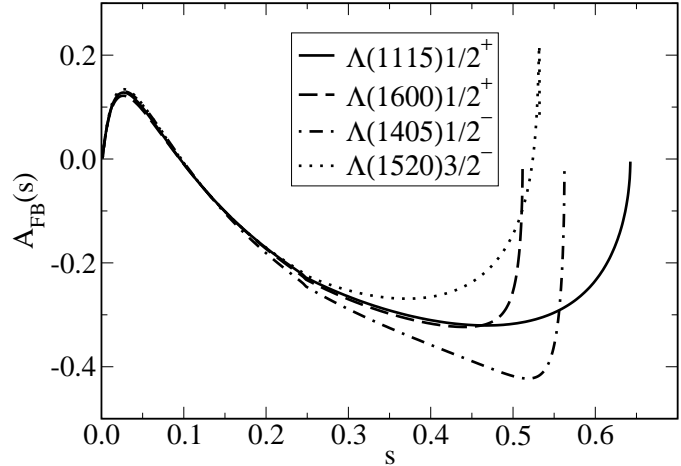}}
\caption{Forward-backward asymmetry for the different final states treated. The solid curve is for the ground state $\Lambda(1115)$, the dashed curve arises from the $\Lambda(1600)$, the dot-dashed curve is for the the $\Lambda(1405)$, and the dotted curve is for the $\Lambda(1520)$. For values of $s$ less than about 0.2, the four curves are indistinguishable.}\label{fig:fba_all}
\end{figure}

We conclude this section by returning to the discussion near the end of Section \ref{sec:hqetasym}. There, it was pointed out that use of the HQET relations among form factors indicated that the shapes of the FBAs would be largely independent of the quantum numbers of the final state, and would be determined largely by the Wilson coefficients and kinematics. In Fig. \ref{fig:fba_all}, the FBAs for all of the states we consider are shown on the same set of axes. The long-distance contributions from the charmonium resonances are omitted, and the results shown are for the muon, using the full form factors from the MCN calculation. The curves are essentially identical with each other for $s$ between 0.0 and about 0.2, after which the curves deviate from each other. However, the curves for the two $\hlf^+$ states remain close beyond $s=0.4$. The deviations should not be too surprising as $\xi_2/\xi_1$ near the non-recoil point is sizeable for some of these states (see table \ref{hqet_ratio}).

\subsubsection{Lepton Polarization Asymmetry}

\subsubsubsection{Longitudinal Lepton Polarization Asymmetries}

The differential longitudinal lepton polarization asymmetry (LLPA) $\P_L^-$ is shown in Figs. \ref{fig:PLq212p}-\ref{fig:PL32m}. In Figs. \ref{fig:PL12p}-\ref{fig:PL32m}, the legends are the same as those for the differential BRs and FBAs in Figs. \ref{fig:br12p}-\ref{fig:asym32m}. In order to characterize the typical values of the LLPA, we introduce the integrated LLPA, which is defined as
\beq
\<\P_L^\pm\>=\int_{4\mlh^2}^{(1-\sqrt{r})^2}\P_L^\pm(\sh)d\sh.
\eeq
Since $\P_L^+=-\P_L^-$, we will only discuss $\P_L^-$. The integrated LLPAs we obtain are shown in Tables \ref{llpa1} and \ref{llpa2}. The column labels have the same meaning as with the branching ratios shown in Tables \ref{br1} and \ref{br2}. We also compare our model predictions with LCSR, QCDSR and PM predictions for transitions to the ground state using SM Wilson coefficients and Eqs. \ref{eq:d12p}, \ref{eq:d0}, \ref{eq:ax}, \ref{eq:dx}, \ref{eq:dl}.

\begin{table}
\caption{Integrated longitudinal lepton polarization asymmetry for $\Lambda_b\rightarrow \Lambda^{(*)}\mu^{+}\mu^{-}$ in units of $10^{-2}$. The columns are labeled as in Table \ref{br1}.}
{\begin{tabular}{cccccccc}
\hline State, $J^{P}$       & LD & SM1 & SM2 & SUSY & LCSR & QCDSR & PM  \\ \hline
$\Lambda(1115)\,1/2^{+}$    & a & $-58.1$ & $-58.3$ & $-53.1$ & $-60.6$ & $-60.1$ & $-59.1$ \\
                            & b & $-51.6$ & $-51.7$ & $-49.0$ & $-53.9$ & $-54.5$ & $-52.6$ \\ \hline
$\Lambda(1600)\,1/2^{+}$    & a & $-45.4$ & $-45.6$ & $-41.3$ & $-$ & $-$ & $-$ \\
                            & b & $-40.1$ & $-40.4$ & $-37.1$ & $-$ & $-$ & $-$ \\ \hline
$\Lambda(1405)\,1/2^{-}$    & a & $-49.7$ & $-50.1$ & $-45.6$ & $-$ & $-$ & $-$ \\
                            & b & $-43.9$ & $-44.3$ & $-41.4$ & $-$ & $-$ & $-$ \\ \hline
$\Lambda(1520)\,3/2^{-}$    & a & $-46.7$ & $-47.2$ & $-42.9$ & $-$ & $-$ & $-$ \\
                            & b & $-41.4$ & $-41.8$ & $-38.7$ & $-$ & $-$ & $-$ \\ \hline
$\Lambda(1890)\,3/2^{+}$    & a & $-38.1$ & $-38.3$ & $-34.9$ & $-$ & $-$ & $-$  \\
                            & b & $-34.1$ & $-34.3$ & $-30.3$ & $-$ & $-$ & $-$ \\ \hline
$\Lambda(1820)\,5/2^{+}$    & a & $-38.7$ & $-39.7$ & $-36.1$ & $-$ & $-$ & $-$  \\
                            & b & $-34.3$ & $-35.1$ & $-31.8$ & $-$ & $-$ & $-$ \\ \hline
\end{tabular}\label{llpa1}}
\end{table}

\begin{table}
\caption{Integrated longitudinal lepton polarization asymmetry for $\Lambda_b\rightarrow \Lambda^{(*)}\tau^{+}\tau^{-}$ in units of $10^{-2}$. The columns are labeled as in Table \ref{br1}.}
{\begin{tabular}{ccccccccc}
\hline State, $J^{P}$      & LD & SM1 & SM2 & SUSY & LCSR & QCDSR & PM \\ \hline
$\Lambda(1115)\,1/2^{+}$   & a & $-10.7$ & $-10.8$ & $-8.2$  & $-11.5$ & $-11.8$ & $-11.0$ \\
                           & b & $-10.2$ & $-10.3$ & $-8.1$  & $-11.0$ & $-11.4$ & $-10.5$ \\ \hline
$\Lambda(1600)\,1/2^{+}$   & a & $-3.4$  & $-3.5$  & $-2.6$  & $-$ & $-$ & $-$ \\
                           & b & $-3.1$  & $-3.2$  & $-2.5$  & $-$ & $-$ & $-$  \\ \hline
$\Lambda(1405)\,1/2^{-}$   & a & $-5.9$  & $-6.0$  & $-4.5$  & $-$ & $-$ & $-$  \\
                           & b & $-5.5$  & $-5.6$  & $-4.4$  & $-$ & $-$ & $-$  \\ \hline
$\Lambda(1520)\,3/2^{-}$   & a & $-4.6$  & $-4.7$  & $-3.4$  & $-$ & $-$ & $-$  \\
                           & b & $-4.2$  & $-4.3$  & $-3.2$  & $-$ & $-$ & $-$  \\ \hline
$\Lambda(1890)\,3/2^{+}$   & a & $-0.81$ & $-0.79$ & $-0.58$ & $-$ & $-$ & $-$  \\
                           & b & $-0.54$ & $-0.52$ & $-0.46$ & $-$ & $-$ & $-$  \\ \hline
$\Lambda(1820)\,5/2^{+}$   & a & $-1.4$  & $-1.4$  & $-0.99$ & $-$ & $-$ & $-$  \\
                           & b & $-1.1$  & $-1.1$  & $-0.89$ & $-$ & $-$ & $-$  \\ \hline
\end{tabular}\label{llpa2}}
\end{table}

\subsubsection*{$\it\jp=1/2^+$}

\begin{figure}
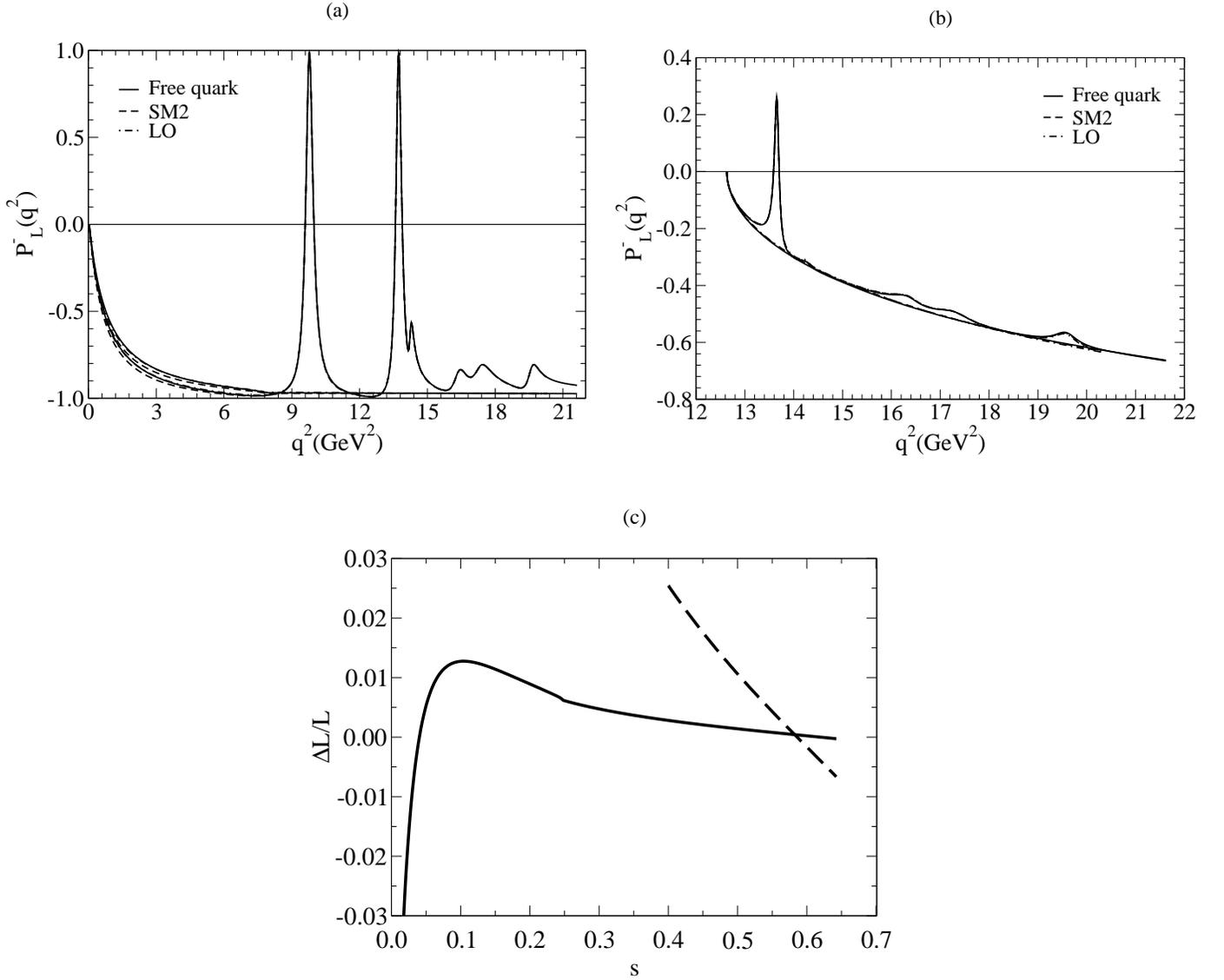

\centerline{\includegraphics[width=3.5in]{mu_slpa_L_12p_LO_inclusive_q2.eps}\,\,\,\,\,\,
\includegraphics[width=3.5in]{tau_slpa_L_12p_LO_inclusive_q2.eps}}\vskip 0.3in
\centerline{\includegraphics[width=3.5in]{L_diff.eps}}
\caption{$\P_L^-(q^2)$ for (a) $\Lambda_{b}\rightarrow\Lambda(1115)\mu^{+}\mu^{-}$ and (b) $\Lambda_{b}\rightarrow\Lambda(1115)\tau^{+}\tau^{-}$ without and with long distance (LD) contributions from charmonium resonances. The solid curves arise from the free quark process $b\to s\ell^+\ell^-$, while the dashed curves are the SM2 results, and the dot-dashed curves are the leading order SM results, $\D_L^{(0)-}/\D_0^{(0)}$.The curves in (c) show the quantity $
\frac{P_{L_{\rm full}}^--P_{L_{\rm HQET}}^-}{P_{L_{\rm HQET}}^-}$. The solid curve is for the $\Lambda_b\to \Lambda\mu^+\mu^-$, while the dashed curve is for $\Lambda_b\to \Lambda\tau^+\tau^-$,}\label{fig:PLq212p}
\end{figure}

In Fig. \ref{fig:PLq212p}, we show the differential LLPA as a functions of $q^2$ for decays to the ground state, $\Ls(1115)$ for the SM2 case. Also shown in that figure are the results obtained for the free-quark process $b\to s\ell^+\ell^-$, and the leading order HQET result with no form factor dependence, $\D_L^{(0)-}/\D_0^{(0)}$. It can be seen that these curves are nearly indistinguishable. The closeness of the results obtained in the different limits is examined by plotting
\beq
\frac{P_{L_{\rm MCN}}^--P_{L_{\rm HQET}}^-}{P_{L_{\rm HQET}}^-}\nonumber,
\eeq
shown as the graph (c) in Fig. \ref{fig:PLq212p}. In that graph, the solid curve is obtained for the $\mu^-$, while the dashed curve is for the $\tau^-$. Except for a small region at low $s$ in the case of the $\mu^-$, the difference between the two results is less than about 1.5\% for most of the kinematic range allowed. For the $\tau$ the difference is larger, growing to about 2.5\%. Note that an analogous quantity calculated for the HQET and LO (no form factor dependence) results shows a maximum deviation of 0.04\%.

\begin{figure}
\centerline{\includegraphics[width=3.5in]{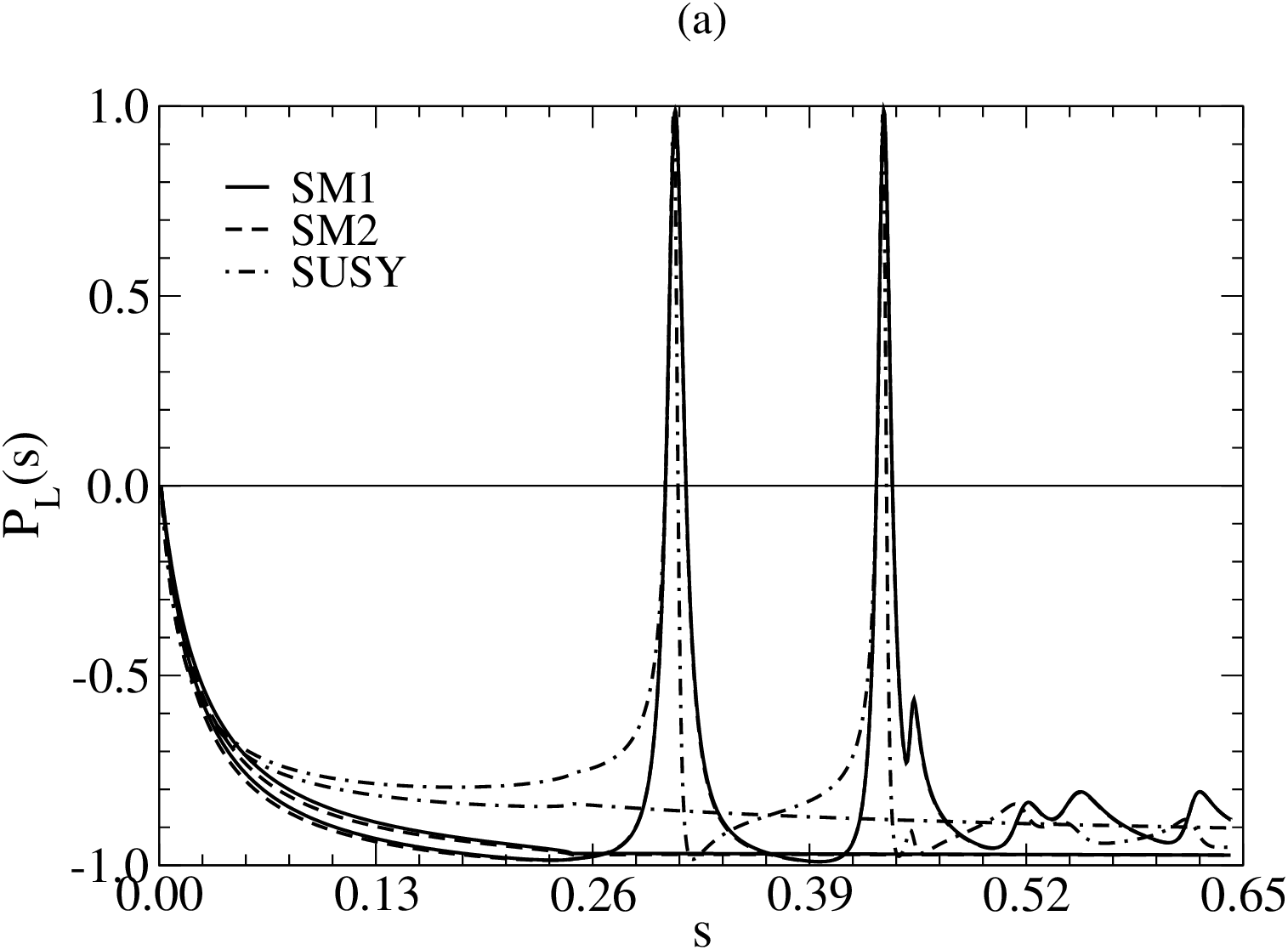}\,\,\,\,\,\,
\includegraphics[width=3.5in]{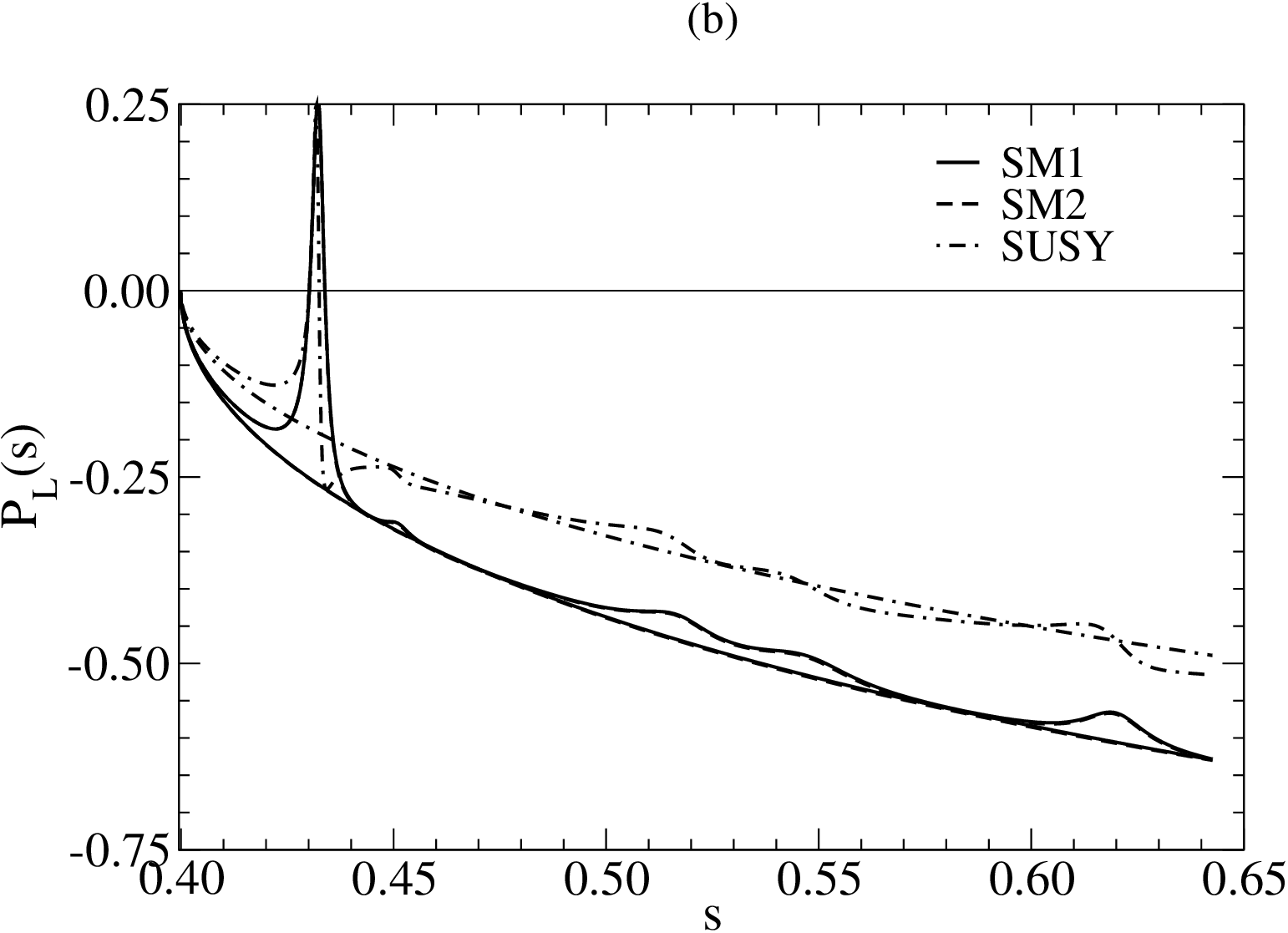}}
\caption{$\P_L^-(\sh)$ for (a) $\Lambda_{b}\rightarrow\Lambda(1115)\mu^{+}\mu^{-}$ and (b) $\Lambda_{b}\rightarrow\Lambda(1115)\tau^{+}\tau^{-}$ without and with long distance (LD) contributions from charmonium resonances. The solid and dashed curves represent SLPAs obtained from SCA and MCN form factors, respectively. The dash-dotted curves represent a SUSY extension to the SM.}\label{fig:PL12p}
\end{figure}

Fig. \ref{fig:PL12p} shows these same LLPAs as a function of $\sh$ for the SM1, SM2, and SUSY cases with and without LD contributions. The solid curves represent the results of SM1, the dashed curves are SM2 results, while SUSY results are the dot-dashed curves. The curves for the SM1 and SM2 cases are nearly indistinguishable over the entire kinematic range for the both the $\mu$ and $\tau$ channels. However, we see that there is a significant difference between the SM predictions and the  SUSY result in both channels. These observations are borne out by the results presented  in Tables \ref{llpa1} and \ref{llpa2}. We also see from these tables that there is general agreement amongst our model results and those of LCSR, QCDSR and PM.

\begin{figure}
\centerline{\includegraphics[width=3.5in]{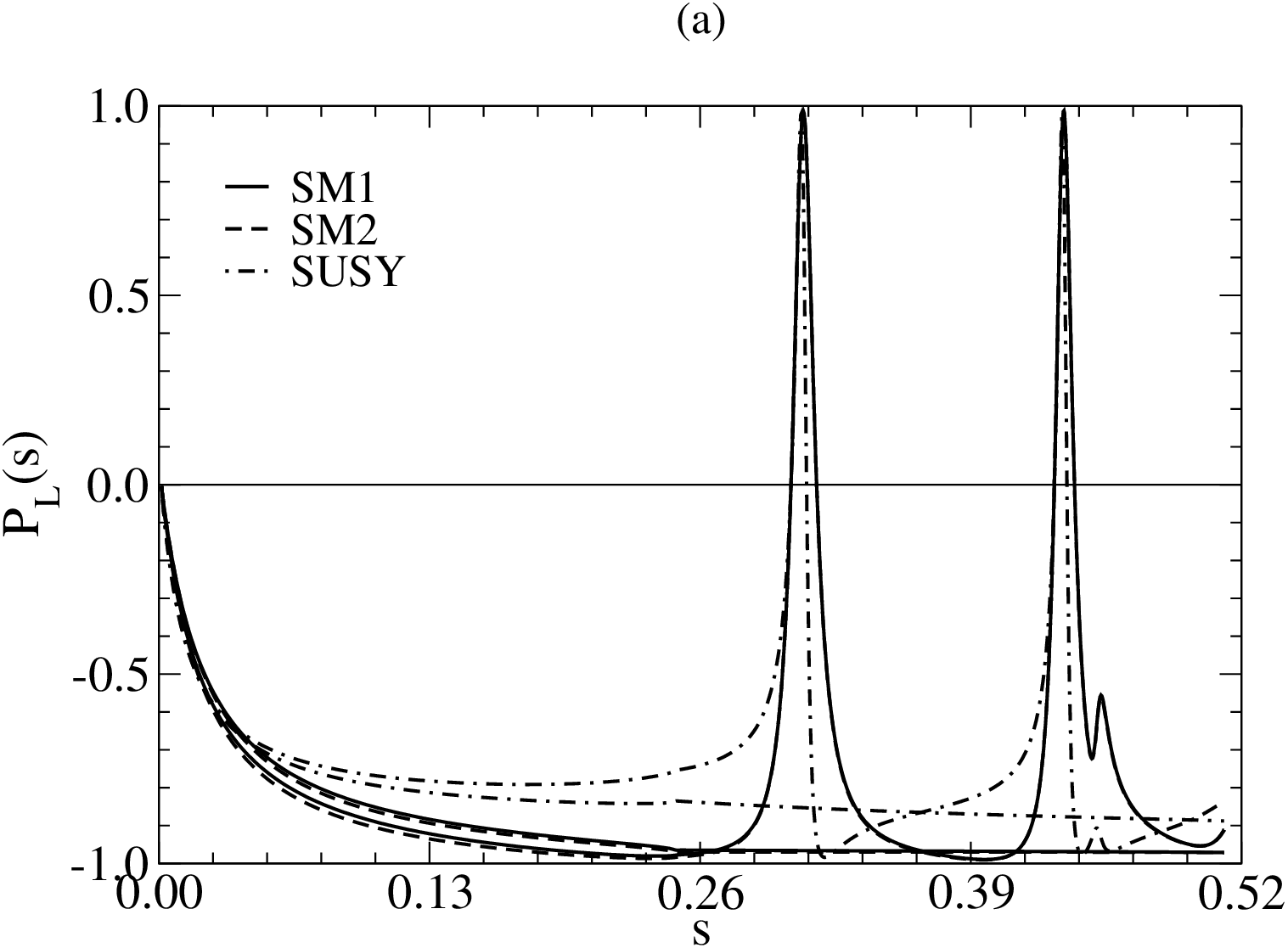}\,\,\,\,\,\, \includegraphics[width=3.5in]{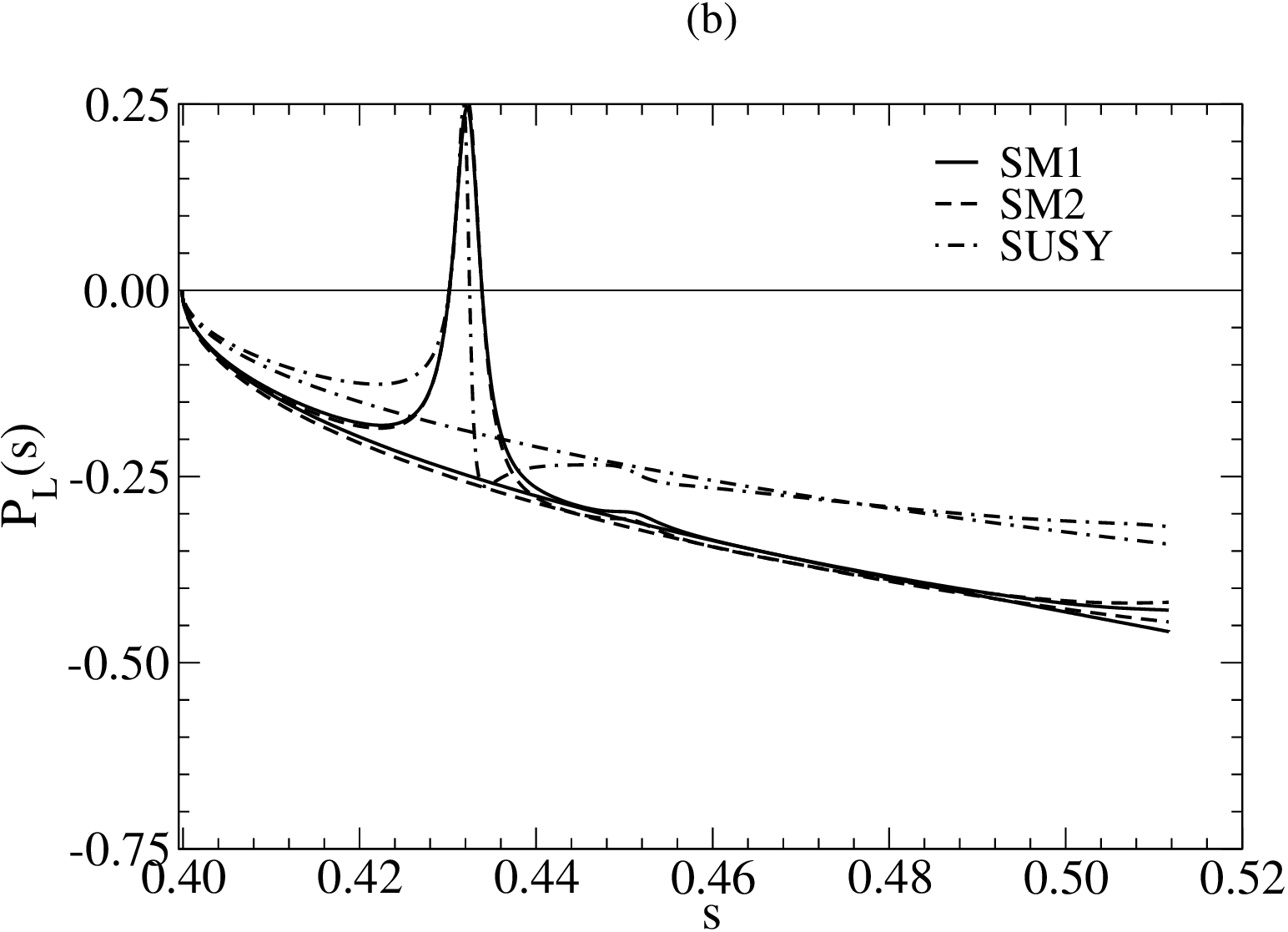}}
\caption{Same as Fig. \ref{fig:PL12p} but for $\Lambda_{b}\rightarrow\Lambda(1600)$.}\label{fig:PL12p1r}
\end{figure}

Fig. \ref{fig:PL12p1r} shows the predictions for LLPAs for decays to $\Ls(1600)$ for SM1, SM2, and SUSY models with both LD scenarios. The SM1 and SM2 results are essentially identical for both channels and this is reflected in the values of the integrated LLPAs. Again, the SM and SUSY results are quite different.

\subsubsection*{$\it\jp=1/2^-$}

\begin{figure}
\centerline{\includegraphics[width=3.5in]{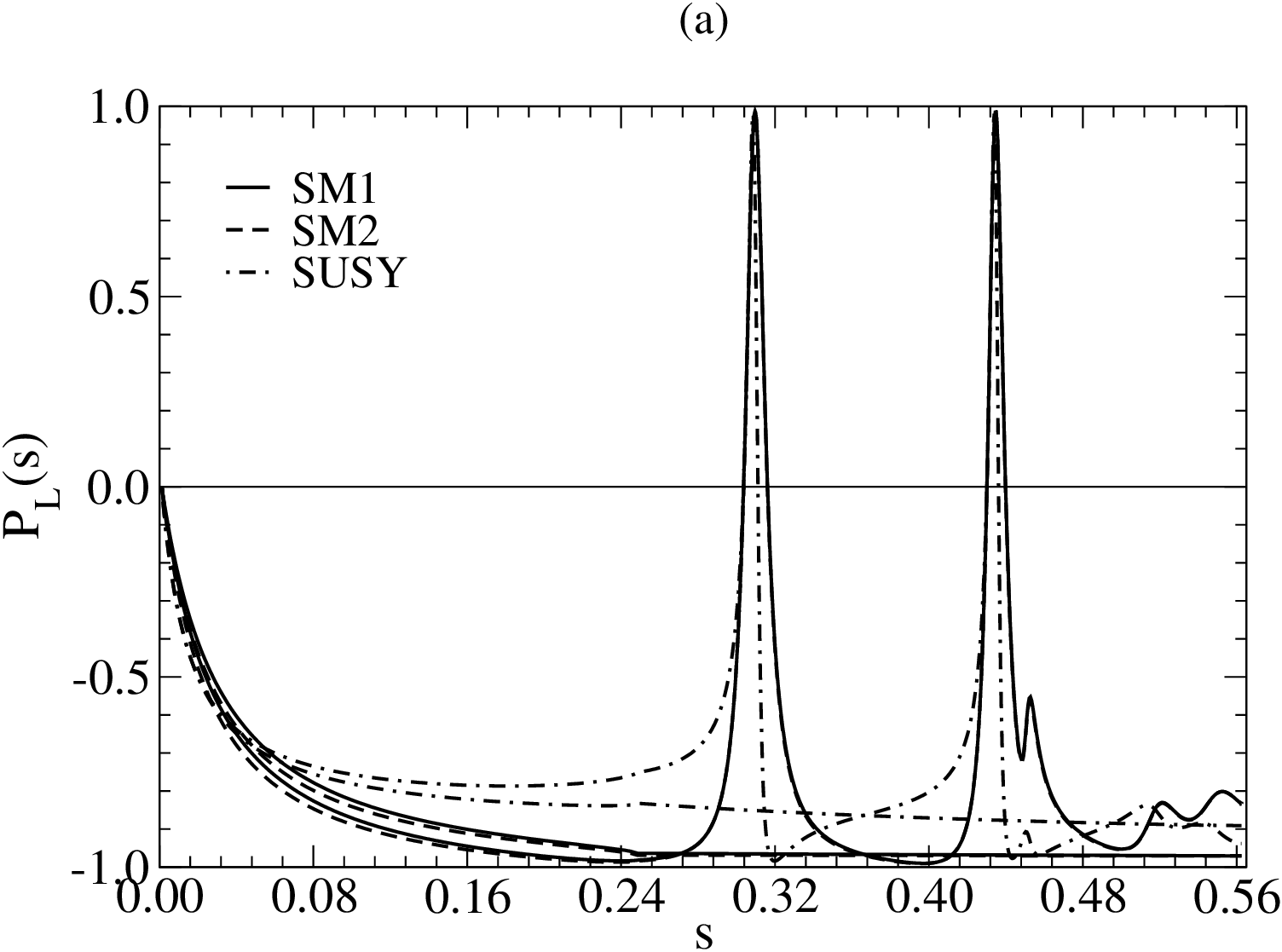}\,\,\,\,\,\, \includegraphics[width=3.5in]{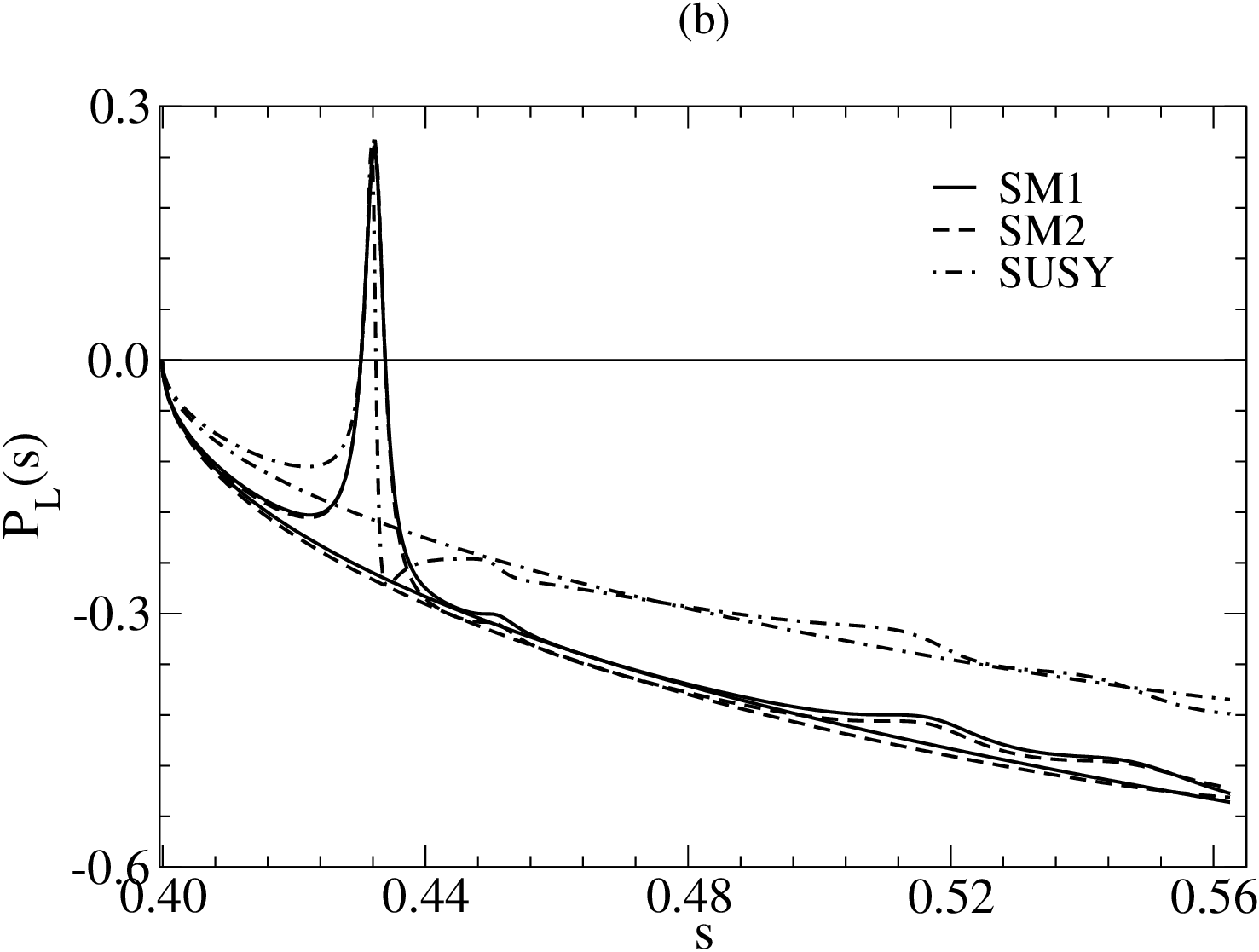}}
\caption{Same as Fig. \ref{fig:PL12p} but for $\Lambda_{b}\rightarrow\Lambda(1405)$.}\label{fig:PL12m}
\end{figure}

The predictions for LLPAs for decays to $\Ls(1405)$ are shown in Fig. \ref{fig:PL12m}. As can be seen, the predictions from SM1 and SM2 are nearly identical in both channels. The SUSY prediction is quite different from the SM predictions in both channels.

\subsubsection*{$\it\jp=3/2^-$}

\begin{figure}
\centerline{\includegraphics[width=3.5in]{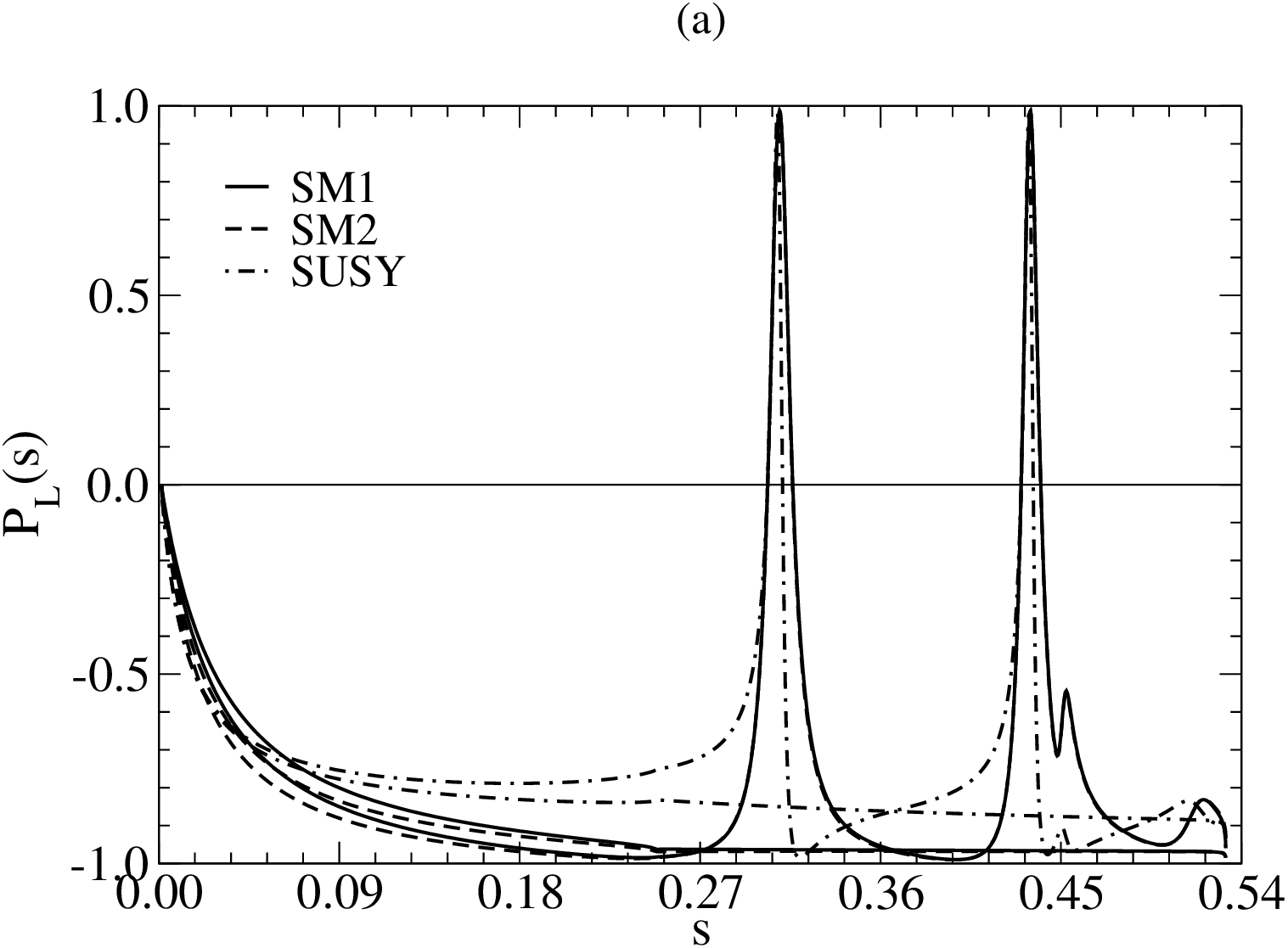}\,\,\,\,\,\, \includegraphics[width=3.5in]{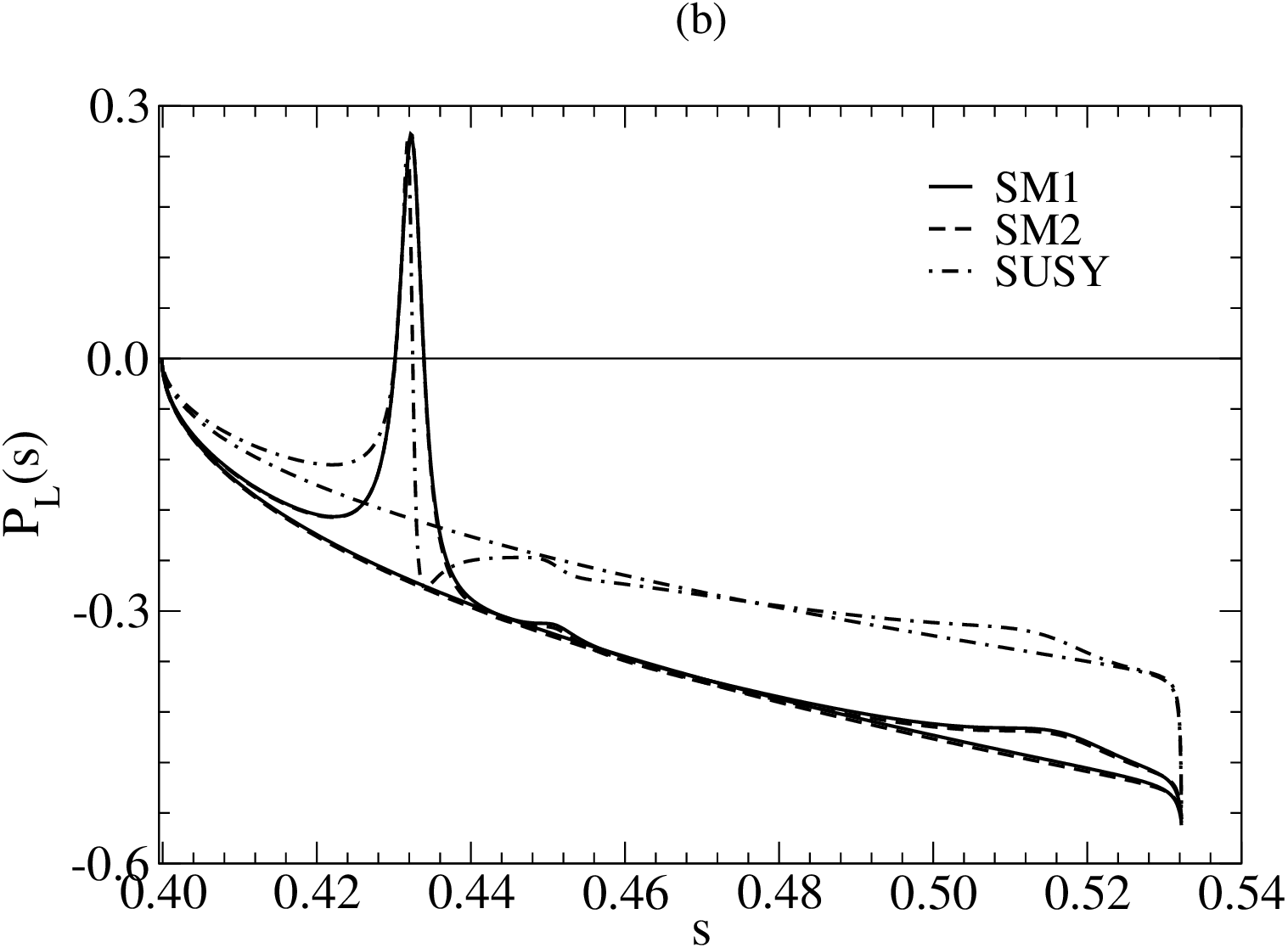}}
\caption{Same as Fig. \ref{fig:PL12p} but for $\Lambda_{b}\rightarrow\Lambda(1520)$.}\label{fig:PL32m}
\end{figure}

In Fig. \ref{fig:PL32m}, the predictions for LLPAs for decays to $\Ls(1520)$ are shown. These figures show that the predictions from SM1 and SM2 are nearly identical while the SUSY prediction is quite different in both channels.

\begin{figure}\vskip 18pt
\centerline{\includegraphics[width=3.5in]{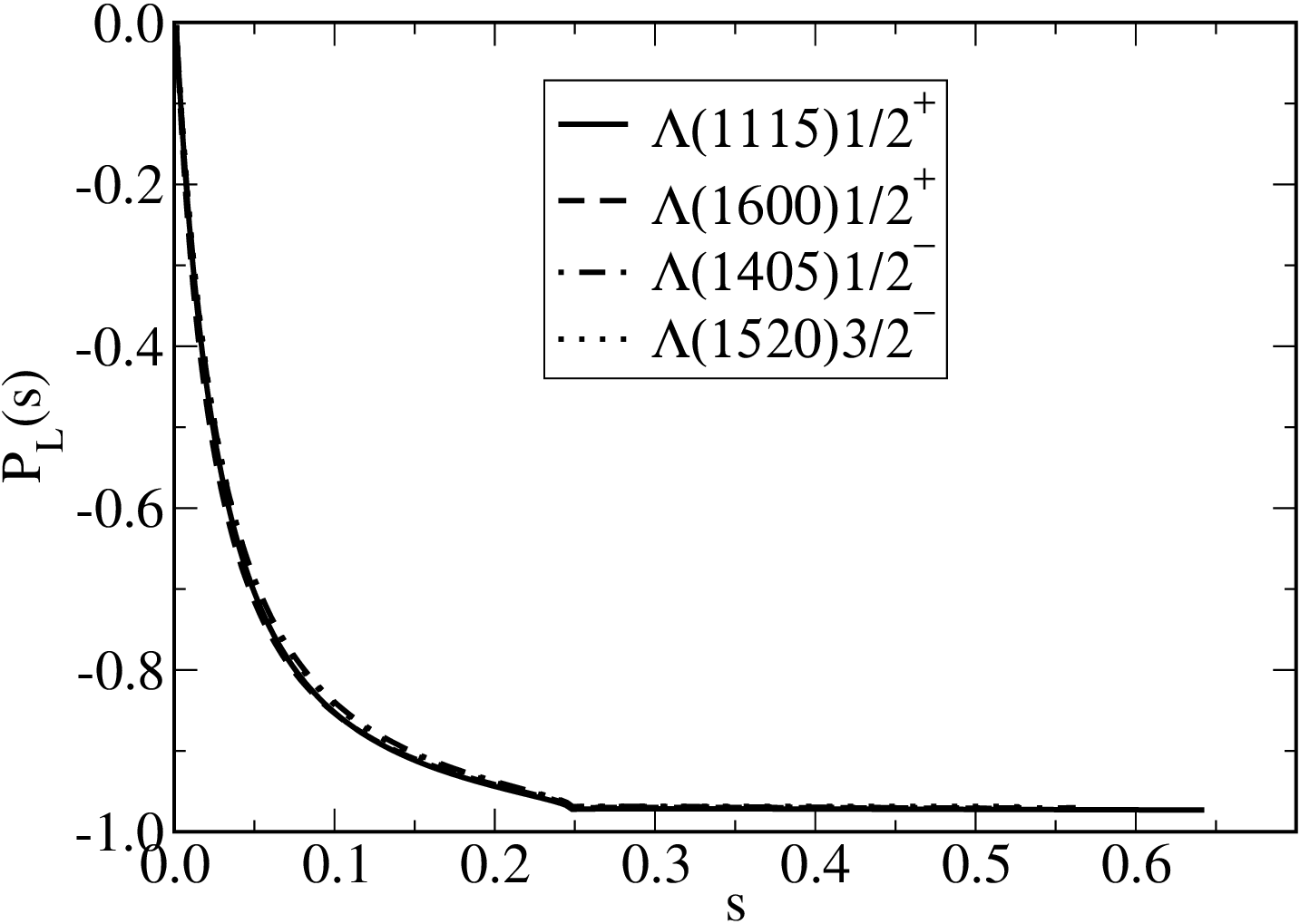}}
\caption{LLPA for all of the states considered in this work, with the long-distance contributions omitted. The solid curve is for the ground state $\Lambda(1115)$, the dashed curve arises from the $\Lambda(1600)$, the dot-dashed curve is for the $\Lambda(1405)$, and the dotted curve is for the $\Lambda(1520)$. The curves for the different states are indistinguishable from each other.}\label{fig:PLall}
\end{figure}

We conclude this section with the following observations. The LLPAs are largely independent of the form factor models; thus, they are free of hadronic uncertainties. This makes them optimal for the extraction of $\Re(C_7^*C_{10})$ and $\Re(C_9^*C_{10})$ or their ratio. Additionally, it has been shown that this observable is quite sensitive to new physics. Because it is free of hadronic uncertainties, any deviation from SM predictions would be a clear signal of physics beyond the SM. Furthermore, Fig. \ref{fig:PLall} shows the prediction for this LLPA, plotted for all of the final states considered in this work, with the long distance contributions omitted. The results shown are obtained using the MCN form factors. The solid curve is for the ground state $\Lambda(1115)$, the dashed curve arises from the $\Lambda(1600)$, the dot-dashed curve is for the $\Lambda(1405)$, and the dotted curve is for the $\Lambda(1520)$. The four curves shown are indistinguishable from each other, indicating that this asymmetry 
is also independent of the final state considered. This conclusion does not change when the long-distance contributions are included. This is an astounding result, given that $\xi_2/\xi_1$ is as large as -0.6 at the kinematic end-point for two of the states considered. 

\subsubsubsection{Transverse Lepton Polarization Asymmetries}

\begin{table}
\caption{Integrated transverse lepton polarization asymmetry for $\Lambda_b\rightarrow \Lambda^{(*)}\mu^{+}\mu^{-}$ in units of $10^{-2}$. The columns are labeled as in Table \ref{br1}.}
{\begin{tabular}{cccccccc}
\hline State, $J^{P}$       & LD & SM1 & SM2 & SUSY & LCSR & QCDSR & PM \\ \hline
$\Lambda(1115)\,1/2^{+}$   & a & $-0.07$ & $-0.07$ & $-0.04$ & $-0.02$ & $-0.08$ & $-0.07$ \\
                           & b & $-0.12$ & $-0.11$ & $-0.07$ & $-0.03$ & $-0.13$ & $-0.12$  \\ \hline
$\Lambda(1600)\,1/2^{+}$   & a & $-0.05$ & $-0.05$ & $-0.03$ & $-$ & $-$ & $-$ \\
                           & b & $-0.06$ & $-0.06$ & $-0.04$ & $-$ & $-$ & $-$  \\ \hline
$\Lambda(1405)\,1/2^{-}$   & a & $-0.06$ & $-0.07$ & $-0.04$ & $-$ & $-$ & $-$  \\
                           & b & $-0.10$ & $-0.11$ & $-0.06$ & $-$ & $-$ & $-$  \\ \hline
$\Lambda(1520)\,3/2^{-}$   & a & $-0.04$ & $-0.04$ & $-0.02$ & $-$ & $-$ & $-$  \\
                           & b & $-0.05$ & $-0.05$ & $-0.03$ & $-$ & $-$ & $-$  \\ \hline
$\Lambda(1890)\,3/2^{+}$   & a & $-0.03$ & $-0.04$ & $-0.02$ & $-$ & $-$ & $-$  \\
                           & b & $-0.02$ & $-0.03$ & $-0.02$ & $-$ & $-$ & $-$  \\ \hline
$\Lambda(1820)\,5/2^{+}$   & a & $-0.02$ & $-0.02$ & $-0.01$ & $-$ & $-$ & $-$  \\
                           & b & $-0.01$ & $-0.02$ & $-0.01$ & $-$ & $-$ & $-$  \\ \hline
\end{tabular}\label{tlpa1}}
\end{table}

\begin{table}
\caption{Integrated transverse lepton polarization asymmetry for $\Lambda_b\rightarrow \Lambda^{(*)}\tau^{+}\tau^{-}$ in units of $10^{-2}$. The columns are labeled as in Table \ref{br1}.}
{\begin{tabular}{ccccccccc}
\hline State, $J^{P}$      & LD & SM1 & SM2 & SUSY & LCSR & QCDSR & PM \\ \hline
$\Lambda(1115)\,1/2^{+}$   & a & $-0.33$ & $-0.29$ & $-0.14$ & $-0.08$ & $-0.38$ & $-0.33$ \\
                           & b & $-0.57$ & $-0.50$ & $-0.24$ & $-0.14$ & $-0.66$ & $-0.57$  \\ \hline
$\Lambda(1600)\,1/2^{+}$   & a & $-0.12$ & $-0.11$ & $-0.05$ & $-$ & $-$ & $-$ \\
                           & b & $-0.16$ & $-0.14$ & $-0.07$ & $-$ & $-$ & $-$  \\ \hline
$\Lambda(1405)\,1/2^{-}$   & a & $-0.22$ & $-0.24$ & $-0.11$ & $-$ & $-$ & $-$  \\
                           & b & $-0.38$ & $-0.41$ & $-0.20$ & $-$ & $-$ & $-$  \\ \hline
$\Lambda(1520)\,3/2^{-}$   & a & $-0.08$ & $-0.08$ & $-0.04$ & $-$ & $-$ & $-$  \\
                           & b & $-0.11$ & $-0.11$ & $-0.05$ & $-$ & $-$ & $-$  \\ \hline
$\Lambda(1890)\,3/2^{+}$   & a & $-0.01$ & $-0.02$ & $-0.01$ & $-$ & $-$ & $-$  \\
                           & b & $\sim-\E{-3}$ & $-0.01$ & $-0.01$ & $-$ & $-$ & $-$  \\ \hline
$\Lambda(1820)\,5/2^{+}$   & a & $0.01$ &  $\sim\E{-3}$ & $\sim\E{-3}$ & $-$ & $-$ & $-$  \\
                           & b & $0.02$ & $0.02$ & $0.01$ & $-$ & $-$ & $-$  \\ \hline
\end{tabular}\label{tlpa2}}
\end{table}

The differential transverse lepton polarization asymmetry (TLPA) $\P_T^-$ is shown in Figs. \ref{fig:PTlo12p}-\ref{fig:PT32m}. The legends In Figs. \ref{fig:PT12p}-\ref{fig:PT32m} are the same as those for the differential BRs and FBAs in Figs. \ref{fig:br12p}-\ref{fig:asym32m}. In order to characterize the typical values of the TLPA, we introduce the integrated TLPA, which is defined as
\beq
\<\P_T^\pm\>=\int_{4\mlh^2}^{(1-\sqrt{r})^2}\P_T^\pm(\sh)d\sh.
\eeq
Since $\P_T^+=\P_T^-$, we will only discuss $\P_T^-$. The integrated TLPAs we obtain are shown in Tables \ref{tlpa1} and \ref{tlpa2}. The column labels have the same meaning as with the branching ratios shown in Tables \ref{br1} and \ref{br2}. We also compare our model predictions with LCSR, QCDSR and PM predictions for transitions to the ground state using SM Wilson coefficients and Eqs. \ref{eq:d12p}, \ref{eq:d0}, \ref{eq:ax}, \ref{eq:dx}, and \ref{eq:dt}.

\subsubsection*{$\it\jp=1/2^+$}

\begin{figure}
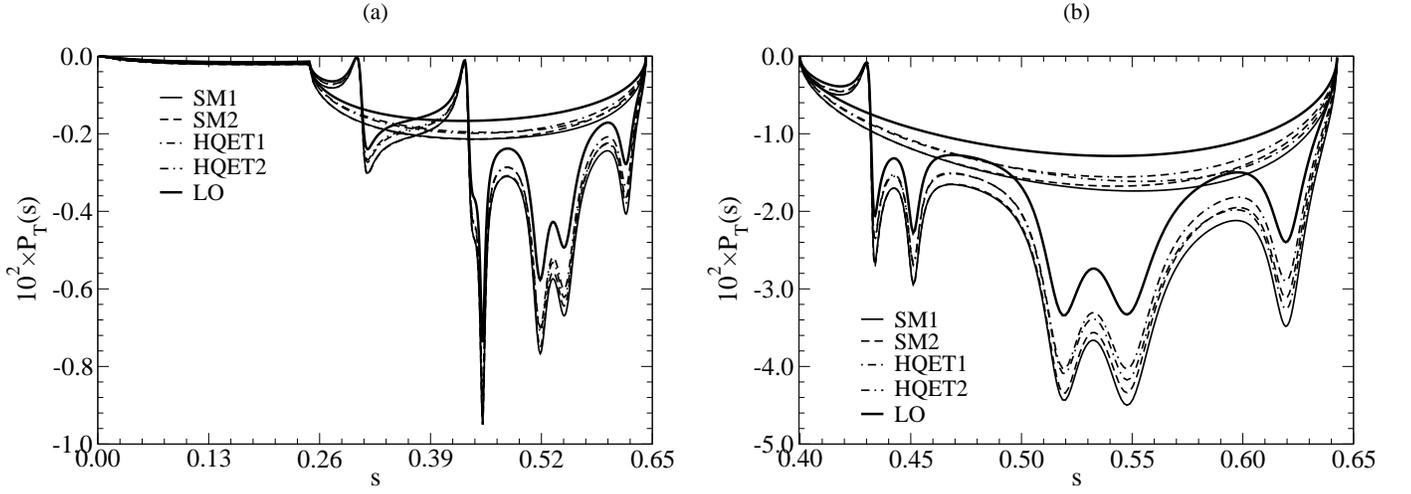

\centerline{\includegraphics[width=3.5in]{mu_slpa_T_12p_mcn_hqet_lo.eps}\,\,\,\,\,\,
\includegraphics[width=3.5in]{tau_slpa_T_12p_mcn_hqet_lo.eps}}
\caption{$\P_T^-(\hat s)$ for (a) $\Lambda_{b}\rightarrow\Lambda(1115)\mu^{+}\mu^{-}$ and (b) $\Lambda_{b}\rightarrow\Lambda(1115)\tau^{+}\tau^{-}$. The 
solid curves arise from the SM1 model, the dashed curves from SM2, the dot-dashed curves from HQET with the vector form factors from SCA model, the dash-dotted curves from HQET with the vector form factors from MCN model, and the thick solid curves from $\D_T^{(0)-}/\D_0^{(0)}$.}\label{fig:PTlo12p}
\end{figure}

In Fig. \ref{fig:PTlo12p}, we show the differential TLPAs for decays to $\Ls(1115)$ for SM1 and SM2 cases. We also show results for HQET using the the SCA (HQET1) and MCN (HQET2) vector form factors, as well as leading order results with no form factor dependence, $\D_T^{(0)-}/\D_0^{(0)}$. It is clear from the figure that this observable depends more strongly on the form factors than do the LLPAs. However, as one can see, there is little distinction between the form factor models. The HQET1 and HQET2 curves are in strong agreement with each other, and the SM1 and SM2 curves are in reasonable agreement with each other as well. Though the TLPAs have a stronger dependence on the form factors, they appear to be essentially independent of the form factor model. This can also be seen from the results in Tables \ref{tlpa1} and \ref{tlpa2}. We see that the SM1 and SM2 predictions are agreement with QCDSR and PM; however, LCSR prediction is different from the other SM predictions.

\begin{figure}
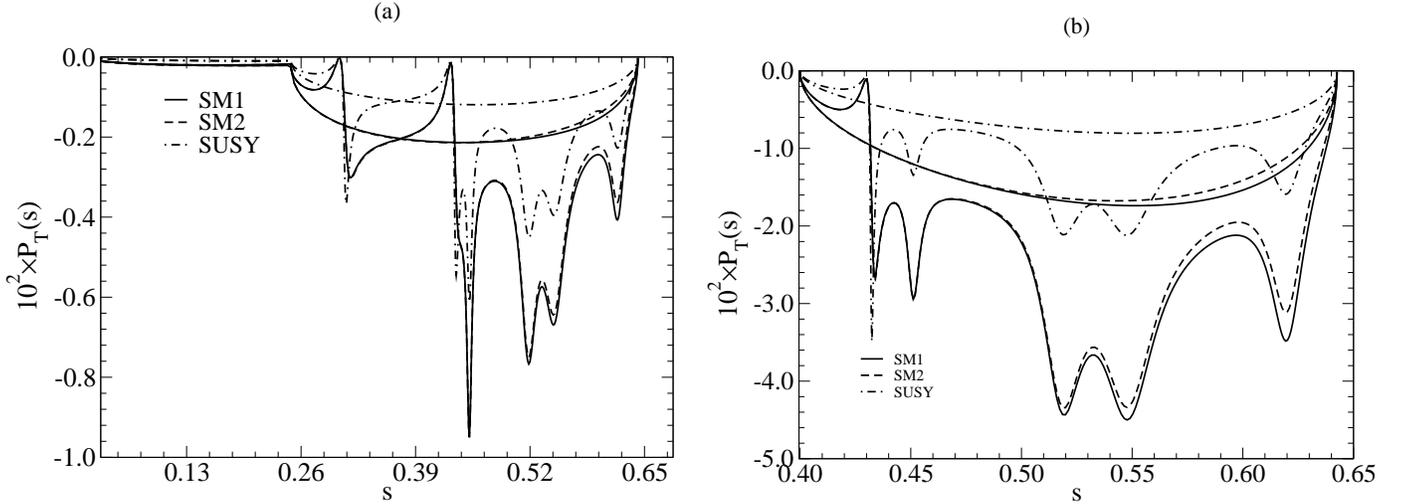

\centerline{\includegraphics[width=3.5in]{mu_slpa_T_12p.eps}\,\,\,\,\,\,
\includegraphics[width=3.5in]{tau_slpa_T_12p.eps}}
\caption{$\P_T^-(\sh)$ for (a) $\Lambda_{b}\rightarrow\Lambda(1115)\mu^{+}\mu^{-}$ and (b) $\Lambda_{b}\rightarrow\Lambda(1115)\tau^{+}\tau^{-}$ without and with long distance (LD) contributions from charmonium resonances. The solid and dashed curves represent SLPAs obtained from SCA and MCN form factors, respectively. The dash-dotted curves represent a SUSY extension to the SM.}\label{fig:PT12p}
\end{figure}

Figure \ref{fig:PT12p} shows the differential TLPAs for decays to the ground state from the SM1, SM2, and SUSY cases with and without LD contributions. Again, we see that both SM predictions are in agreement in both channels. However, the SUSY curves are significantly different from the SM curves.  This is supported by Tables \ref{tlpa1} and \ref{tlpa2}, where we see that the SUSY predictions are quite different from the various SM predictions for the integrated TLPAs. 

\begin{figure}
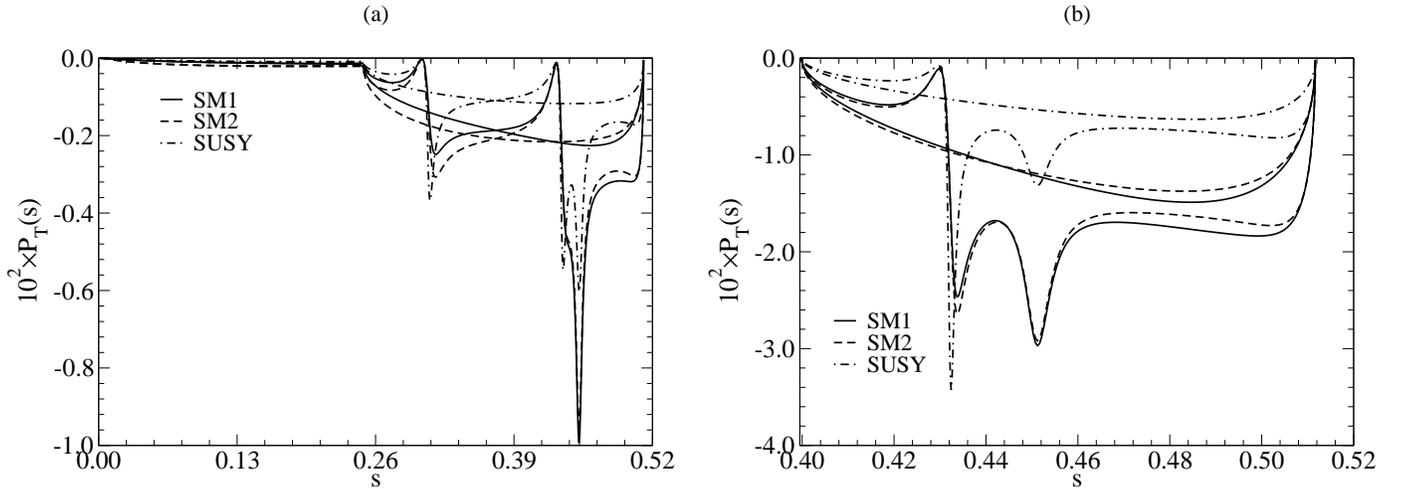

\centerline{\includegraphics[width=3.5in]{mu_slpa_T_12p1r.eps}\,\,\,\,\,\, \includegraphics[width=3.5in]{tau_slpa_T_12p1r.eps}}
\caption{Same as Fig. \ref{fig:PT12p} but for $\Lambda_{b}\rightarrow\Lambda(1600)$.}\label{fig:PT12p1r}
\end{figure}

In Fig. \ref{fig:PT12p1r}, we show differential TLPAs for transitions to $\Ls(1600)$. The story here is the same as with decays to the ground state. The SM1 and SM2 results are in agreement with each other, but the SUSY values are quite different. This is borne out by the values of the integrated asymmetries that are presented in Tables \ref{tlpa1} and \ref{tlpa2}.

\subsubsection*{$\it\jp=1/2^-$}

\begin{figure}
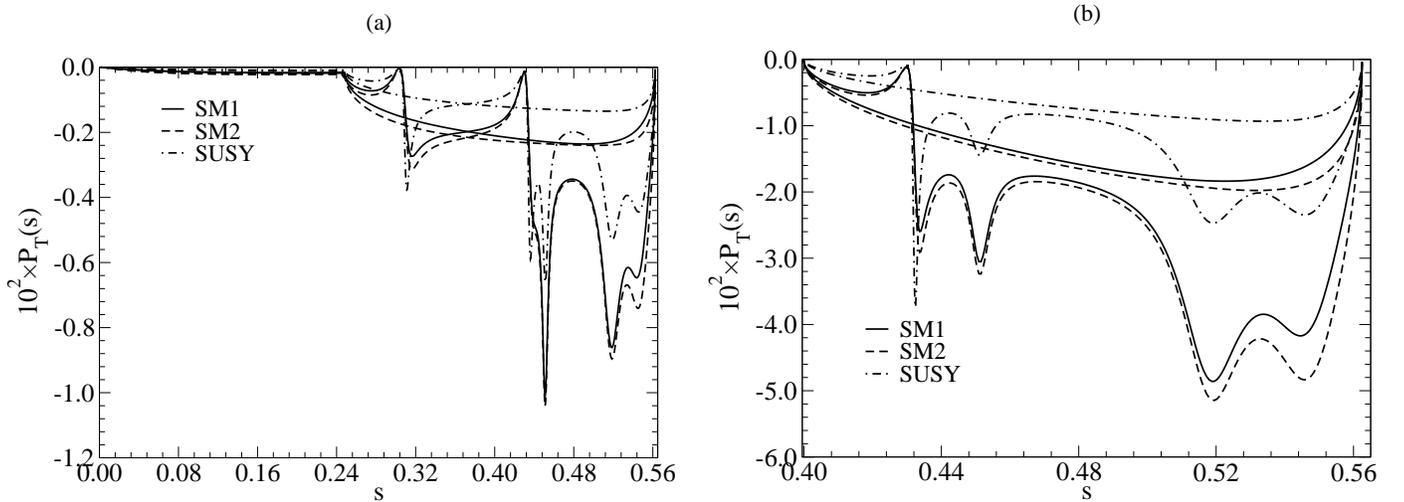

\centerline{\includegraphics[width=3.5in]{mu_slpa_T_12m_00.eps}\,\,\,\,\,\, \includegraphics[width=3.5in]{tau_slpa_T_12m_00.eps}}
\caption{Same as Fig. \ref{fig:PT12p} but for $\Lambda_{b}\rightarrow\Lambda(1405)$.}\label{fig:PT12m}
\end{figure}

Fig. \ref{fig:PT12m} shows the differential TLPAs for transitions to $\Ls(1405)$. As can be seen, SM1 and SM2 are essentially the same over the entire kinematic range; however, the SUSY values are quite different. This is confirmed by the values presented in Tables \ref{tlpa1} and \ref{tlpa2}.

\subsubsection*{$\it\jp=3/2^-$}

\begin{figure}
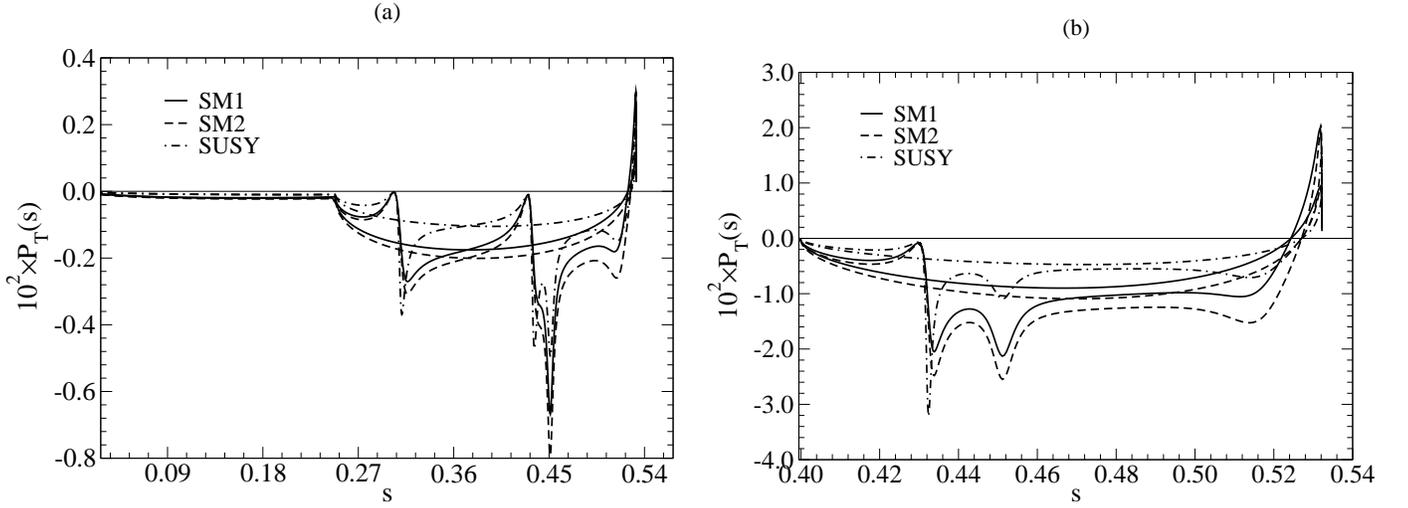

\centerline{\includegraphics[width=3.5in]{mu_slpa_T_32m.eps}\,\,\,\,\,\, \includegraphics[width=3.5in]{tau_slpa_T_32m.eps}}
\caption{Same as Fig. \ref{fig:PT12p} but for $\Lambda_{b}\rightarrow\Lambda(1520)$.}\label{fig:PT32m}
\end{figure}

Differential TLPAs for transitions to $\Ls(1520)$ are shown in Fig. \ref{fig:PT32m}. The observations here are the same as with the previous decay modes. The SM1 and SM2 results are in agreement with each other, but the SUSY values are quite different. This is supported by the numerical results for the integrated asymmetries in Tables \ref{tlpa1} and \ref{tlpa2}.

There is a difference between the differential asymmetries in this mode and the previous ones; there is a zero in both the $\mu$ and $\tau$ channels for this mode that does not appear for states with $J=1/2$. This is a situation that is similar to what was shown for the FBAs. Since the TLPAs are quite small, $\O(\E{-4})$ for the $\mu$ and $\O(\E{-3})$ for the $\tau$, there is little hope of them being measured in the near (or even distant) future. We will therefore explore the locations of the zeroes no further.

\begin{figure}\vskip 18pt
\centerline{\includegraphics[width=3.5in]{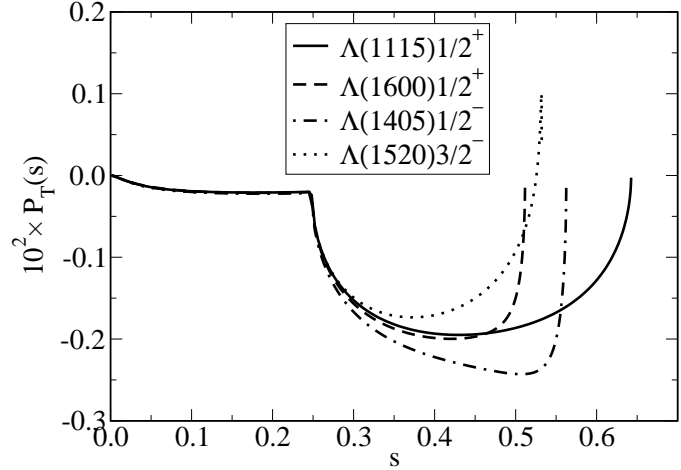}}
\caption{TLPAs for all of the states considered in this work, with the long-distance contributions omitted. The solid curve is for the ground state $\Lambda(1115)$, the dashed curve arises from the $\Lambda(1600)$, the dot-dashed curve is for the $\Lambda(1405)$, and the dotted curve is for the $\Lambda(1520)$. The curves for the different states are indistinguishable from each other for $s\lesssim$ 0.25.}\label{fig:PTall}
\end{figure}

In Fig. \ref{fig:PTall} we show $P_T$ for all of the states considered. The solid curve is for the ground state $\Lambda(1115)$, the dashed curve arises from the $\Lambda(1600)$, the dot-dashed curve is for the $\Lambda(1405)$, and the dotted curve is for the $\Lambda(1520)$. For $s\lesssim0.25$, the four sets of curves are indistinguishable. Form factor effects, particularly the large values of $\xi_2/\xi_1$ near the kinematic end-point for some of the states, give rise to the deviations shown.

\subsubsubsection{Normal Lepton Polarization Asymmetries}

\begin{table}
\caption{Integrated normal lepton polarization asymmetry for $\Lambda_b\rightarrow \Lambda^{(*)}\mu^{+}\mu^{-}$ in units of $10^{-2}$. The columns are labeled as in Table \ref{br1}.}
{\begin{tabular}{c|c|ccc|ccc|ccc|ccc|ccc|ccc}
\hline State, $J^{P}$       & LD & & SM1 & & & SM2 & & & SUSY & & & LCSR & & & QCDSR & & & PM & \\
& & $-$ & & $+$ & $-$ & & $+$ & $-$ & & $+$ & $-$ & & $+$ & $-$ & & $+$ & $-$ & & $+$ \\ \hline
$\Lambda(1115)\,1/2^{+}$   & a & $5.1$ & & $2.9$ & $5.0$ & & $2.8$ & $5.4$ & & $-0.05$ & $4.1$ & & $2.9$ & $5.4$ & & $3.1$ & $5.3$ & & $3.1$ \\
                           & b & $5.0$ & & $2.7$ & $4.8$ & & $2.6$ & $5.2$ & & $-0.15$ & $4.0$ & & $2.8$ & $5.2$ & & $2.9$ & $5.1$ & & $2.9$  \\ \hline
$\Lambda(1600)\,1/2^{+}$   & a & $4.9$ & & $3.0$ & $4.7$ & & $2.8$ & $5.0$ & & $0.02$   & $-$ & & $-$ & $-$ & & $-$ & $-$ & & $-$ \\
                           & b & $4.7$ & & $2.8$ & $4.6$ & & $2.6$ & $4.8$ & & $-0.09$ & $-$ & & $-$ & $-$ & & $-$ & $-$ & & $-$  \\ \hline
$\Lambda(1405)\,1/2^{-}$   & a & $4.9$ & & $2.8$ & $4.9$ & & $2.7$ & $5.3$ & & $-0.17$ & $-$ & & $-$ & $-$ & & $-$ & $-$ & & $-$  \\
                           & b & $4.8$ & & $2.6$ & $4.8$ & & $2.5$ & $5.1$ & & $-0.28$ & $-$ & & $-$ & $-$ & & $-$ & $-$ & & $-$  \\ \hline
$\Lambda(1520)\,3/2^{-}$   & a & $4.6$ & & $2.7$ & $4.5$ & & $2.7$ & $4.8$ & & $-0.11$ & $-$ & & $-$ & $-$ & & $-$ & $-$ & & $-$  \\
                           & b & $4.5$ & & $2.5$ & $4.4$ & & $2.5$ & $4.6$ & & $-0.21$ & $-$ & & $-$ & $-$ & & $-$ & $-$ & & $-$  \\ \hline
$\Lambda(1890)\,3/2^{+}$   & a & $4.4$ & & $2.7$ & $4.4$ & & $2.5$ & $4.7$ & & $-0.12$ & $-$ & & $-$ & $-$ & & $-$ & $-$ & & $-$  \\
                           & b & $4.3$ & & $2.5$ & $4.4$ & & $2.4$ & $4.5$ & & $-0.26$ & $-$ & & $-$ & $-$ & & $-$ & $-$ & & $-$  \\ \hline
$\Lambda(1820)\,5/2^{+}$   & a & $4.3$ & & $2.6$ & $4.2$ & & $2.6$ & $4.4$ & & $-0.07$ & $-$ & & $-$ & $-$ & & $-$ & $-$ & & $-$  \\
                           & b & $4.2$ & & $2.4$ & $4.1$ & & $2.4$ & $4.3$ & & $-0.17$ & $-$ & & $-$ & $-$ & & $-$ & $-$ & & $-$  \\ \hline
\end{tabular}\label{nlpa1}}
\end{table}

\begin{table}
\caption{Integrated normal lepton polarization asymmetry for $\Lambda_b\rightarrow \Lambda^{(*)}\tau^{+}\tau^{-}$ in units of $10^{-2}$. The columns are labeled as in Table \ref{br1}.}
{\begin{tabular}{c|c|ccc|ccc|ccc|ccc|ccc|ccc}
\hline State, $J^{P}$      & LD & & SM1 & & & SM2 & & & SUSY & & & LCSR & & & QCDSR & & & PM & \\
& & $-$ & & $+$ & $-$ & & $+$ & $-$ & & $+$ & $-$ & & $+$ & $-$ & & $+$ & $-$ & & $+$ \\ \hline
$\Lambda(1115)\,1/2^{+}$   & a & $13.6$ & & $4.9$  & $12.4$ & & $4.8$ & $12.5$ & & $-0.70$ & $6.5$ & & $5.2$ & $15.5$ & & $6.0$ & $14.2$ & & 
                           $5.9$ \\
                           & b & $12.9$ & & $4.1$  & $11.8$ & & $4.0$ & $12.1$ & & $-0.89$ & $6.1$ & & $4.7$ & $14.8$ & & $5.2$ & $13.5$ & & $5.0$  \\ \hline
$\Lambda(1600)\,1/2^{+}$   & a & $7.5$  & & $3.3$  & $6.8$  & & $2.9$  & $6.7$  & & $-0.20$ & $-$ & & $-$ & $-$ & & $-$ & $-$ & & $-$ \\
                           & b & $6.9$  & & $2.5$  & $6.3$  & & $2.2$  & $6.3$  & & $-0.42$ & $-$ & & $-$ & $-$ & & $-$ & $-$ & & $-$  \\ \hline
$\Lambda(1405)\,1/2^{-}$   & a & $10.7$ & & $4.2$  & $11.0$ & & $3.8$  & $11.0$ & & $-0.96$ & $-$ & & $-$ & $-$ & & $-$ & $-$ & & $-$  \\
                           & b & $10.1$ & & $3.3$  & $10.4$ & & $2.9$  & $10.6$ & & $-1.2$  & $-$ & & $-$ & $-$ & & $-$ & $-$ & & $-$  \\ \hline
$\Lambda(1520)\,3/2^{-}$   & a & $6.4$  & & $3.9$  & $6.2$  & & $3.5$  & $5.8$  & & $0.14$  & $-$ & & $-$ & $-$ & & $-$ & $-$ & & $-$  \\
                           & b & $5.9$  & & $3.2$  & $5.7$  & & $2.8$  & $5.5$  & & $-0.06$ & $-$ & & $-$ & $-$ & & $-$ & $-$ & & $-$  \\ \hline
$\Lambda(1890)\,3/2^{+}$   & a & $1.7$  & & $1.4$  & $2.3$  & & $1.3$  & $2.2$  & & $0.11$  & $-$ & & $-$ & $-$ & & $-$ & $-$ & & $-$  \\
                           & b & $1.4$  & & $0.94$ & $2.1$  & & $0.71$ & $1.9$  & & $-0.15$ & $-$ & & $-$ & $-$ & & $-$ & $-$ & & $-$  \\ \hline
$\Lambda(1820)\,5/2^{+}$   & a & $1.9$  & & $2.9$  & $1.8$  & & $2.4$  & $1.5$  & & $0.74$  & $-$ & & $-$ & $-$ & & $-$ & $-$ & & $-$  \\
                           & b & $1.6$  & & $2.3$  & $1.6$  & & $1.9$  & $1.3$  & & $0.57$  & $-$ & & $-$ & $-$ & & $-$ & $-$ & & $-$  \\ \hline
\end{tabular}\label{nlpa2}}
\end{table}

The differential normal lepton polarization asymmetry (NLPA) $\P_N^\pm$ is shown in Figs. \ref{fig:PNlo12p}-\ref{fig:PN32m}. The legends in Figs. \ref{fig:PN12p}-\ref{fig:PN32m} are the same as those for the differential BRs and FBAs in Figs. \ref{fig:br12p}-\ref{fig:asym32m}. As before, we characterize the typical values of the NLPA by introducing the integrated NLPA, defined as
\beq
\<\P_N^\pm\>=\int_{4\mlh^2}^{(1-\sqrt{r})^2}\P_N^\pm(\sh)d\sh.
\eeq
The integrated NLPAs we obtain are shown in Tables \ref{nlpa1} and \ref{nlpa2}. The column labels have the same meaning as with the branching ratios shown in Tables \ref{br1} and \ref{br2}. We also compare our model predictions with LCSR, QCDSR and PM predictions for transitions to the ground state using SM Wilson coefficients and Eqs. \ref{eq:d12p}, \ref{eq:d0}, \ref{eq:ax}, \ref{eq:dx}, and \ref{eq:dn}.

\subsubsection*{$\it\jp=1/2^+$}

\begin{figure}
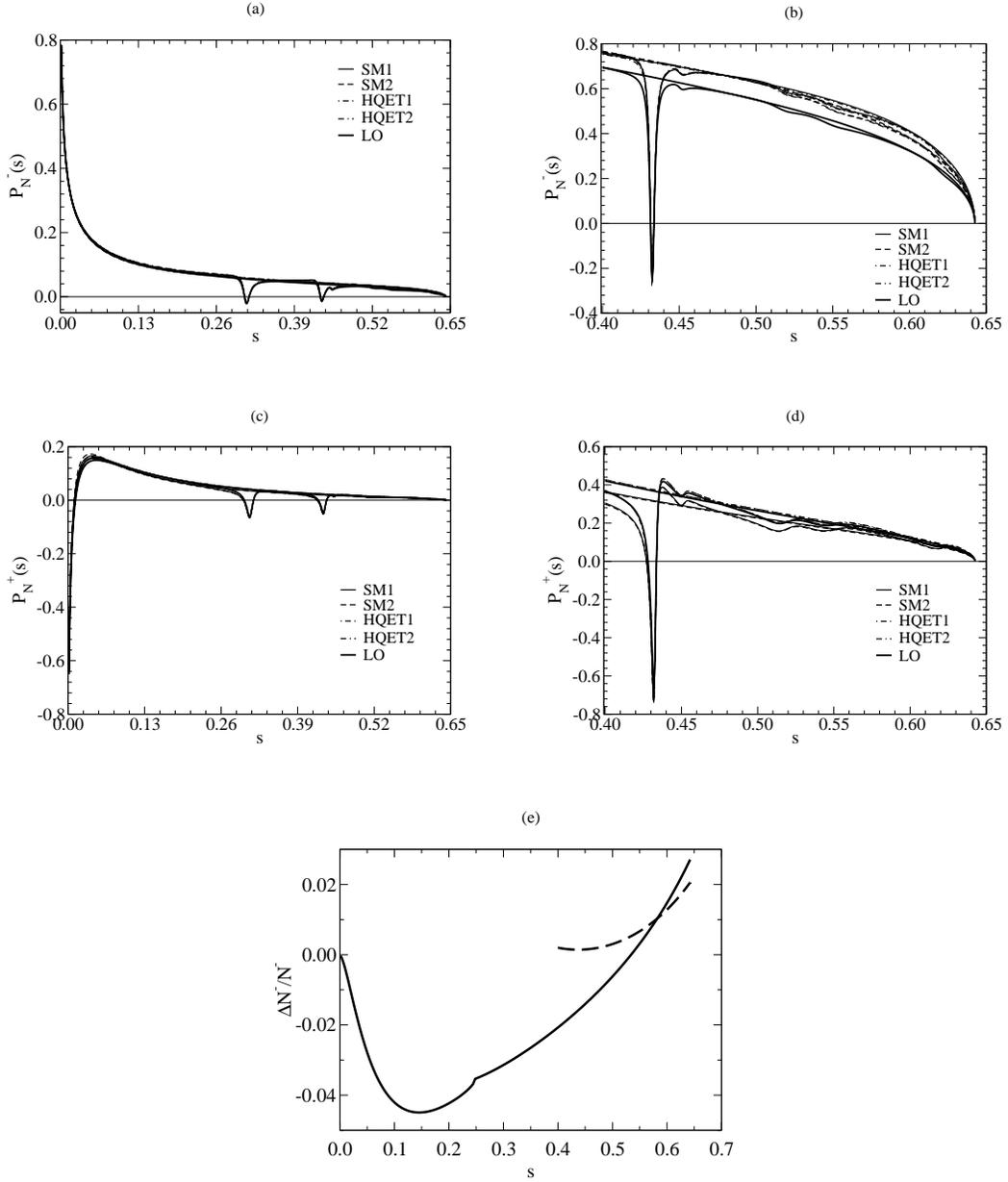

\centerline{\includegraphics[width=2.5in]{mu_slpa_N_12p_mcn_hqet_lo.eps}\hskip30pt \includegraphics[width=2.5in]{tau_slpa_N_12p_mcn_hqet_lo.eps}}\vskip25pt
\centerline{\includegraphics[width=2.5in]{mu_slpa_N2_12p_mcn_hqet_lo.eps}\hskip30pt \includegraphics[width=2.5in]{tau_slpa_N2_12p_mcn_hqet_lo.eps}}\vskip 25pt
\centerline{\includegraphics[width=2.5in]{N_diff.eps}}
\caption{$\P_N^-(\sh)$ for (a) $\Lambda_{b}\rightarrow\Lambda(1115)\mu^{+}\mu^{-}$ and (b) $\Lambda_{b}\rightarrow\Lambda(1115)\tau^{+}\tau^{-}$ without and with long distance (LD) contributions from charmonium resonances. $\P_N^+(\sh)$ for the $\mu$ and $\tau$ channels are also shown in (c) and (d), respectively. The solid curves arise from the SM1 model, the dashed curves from SM2, the dot-dashed curves from HQET with the vector form factors from SCA model, the dash-dotted curves from HQET with the vector form factors from MCN model, and the thick solid curves from $\D_N^{(0)\pm}/\D_0^{(0)}$. In graph (e) is plotted 
$\frac{P_{N_{\rm full}}^--P_{N_{\rm HQET}}^-}{P_{N_{\rm HQET}}^-}$}\label{fig:PNlo12p}
\end{figure}

In Fig. \ref{fig:PNlo12p}, we show the differential NLPAs for decays to $\Ls(1115)$ for SM1 and SM2 cases. We also show results for HQET using the the SCA (HQET1) and MCN (HQET2) vector form factors, as well as leading order results with no form factor dependence, $\D_N^{(0)\pm}/\D_0^{(0)}$. In the $\mu$ channel, the curves for five different cases are nearly indistinguishable. This could lead one to conclude that $\P_N^\pm$ has little or no form factor dependence. This can also be seen from the results in Tables \ref{tlpa1} and \ref{tlpa2}. We see that the SM1 and SM2 predictions are agreement with QCDSR and PM; however, LCSR prediction for $\PNm$ is different from the other SM predictions in both channels. 

As with $P_L$, we examine the ratio
\beq
\frac{P_{N_{\rm MCN}}^--P_{N_{\rm HQET}}^-}{P_{N_{\rm HQET}}^-}\nonumber,
\eeq
shown as the graph (e) in Fig. \ref{fig:PNlo12p}. In that graph, the solid curve is obtained for the $\mu^-$, while the dashed curve is for the $\tau^-$. For the entire kinematically allowed region, the ratio is less that 4.5\% for the $\mu^-$, and less than 2\% for the $\tau^+$

\begin{figure}
\centerline{\includegraphics[width=2.5in]{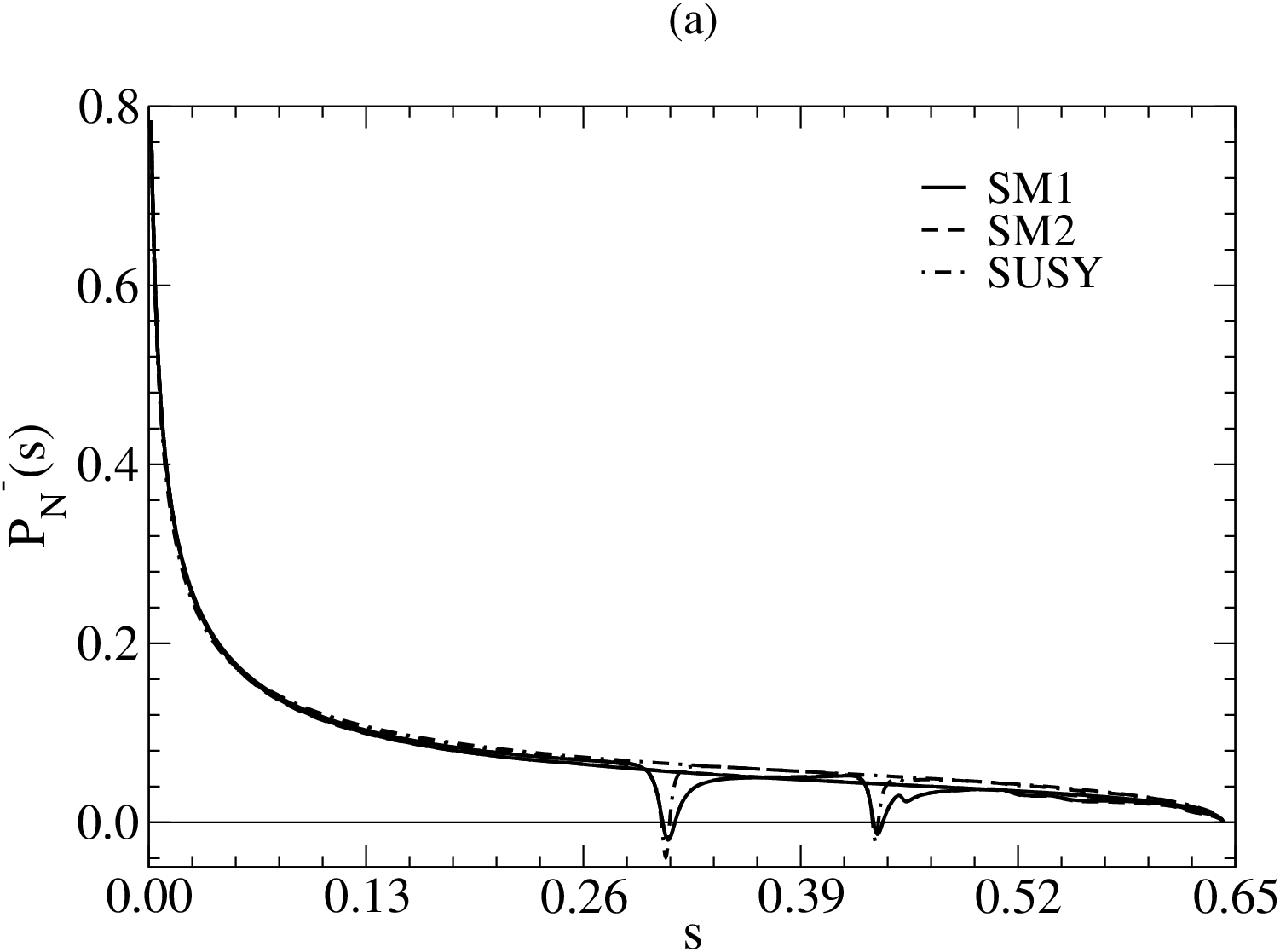}\hskip30pt \includegraphics[width=2.5in]{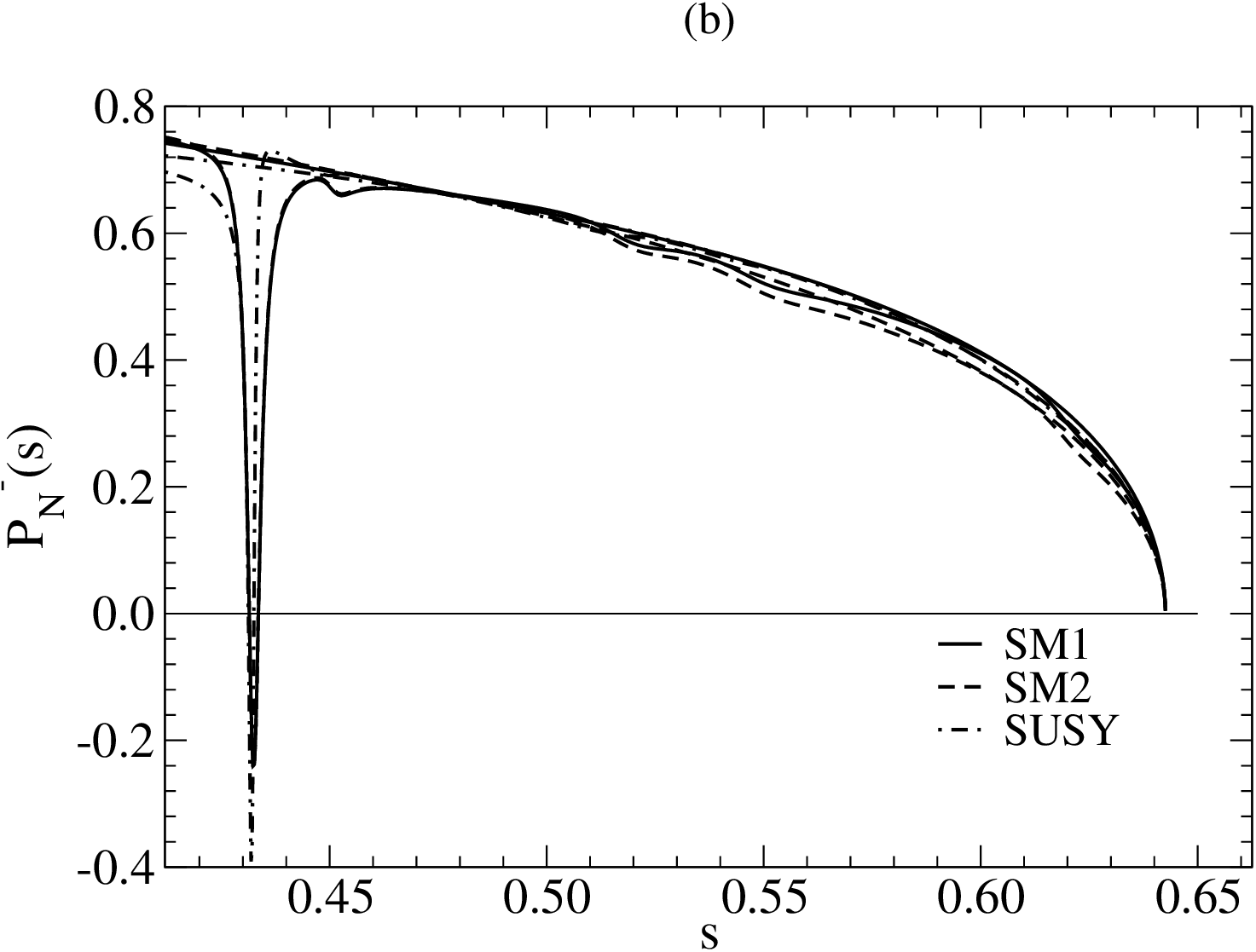}}\vskip25pt
\centerline{\includegraphics[width=2.5in]{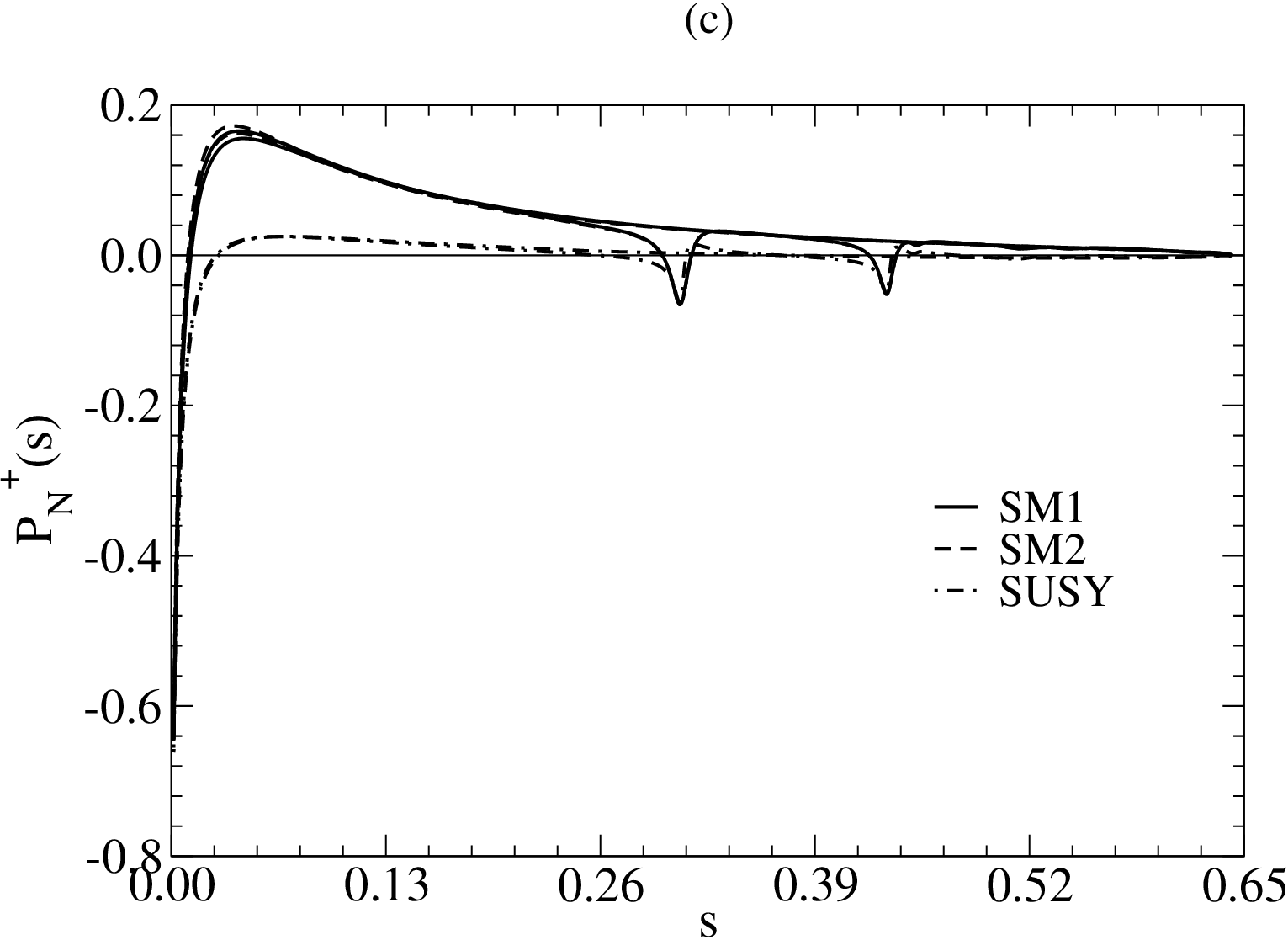}\hskip30pt \includegraphics[width=2.5in]{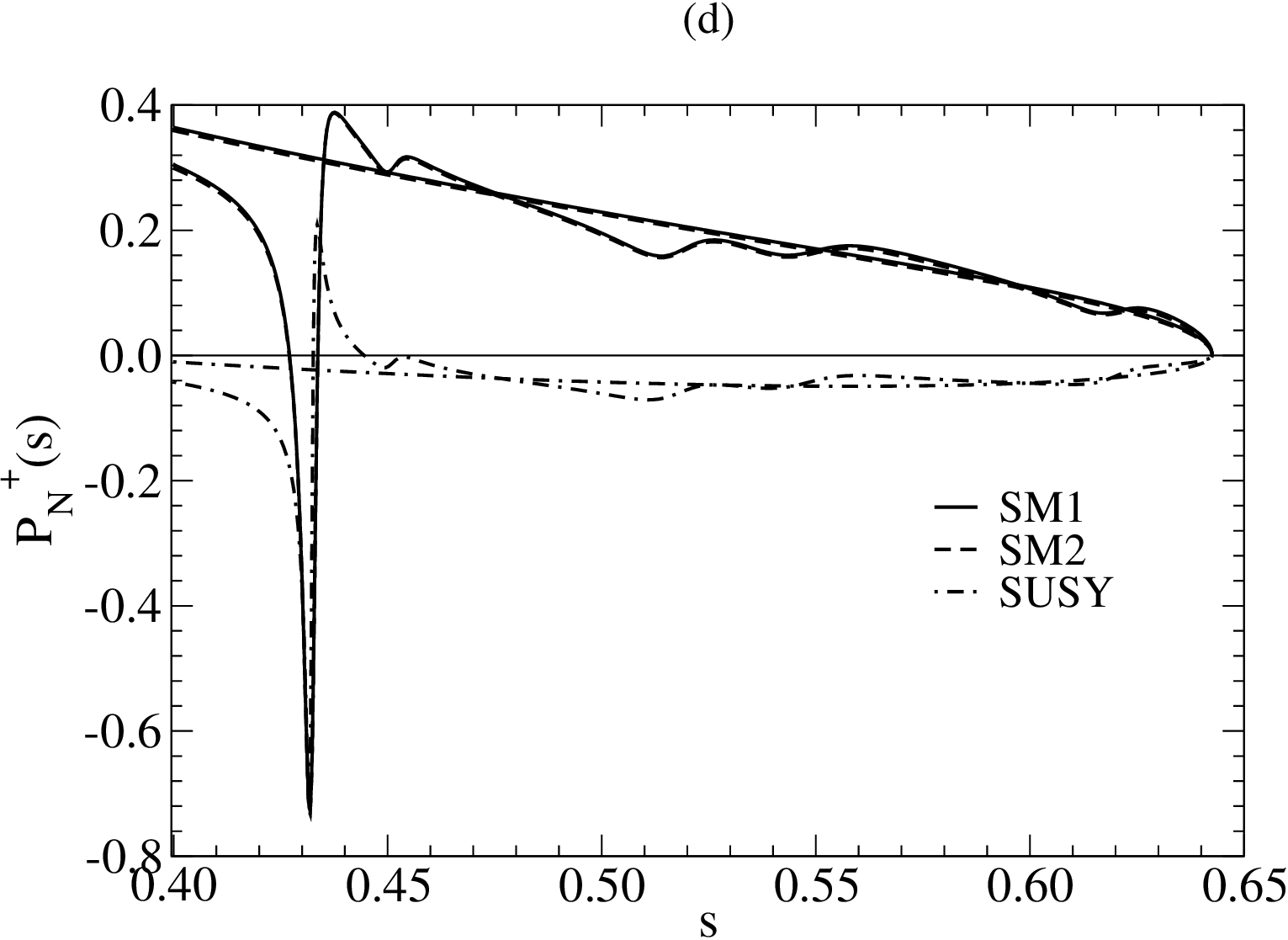}}
\caption{$\P_N^-(\sh)$ for (a) $\Lambda_{b}\rightarrow\Lambda(1115)\mu^{+}\mu^{-}$ and (b) $\Lambda_{b}\rightarrow\Lambda(1115)\tau^{+}\tau^{-}$ without and with long distance (LD) contributions from charmonium resonances. $\P_N^+(\sh)$ for the $\mu$ and $\tau$ channels are also shown in (c) and (d), respectively. The solid and dashed curves represent SLPAs obtained from SCA and MCN form factors, respectively. The dash-dotted curves represent a SUSY extension to the SM.}\label{fig:PN12p}
\end{figure}

Figure \ref{fig:PN12p} shows the differential NLPAs for decays to the ground state from the SM1, SM2, and SUSY cases with and without LD contributions. Again, we see that both SM predictions are in agreement in both channels. For $\PNm$, apart from the resonance regions, the SUSY curves are essentially the same as the SM curves. However, the SUSY curves for $\PNp$ are significantly different from the SM curves.  This is supported by the values for the integrated asymmetries in Tables \ref{nlpa1} and \ref{nlpa2}. 

\begin{figure}
\centerline{\includegraphics[width=2.5in]{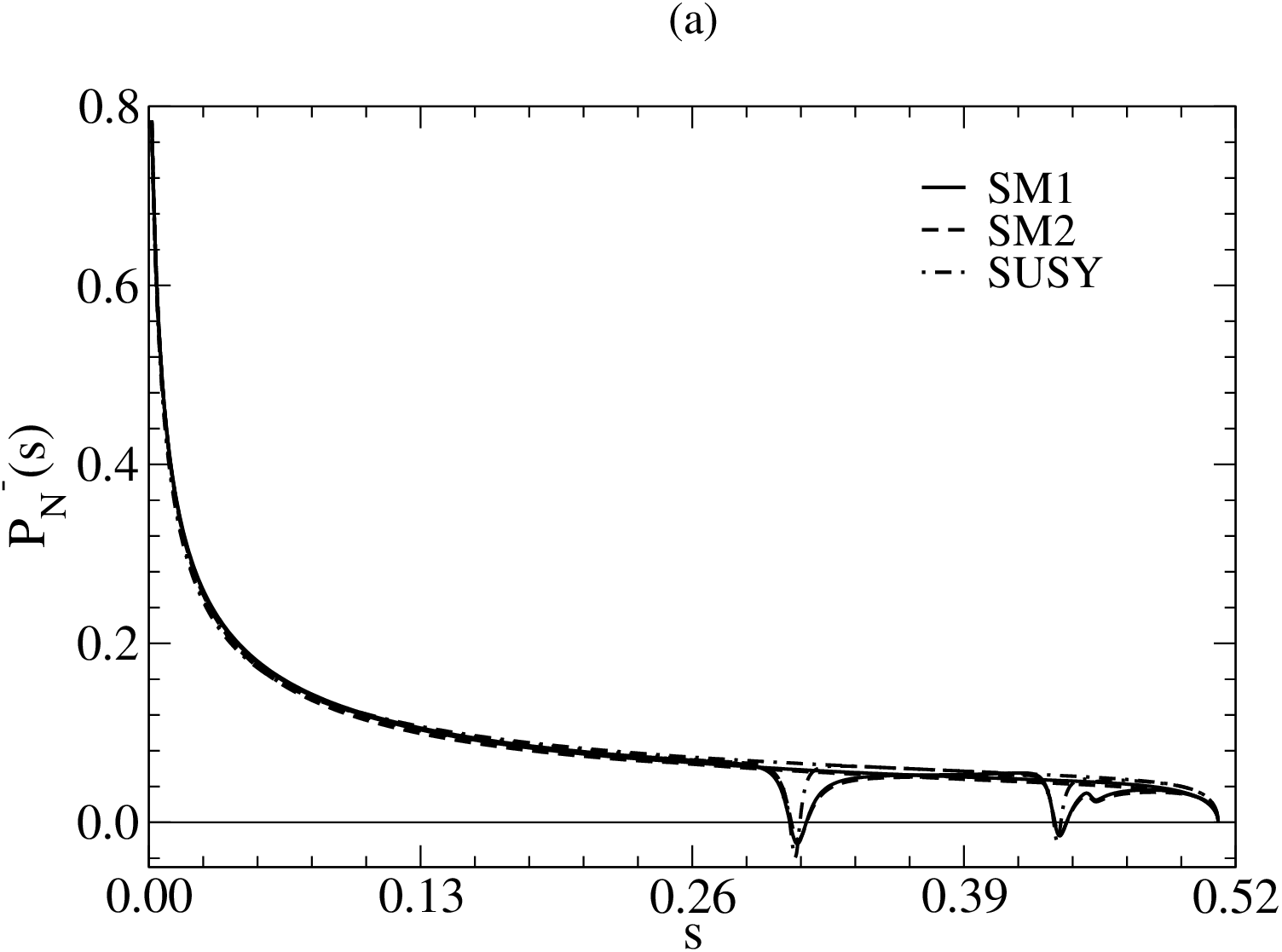}\hskip30pt \includegraphics[width=2.5in]{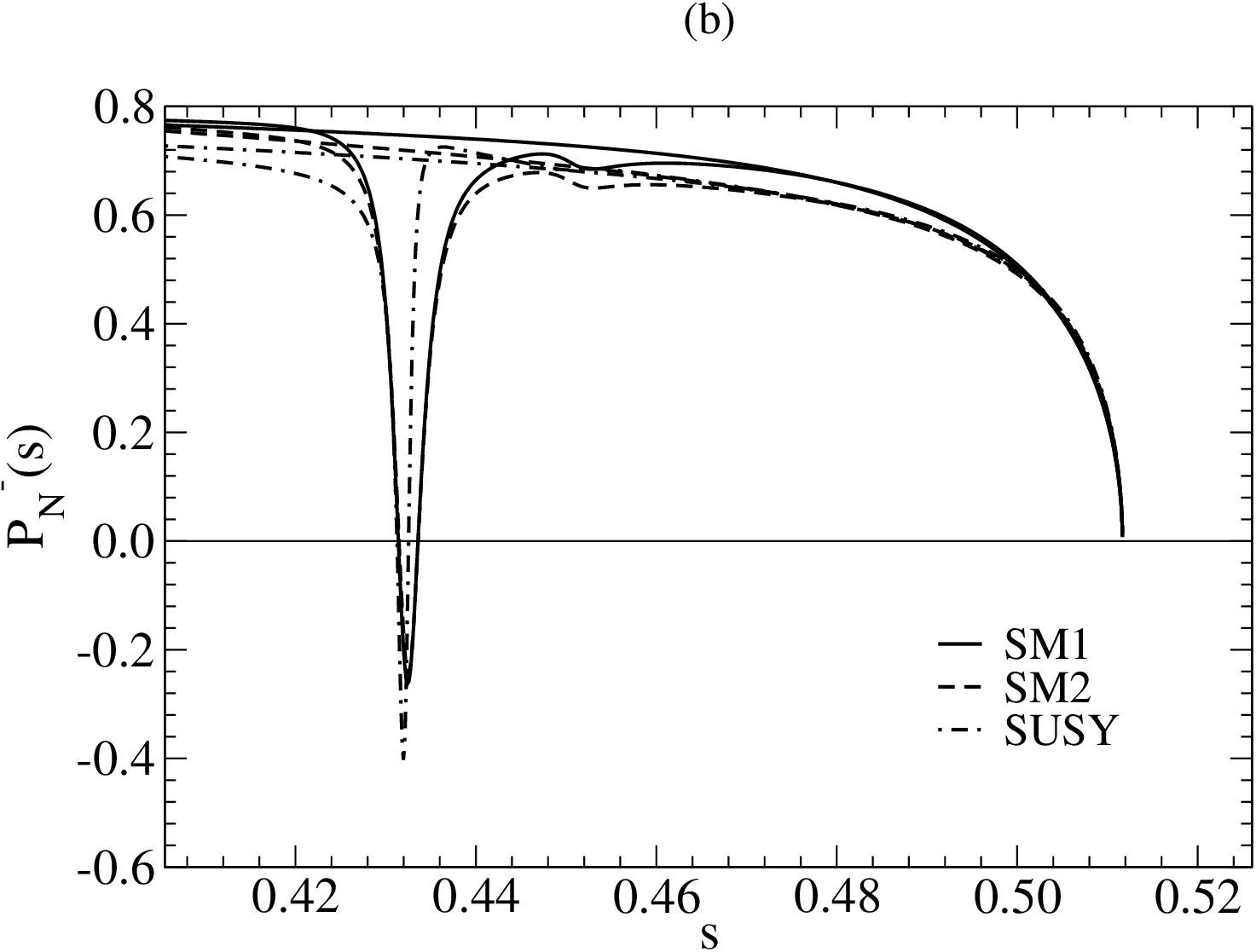}}\vskip25pt
\centerline{\includegraphics[width=2.5in]{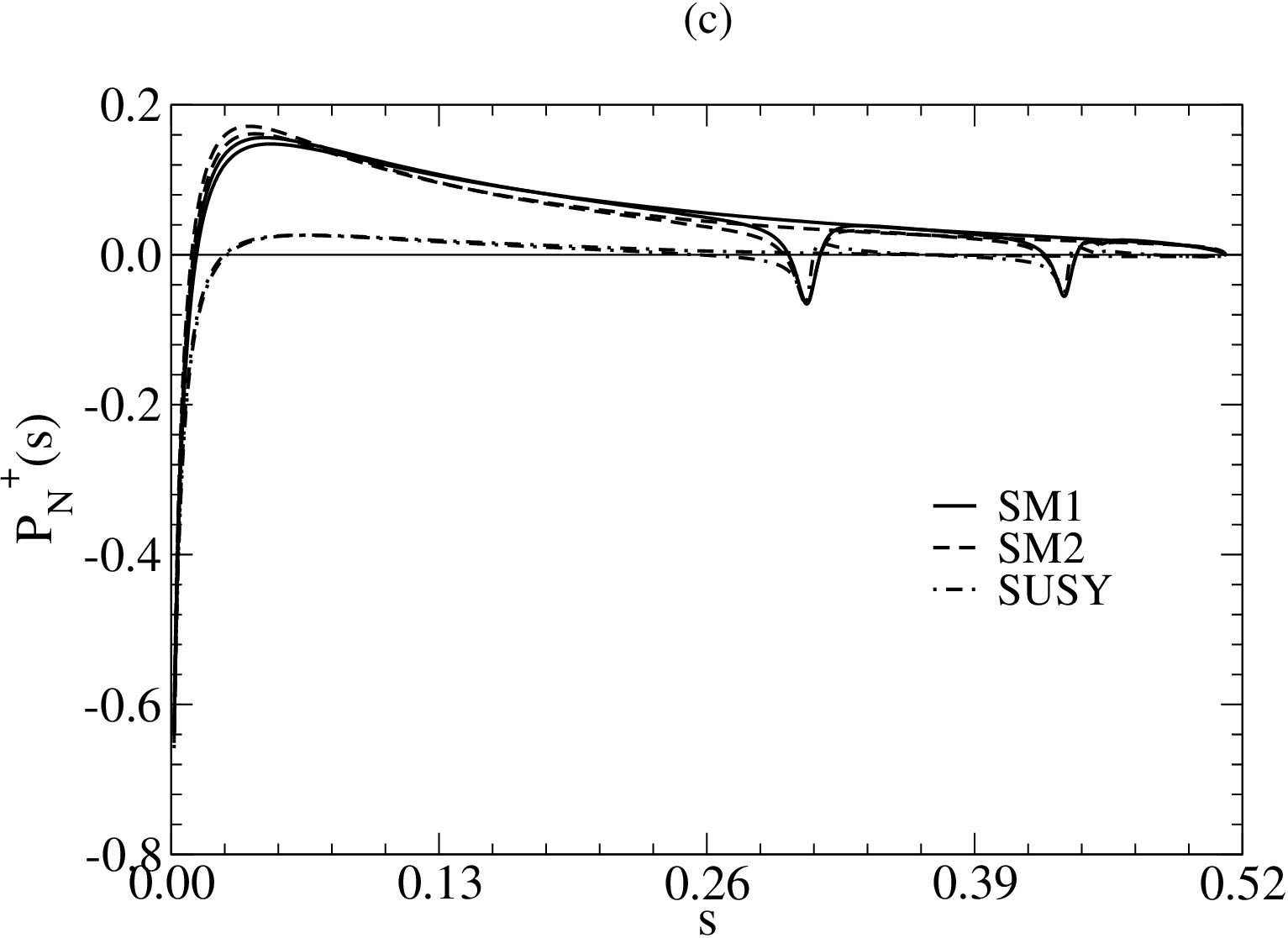}\hskip30pt \includegraphics[width=2.5in]{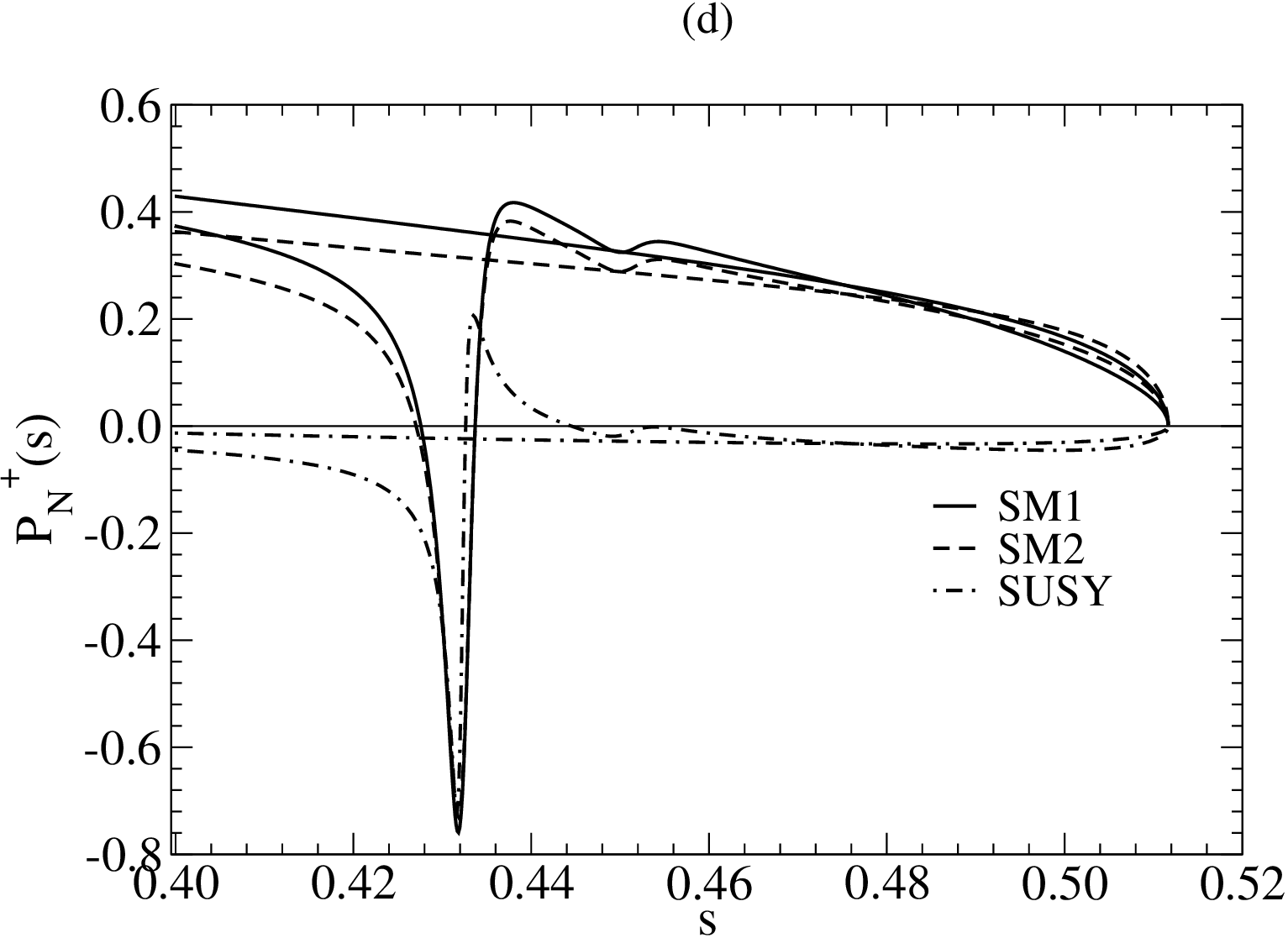}}
\caption{Same as Fig. \ref{fig:PN12p} but for $\Lambda_{b}\rightarrow\Lambda(1600)$.}\label{fig:PN12p1r}
\end{figure}

In Fig. \ref{fig:PN12p1r}, we show differential NLPAs for transitions to $\Ls(1600)$. What we find here is the same as with decays to the ground state. The SM1 and SM2 results are in agreement with each other, while the SUSY prediction, excluding the resonance regions, for $\PNm$ agrees with both SM predictions. However, the SUSY values are quite different for $\PNp$. This is borne out by the values of the integrated asymmetries that are presented in Tables \ref{nlpa1} and \ref{nlpa2}.

\subsubsection*{$\it\jp=1/2^-$}

\begin{figure}
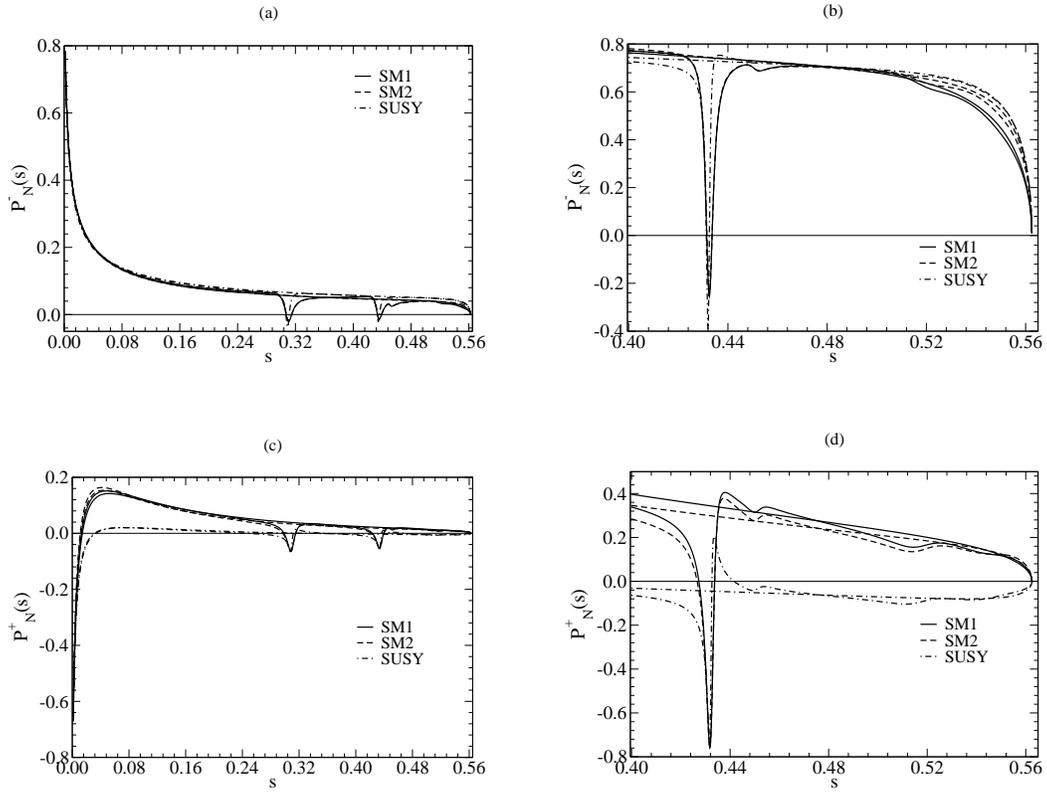

\centerline{\includegraphics[width=2.5in]{mu_slpa_N_12m_00.eps}\hskip30pt \includegraphics[width=2.5in]{tau_slpa_N_12m_00.eps}}\vskip25pt
\centerline{\includegraphics[width=2.5in]{mu_slpa_N2_12m_00.eps}\hskip30pt \includegraphics[width=2.5in]{tau_slpa_N2_12m_00.eps}}
\caption{Same as Fig. \ref{fig:PN12p} but for $\Lambda_{b}\rightarrow\Lambda(1405)$.}\label{fig:PN12m}
\end{figure}

Fig. \ref{fig:PN12m} shows the differential NLPAs for transitions to $\Ls(1405)$. As can be seen, SM1 and SM2 are essentially the same over the entire kinematic range. Excluding the resonance regions, the SUSY curves for $\PNm$ are virtually identical to those for the two SM cases; however, the SUSY values for $\PNp$ are quite different. This is confirmed by the values presented in Tables \ref{nlpa1} and \ref{nlpa2}.

\subsubsection*{$\it\jp=3/2^-$}

\begin{figure}
\centerline{\includegraphics[width=2.5in]{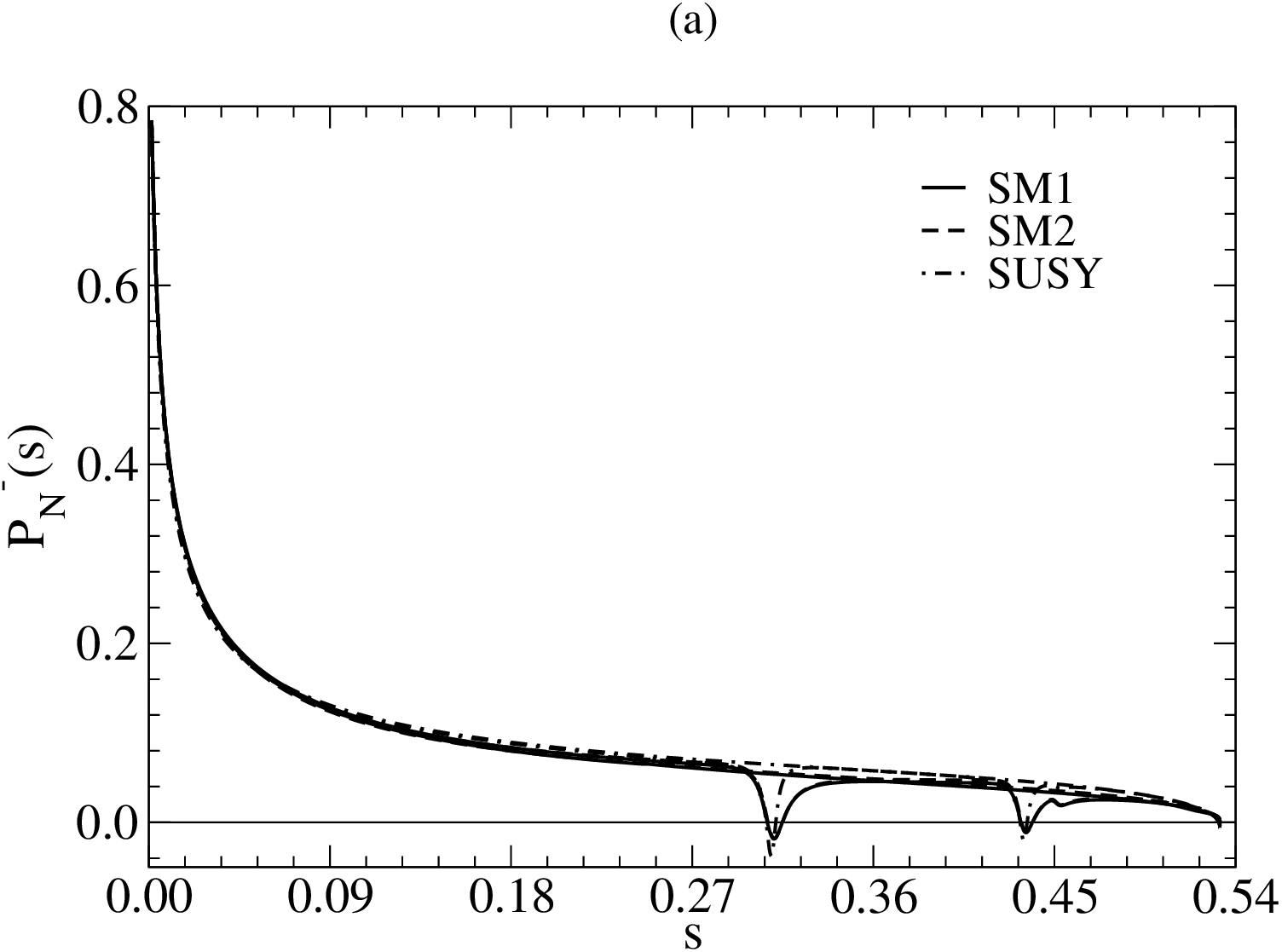}\hskip30pt \includegraphics[width=2.5in]{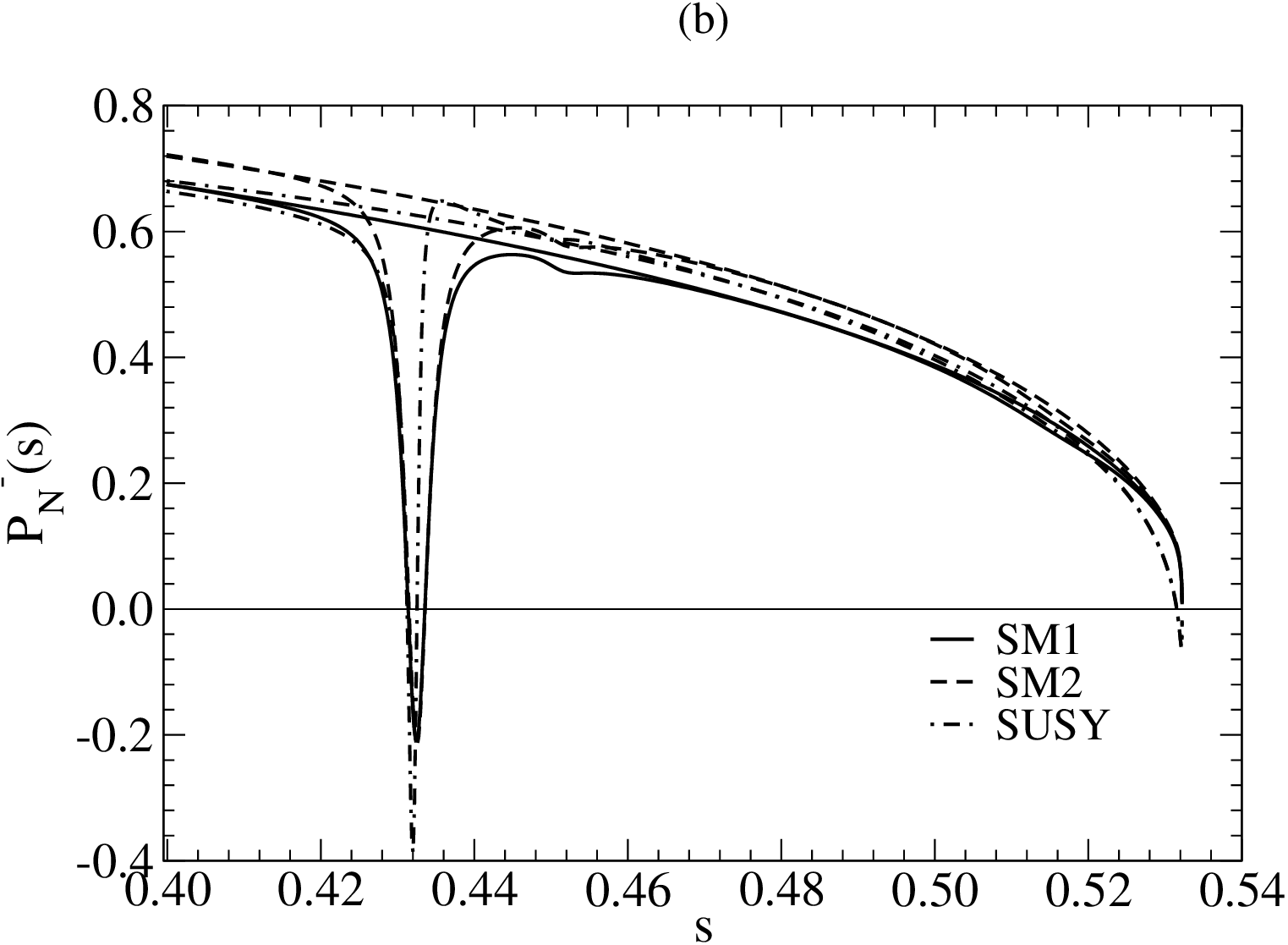}}\vskip25pt
\centerline{\includegraphics[width=2.5in]{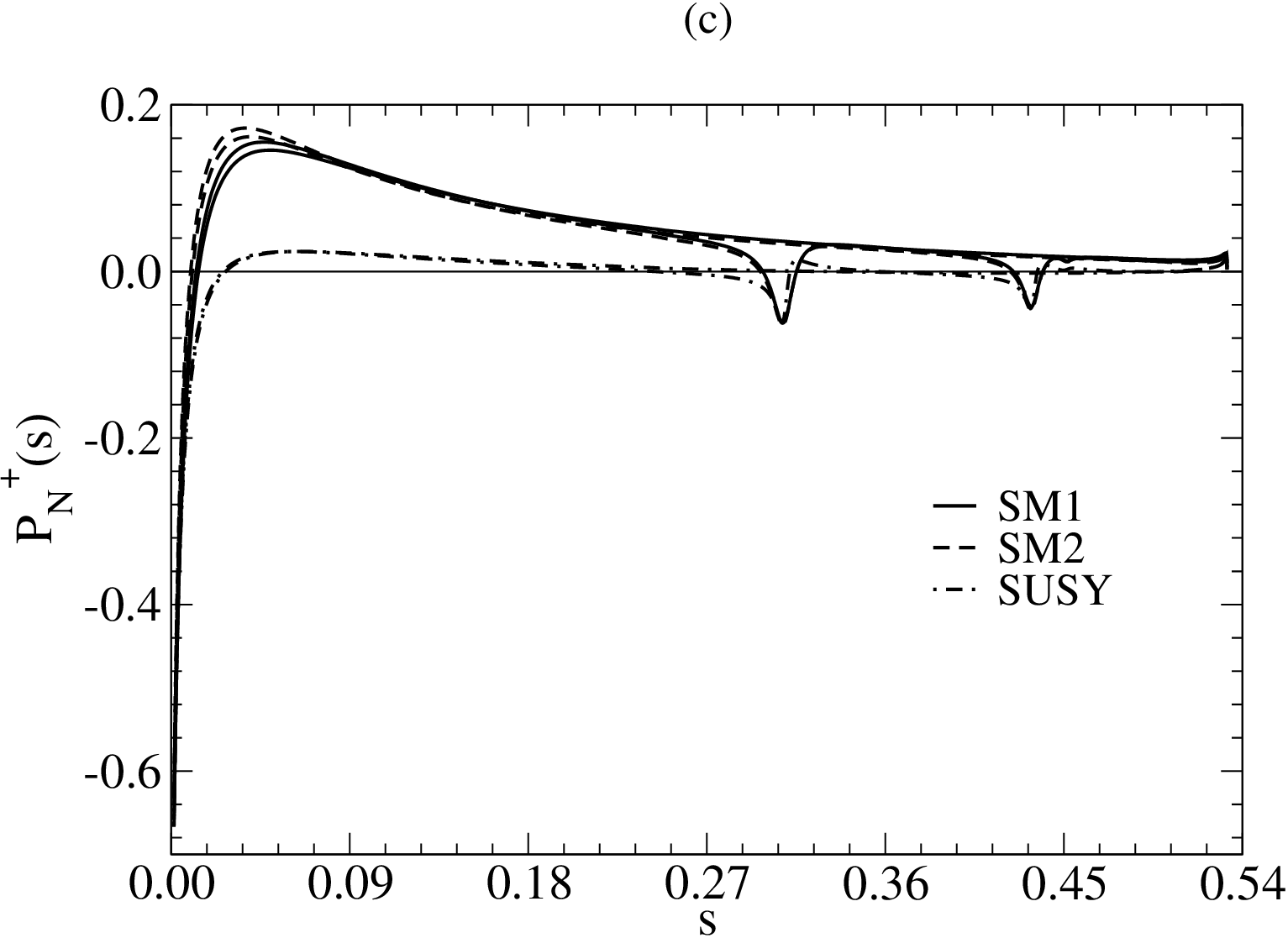}\hskip30pt \includegraphics[width=2.5in]{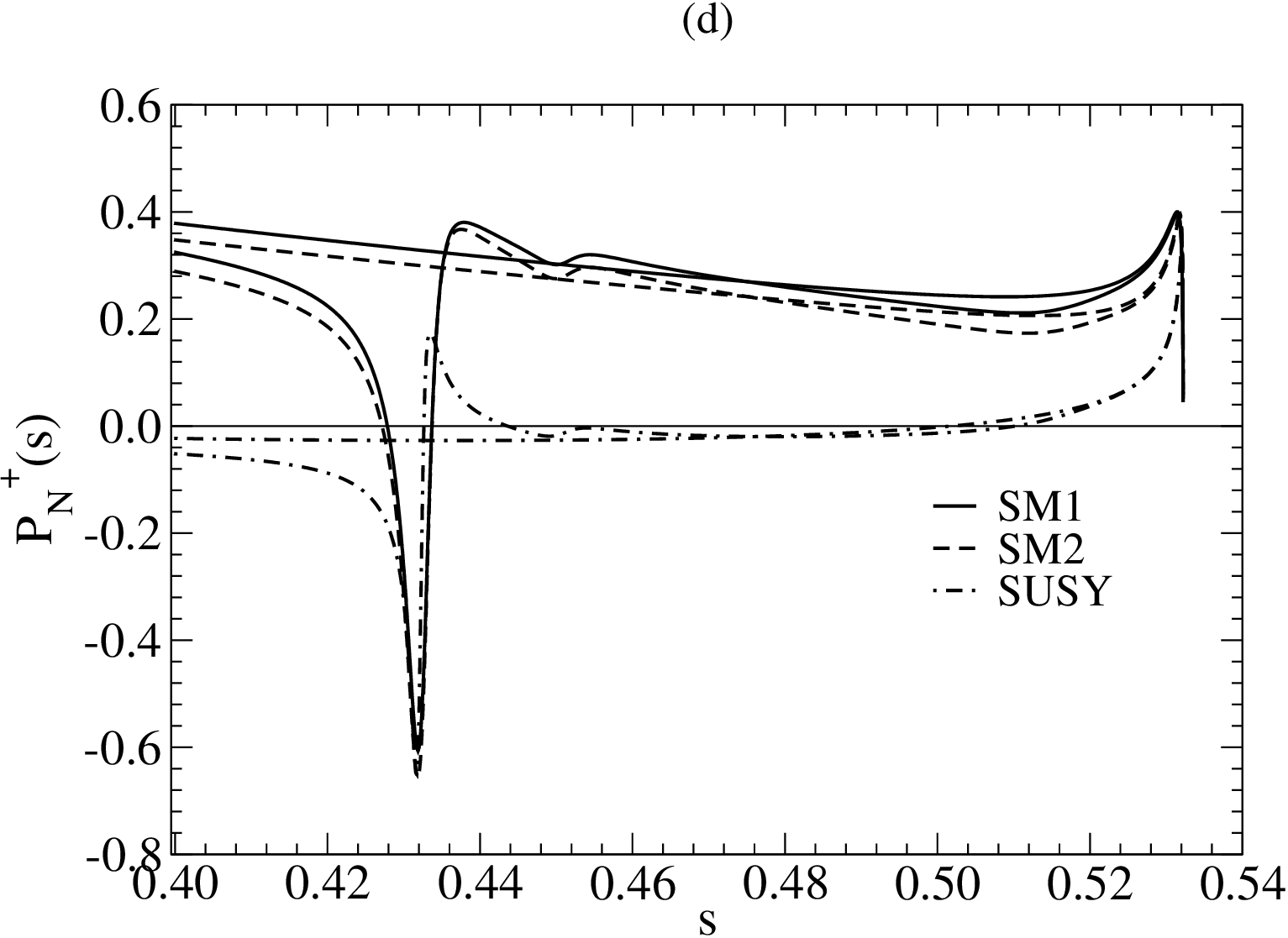}}
\caption{Same as Fig. \ref{fig:PN12p} but for $\Lambda_{b}\rightarrow\Lambda(1520)$.}\label{fig:PN32m}
\end{figure}

In Fig. \ref{fig:PN32m}, we show differential NLPAs for transitions to $\Ls(1520)$. The SM1 and SM2 results are in agreement with each other, while the SUSY prediction, excluding the resonance regions, for $\PNm$ agrees with both SM predictions. However, the SUSY values are quite different for $\PNp$. This is borne out by the values of the integrated asymmetries that are presented in Tables \ref{nlpa1} and \ref{nlpa2}.

\begin{figure}
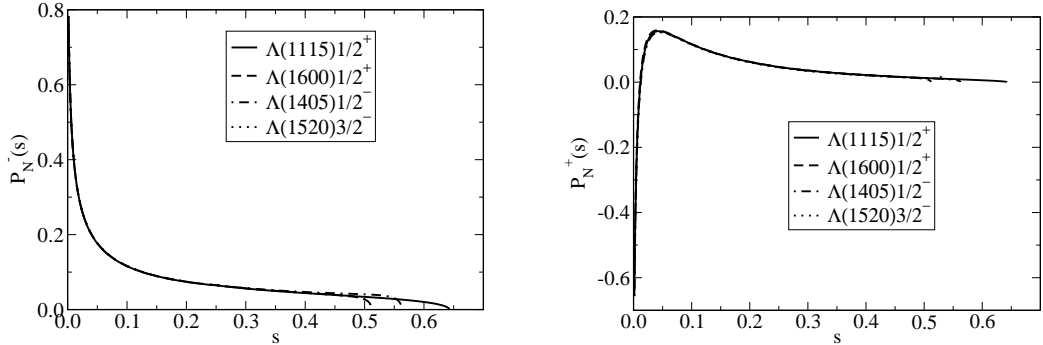
\vskip 6pt
\centerline{\includegraphics[width=2.5in]{slpa_N_all.eps}\hskip30pt\includegraphics[width=2.5in]{slpa_N+_all.eps}}
\caption{NLPAs for all of the states considered in this work, with the long-distance contributions omitted. The solid curve is for the ground state $\Lambda(1115)$, the dashed curve arises from the $\Lambda(1600)$, the dot-dashed curve is for the the $\Lambda(1405)$, and the dotted curve is for the $\Lambda(1520)$. The panel on the left is for the $\mu^-$, while that on the right is for the $\mu^+$. The curves for the different states are indistinguishable from each other, in both panels.}\label{fig:PNall}
\end{figure}

In Fig. \ref{fig:PNall} we show $P_N$ for all of the states considered. The solid curve is for the ground state $\Lambda(1115)$, the dashed curve arises from the $\Lambda(1600)$, the dot-dashed curve is for the the $\Lambda(1405)$, and the dotted curve is for the $\Lambda(1520)$. As with $P_L$, the four curves are indistinguishable despite the large values of $\xi_2/\xi_1$ near the kinematic end-point for some of the states. As with $P_L$, inclusion of the long-distance contributions of the charmonium resonances doesn't change the indistinguishability of the curves.

\section{Conclusions and Outlook\label{sec:concl}}

In this work, we have investigated the rare weak dileptonic decays of the $\Lb$ baryon using the SCA and MCN form factors extracted from a constituent quark model. We have examined the BRs, FBAs, and SLPAs for decays to a number of $\Lambda$ final states, in a number of scenarios. We have compared our results with the predictions of LCSR, QCDSR and PM form factor models, as well as the $\mu$-channel measurements reported by the CDF and LHCb collaborations. Our SM predictions for decays to the ground state are smaller than the CDF measurement and the predictions of other models in the $\mu$ channel, but are in agreement with the more precise measurement of LHCb. Additionally, our SUSY values are well within the margins of error of both the CDF and LHCb measurements. In the $\tau$ channel, our model results are in agreement with QCD sum rules and pole model predictions.

SM1 predicts that decays to the ground state are dominant in the $\mu$ channel. In SM2a, this mode is also dominant, but in SM2b, decays to $\Ls(1600)$ are dominant. SM2b also predicts that the decay rate to the $\Ls(1520)$ is comparable with the decay rate to the ground state, with the decay rate to the $\Ls(1405)$ being only slightly smaller. These results are consistent with the current status of the rare dileptonic decays of $B$ mesons, where decays to the two lowest lying kaons are insufficient to saturate the inclusive rate, and in fact account for less than 50\% of the inclusive rate. These results also suggest that it might be prudent for the LHCb collaboration to search for rare decays in modes other than the $\Ls(1115)$.

There is, however, a very important caveat. Decays in which the $q^2$ of the dileptons is near the mass-squared of the two lowest-lying vector charmonium resonances are inaccessible experimentally, as they are embedded in the much larger background coming from (tree-level) nonleptonic decays with vector charmonia in the final state. Our results obtained omitting the vector charmonia, the SM1a and SM2a models, are therefore closer to experimental reality, and these scenarios suggest that decays to the $\Lambda(1115)$ are indeed the dominant rare decay mode of the $\Lambda_b$. Nevertheless, the SM2a model that uses the more precise calculation of the form factors indicates that there will be a sizable fraction of rare decays into excited states of the $\Lambda$.

For some decay modes, such as decays to the $\Ls(1115)$, the FBAs are largely independent of the model choice for much of the kinematic range. For other modes, particularly the $\Ls(1600)$, there are significant differences between the predictions of the SM1 and SM2 models. The zero that occurs at lower values of $\sh$ in these FBAs turns out to be largely independent of the angular momentum of the daughter baryon, but with some dependence on the form factors. Nevertheless, this form factor dependence is surprisingly small. For decays to states with $J\geq3/2$, these FBAs in general have more than one zero (for the states we have considered, there is one additional zero). At leading order in HQET, the second zero does not exist. For decays to the ground state, it is known that the positions of the zeroes are modified in many scenarios that arise beyond the SM. This is also found to be true of the excited states, not surprisingly. Thus, it would be crucial to know the number of zeroes and their positions reliably (assuming that there will ever be sufficient statistics for the differential FBAs to be extracted with high precision). Our results indicate that the leading order predictions of HQET are misleading in this regard.

Additionally, we have shown that the SLPAs are largely independent of the form factor model chosen. Thus, SLPAs can be useful in determining the values of the Wilson coefficients and in looking for new physics beyond the SM. Furthermore, two of the SLPAs are independent of the final state examined, suggesting that measurement of these SLPAs for inclusive processes may offer a useful way of extracting information on the Wilson coefficients. In addition to the lepton asymmetries, there are also baryon polarization asymmetries that can be studied. Since baryonic decays could maintain the helicity structure of the effective Hamiltonian, baryon asymmetries may also be sensitive to beyond the SM scenarios.

\appendix

\section{Matrix Elements and Form Factors\label{sec:hme}}

In this appendix, we present the matrix elements of the hadronic currents used in this work in terms of the full set form factors for each transition considered. The hadronic amplitudes are expressed in terms of a set of auxiliary form factors and the relationships among the hadronic form factors and these auxiliary form factors are also given.

Recall that the amplitude for the dileptonic decay of the $\Lb$ baryon is given by
\begin{equation}
i{\M}(\Lb\to\Ls\ell^+\ell^-)=\frac{G_{F}}{\sqrt{2}}\frac{\alpha_{em}}{2\pi}V_{tb}V^{*}_{ts}(H_{1}^{\mu}L_{\mu}^{(V)}+H_{2}^{\mu}L_{\mu}^{(A)}),
\end{equation}
where $L_{\mu}^{(V)}$ and $L_{\mu}^{(A)}$ are the vector and axial vector leptonic currents, respectively. The hadronic amplitudes, $H_{1}^{\mu}$ and $H_{2}^{\mu}$, contain the hadronic matrix elements and are given by
\begin{eqnarray}
H_{1}^{\mu}&=&-\frac{2 m_{b}}{q^{2}}C_{7}(m_{b})T_R^{\mu}+C_{9}(m_{b})J_{L}^{\mu},
\label{eq:h1mu} \\
H_{2}^{\mu}&=&C_{10}(m_{b})J_{L}^{\mu},
\label{eq:h2mu}
\end{eqnarray}
where the $C_i$ are the Wilson coefficients, $T_R$ is the matrix element of the right-handed tensor current
\beq
T_R^\mu=\<\Ls(\pls,\sls)\mid\bar{s}i\sigma^{\mu\nu}q_{\nu}(1+\gamma_{5})b\mid\Lb(\plb,\slb)\>,
\label{eq:trmu}
\eeq
and $J_L$ is the matrix element of the standard $V-A$ current
\begin{equation}
J_{L}^{\mu}=\<\Ls(\pls,\sls)\mid \bar{s}\gamma^{\mu}(1-\gamma_{5})b\mid\Lb(\plb,\slb)
\rangle.
\label{eq:jlmu}
\end{equation}

The matrix elements in Eqs. \ref{eq:trmu} and \ref{eq:jlmu} contain the four basic currents $\bar{s}\gamma^\mu b$, $\bar{s}\gamma^\mu\gamma_5 b$, $\bar{s}i\sigma^{\mu\nu} b$ and $\bar{s}i\sigma^{\mu\nu}\gamma_5 b$. We will now present the matrix elements for these currents for each of the transitions we consider here.

\subsection{$J=1/2$}

For transitions between the ground state and any state with $\jp=1/2^+$, the matrix elements of the hadronic currents are
\begin{eqnarray}
\langle \Lambda\mid \bar{s}\gamma^{\mu}b\mid
\Lambda_{b}\rangle&=&\bar{u}(p_{\Lambda},s_{\Lambda})\bigg[F_{1}(q^{2}
)\gamma^{\mu}+F_{2}(q^{2})v^\mu +F_{3}(q^{2})v'^\mu\bigg]u(p_{\Lambda_{b}},s_{\Lambda_{b}}),
\label{eq:f12p} \\
\langle \Lambda\mid \bar{s}\gamma^{\mu}\gamma_{5}b\mid
\Lambda_{b}\rangle&=&\bar{u}(p_{\Lambda},s_{\Lambda})\bigg[G_{1}(q^{2}
)\gamma^{\mu}+G_{2}(q^{2})v^\mu +G_{3}(q^{2})v'^\mu\bigg]\gamma_{5}u(p_{\Lambda_{b}},s_{\Lambda_{b}}), \nn \\
\label{eq:g12p} \\
\langle \Lambda\mid \bar{s}i\sigma^{\mu\nu}b\mid
\Lambda_{b}\rangle&=&\bar{u}(p_{\Lambda},s_{\Lambda}){\cal T}^{\mu\nu}u(p_{\Lambda_{b}},s_{\Lambda_{b}}), 
\label{eq:h12p}
\end{eqnarray}
where we have used $v=\plb/\mlb$, $v'=\pls/\mls$, and
\beqy
{\cal T}^{\mu\nu}&=&H_{1}(q^{2}
)i\sigma^{\mu\nu} +H_{2}(q^{2})(v^\mu\gamma^\nu-v^\nu\gamma^\mu)+H_{3}(q^{2})(v'^\mu\gamma^\nu-v'^\nu\gamma^\mu)
+ \nonumber \\ && H_{4}(q^{2})(v^\mu v'^\nu-v^\nu v'^\mu).
\eeqy
The $F_i$, $G_i$, and $H_i$ are the form factors which are functions of the square of the 4 momentum transfer $q^2=(\plb-\pls)^2$ between the initial and final baryons. Since
\beq
\sigma^{\mu\nu}\gamma_{5}=\frac{i}{2}\varepsilon^{\mu\nu\alpha\beta}\sigma_{\alpha\beta},
\label{eq:tid}
\eeq
the matrix elements involving the current $\bar{s}i\sigma^{\mu\nu}\gamma_{5}b$ can be related to those involving $\bar{s}i\sigma^{\mu\nu}b$.

The above matrix elements involve transitions to a state with natural parity, i.e. states with parity $(-1)^{J-1/2}$. The equations involving transitions to states with unnatural parity can be found by inserting $\gamma_5$ to left of the parent baryon spinor in the equations for natural parity.

The matrix elements for the tensor currents can be written in a more convenient form. By contracting Eq. \ref{eq:h12p} on both sides with the 
four-momentum transfer $q_\nu$ and using the equations of motion, for transitions to states with $\jp=1/2^+$ we have
\begin{equation}
\langle \Lambda\mid \bar{s}i\sigma^{\mu\nu}q_{\nu}b\mid
\Lambda_{b}\rangle=\bar{u}(p_{\Lambda},s_{\Lambda})\bigg[F^{T}_{1}(q^{2}
)\gamma^{\mu}+F^{T}_{2}(q^{2})v^\mu +F^{T}_{3}(q^{2})v'^\mu\bigg]u(p_{\Lambda_{b}},s_{\Lambda_{b}}),
\label{eq:ft12p} \\
\end{equation}
where the effective tensor form factors are 
\begin{eqnarray}
F^{T}_1&=&-\left(\mlb+\mls\right)H_1-(\vq)H_2-(\vpq)H_3, \nn \\
F^{T}_2&=&\mlb H_1+\left(\mlb-\mls\right) H_2+(\vpq) H_4, \nn \\
F^{T}_3&=&\mls H_1+\left(\mlb-\mls\right) H_3-(\vq) H_4,
\end{eqnarray}
For the axial tensor current, the matrix elements are given by
\begin{equation}
\langle \Lambda\mid \bar{s}i\sigma^{\mu\nu}\gm_5 q_{\nu}b\mid
\Lambda_{b}\rangle=\bar{u}(\pls,s_{\Lambda})\bigg[G^{T}_{1}(q^{2}
)\gm^{\mu}+G^{T}_{2}(q^{2})v^\mu +G^{T}_{3}(q^{2})v'^\mu\bigg]\gm_5u(\plb,s_{\Lambda_{b}}),
\label{eq:gt12p} \\
\end{equation}
with
\begin{eqnarray}
G^{T}_1&=&\left(\mlb-\mls\right)H_1-\mls(1-\vvp)H_2-\mlb(1-\vvp)H_3, \nn \\
G^{T}_2&=&\mlb H_1-\mls H_2-\mlb H_3, \nn \\
G^{T}_3&=&\mls H_1+\mls H_2+\mlb H_3.
\end{eqnarray}
Similarly, for transitions to states with $\jp=1/2^-$, the matrix elements for the tensor and axial tensor currents are
\beqy
\<\Ls\mid\bar{s}i\sigma^{\mu\nu}q_\nu b\mid\Lb\>\=\bar{u}(\pls,\sls)\bigg[F^{T}_{1}(q^{2}
)\gamma^{\mu}+F^{T}_{2}(q^{2})v^\mu +F^{T}_{3}(q^{2})v'^\mu\bigg]\gm_5 u(\plb,\slb),\nn\\
\label{eq:ft12m} \\
\<\Ls\mid\bar{s}i\sigma^{\mu\nu}\gm_5 q_\nu b\mid\Lb\>\=\bar{u}(\pls,\pls)\bigg[G^{T}_1(q^2)\gm^\mu +G^{T}_2(q^2)v^\mu+
G^{T}_3(q^2)v'^\mu\bigg]u(\plb,\slb), \nn \\
\label{eq:gt12m}
\eeqy
respectively, where
\beqy
F^{T}_1\=\left(\mlb-\mls\right)H_1-(\vq)H_2-(\vpq)H_3,\nn\\
F^{T}_2\=\mlb H_1-\left(\mlb+\mls\right)H_2+(\vpq)H_4,\nn\\
F^{T}_3\=\mls H_1-\left(\mlb+\mls\right)H_3-(\vq)H_4,
\eeqy
and
\beqy
G^{T}_1\=-\left(\mlb+\mls\right)H_1+\mls(1+\vvp)H_2+\mlb(1+\vvp)H_3,\nn\\
G^{T}_2\=\mlb H_1-\mls H_2-\mlb H_3,\nn\\
G^{T}_3\=\mls H_1-\mls H_2-\mlb H_3.
\eeqy

For transitions to states with $J=1/2$, we can now write Eqs. \ref{eq:h1mu} and \ref{eq:h2mu} in the form
\beqy
H_1^\mu&=&\bar{u}(\pls,\sls)\bigg[\gm^\mu\bigg(A_1+B_1\gm_5\bigg)+v^\mu\bigg(A_2+B_2\gm_5\bigg)+v'^\mu\bigg(A_3+B_3\gm_5\bigg)\bigg]u(\plb,\slb),\nn\\
\\
H_2^\mu&=&\bar{u}(\pls,\sls)\bigg[\gm^\mu\bigg(D_1+E_1\gm_5\bigg)+v^\mu\bigg(D_2+E_2\gm_5\bigg)+v'^\mu\bigg(D_3+E_3\gm_5\bigg)\bigg]u(\plb,\slb),\nn\\
\eeqy
where, for transitions to states with natural parity, the auxiliary form factors are given by
\beqy
A_i&=&-\frac{2m_b}{q^2}C_7F^{T}_i+C_9F_i,\nn\\
B_i&=&-\frac{2m_b}{q^2}C_7G^{T}_i-C_9G_i,\nn\\
D_i&=&C_{10}F_i, \,\,\,\,\ E_i=-C_{10}G_i,
\label{eq:natpar_dilep}
\eeqy
while for transitions to states with unnatural parity,
\beqy
A_i&=&-\frac{2m_b}{q^2}C_7G^{T}_i-C_9G_i,\nn\\
B_i&=&-\frac{2m_b}{q^2}C_7F^{T}_i+C_9F_i,\nn\\
D_i&=&-C_{10}G_i, \,\,\,\,\ E_i=C_{10}F_i.
\label{eq:unnatpar_dilep}
\eeqy

\subsection{$J=3/2$}

The matrix elements for decays to daughter baryons with $J^P=3/2^{-}$, a state with natural parity, are given by
\begin{eqnarray}
\langle \Lambda\mid \bar{s}\gamma^{\mu}b\mid
\Lambda_{b}\rangle&=&\bar{u}_\alpha (p_{\Lambda},s_{\Lambda})\bigg[v^\alpha\bigg(F_{1}\gamma^{\mu}+F_{2}v^\mu +F_{3}v'^\mu\bigg)+
F_4 g^{\alpha\mu}\bigg]u(p_{\Lambda_{b}},s_{\Lambda_{b}}),\nonumber\\
\label{eq:f32m} \\
\langle \Lambda\mid \bar{s}\gamma^{\mu}\gamma_{5}b\mid
\Lambda_{b}\rangle&=&\bar{u}_\alpha (p_{\Lambda},s_{\Lambda})\bigg[v^\alpha\bigg(G_{1}\gamma^{\mu}+G_{2}v^\mu +G_{3}v'^\mu\bigg)+
G_4 g^{\alpha\mu}\bigg]\gamma_{5}u(p_{\Lambda_{b}},s_{\Lambda_{b}}),\nonumber\\
\label{eq:g32m} \\
\langle \Lambda\mid \bar{s}i\sigma^{\mu\nu}b\mid
\Lambda_{b}\rangle&=&\bar{u}_\alpha (p_{\Lambda},s_{\Lambda}){\cal T}^{\alpha\mu\nu}u(p_{\Lambda_{b}},s_{\Lambda_{b}}),
\label{eq:h32m}
\end{eqnarray}
where
\begin{eqnarray}
{\cal T}^{\alpha\mu\nu}&=&v^\alpha\bigg(H_{1}i\sigma^{\mu\nu} +H_{2}(v^\mu\gamma^\nu-v^\nu\gamma^\mu) +H_{3}(v'^\mu\gamma^\nu-v'^\nu\gamma^\mu) 
\nonumber \\ &&
+H_{4}(v^\mu v'^\nu-v^\nu v'^\mu)\bigg)+H_5 (g^{\alpha\mu}\gamma^\nu-g^{\alpha\nu}\gamma^\mu)+ \nonumber \\ && H_6
(g^{\alpha\mu}v^\nu-g^{\alpha\nu}v^\mu).
\end{eqnarray}
The spinor $\bar{u}_\alpha$ is a Rarita-Schwinger spinor and satisfies the conditions
\begin{equation}
p_{\Lambda}^\alpha\bar{u}_\alpha(p_{\Lambda},s)=0,\,\,\bar{u}_\alpha(p_{\Lambda},s)\gamma^\alpha=0,\,\,
\bar{u}_\alpha(p_{\Lambda},s)\slash{p}_{\Lambda}=m_{\Lambda}\bar{u}_\alpha(p_{\Lambda},s).
\end{equation}
Again, the matrix elements for transitions to states with unnatural parity can be found by inserting $\gm_5$ to the left of the parent baryon spinor in the equations for natural parity.

For transitions to states with $\jp=3/2^-$, the tensor and axial-tensor current in terms of their effective form factors are given by
\beqy
\<\Lambda(\pls)\mid\bar{s}i\sigma^{\mu\nu}q_\nu b\mid\Lambda_b(\plb)\> &=& \bar{u}_\alpha(\pls,\sls)\bigg[v^\alpha\bigg(F^{T}_1\gm^\mu+F^{T}_2v^\mu+
F^{T}_3v'^\mu\bigg) \nn \\ && +F^{T}_4g^{\alpha\mu}\bigg]u(\plb,\slb),
\label{eq:ft32m} \\
\<\Lambda(\pls)\mid\bar{s}i\sigma^{\mu\nu}\gm_5q_\nu b\mid\Lambda_b(\plb)\> &=& \bar{u}_\alpha(\pls,\sls)\bigg[v^\alpha\bigg(G^{T}_1\gm^\mu+
G^{T}_2v^\mu+G^{T}_3v'^\mu\bigg) \nn \\ && +G^{T}_4g^{\alpha\mu}\bigg]\gm_5u(\plb,\slb),
\label{eq:gt32m}
\eeqy
where
\beqy
F^{T}_1&=&-\left(\mlb+\mls\right) H_1-(\vq) H_2-(\vpq) H_3-\mlb H_5, \nn \\
F^{T}_2&=&\mlb H_1+\left(\mlb-\mls\right) H_2+(\vpq) H_4-\mlb H_6, \nn \\
F^{T}_3&=&\mls H_1+\left(\mlb-\mls\right) H_3-(\vq) H_4, \nn \\
F^{T}_4&=&\left(\mlb-\mls\right) H_5+(\vq) H_6, \label{eq:ft_nat}
\eeqy
and
\beqy
G^{T}_1&=&\left(\mlb-\mls\right) H_1-\mls(1-\vvp)H_2-\mlb(1-\vvp)H_3+\mlb H_5+\mls H_6, \nn \\
G^{T}_2&=&\mlb H_1-\mls H_2-\mlb H_3, \nn \\
G^{T}_3&=&\mls H_1+\mls H_2+\mlb H_3-\mls H_6, \nn \\
G^{T}_4&=&\left(\mlb+\mls\right) H_5+\mls(1+\vvp) H_6. \label{eq:gt_nat}
\eeqy
For $\jp=3/2^+$, a state with unnatural parity, we have
\beqy
\<\Lambda(\pls)\mid\bar{s}i\sigma^{\mu\nu}q_\nu b\mid\Lambda_b(\plb)\> &=& \bar{u}_\alpha(\pls,\sls)\bigg[v^\alpha\bigg(F^{T}_1\gm^\mu+F^{T}_2v^\mu+
F^{T}_3v'^\mu\bigg) \nn \\ && +F^{T}_4g^{\alpha\mu}\bigg]\gm_5u(\plb,\slb),
\label{eq:ft32p} \\
\<\Lambda(\pls)\mid\bar{s}i\sigma^{\mu\nu}\gm_5q_\nu b\mid\Lambda_b(\plb)\> &=& \bar{u}_\alpha(\pls,\sls)\bigg[v^\alpha\bigg(G^{T}_1\gm^\mu+
G^{T}_2v^\mu+G^{T}_3v'^\mu\bigg) \nn \\ && +G^{T}_4g^{\alpha\mu}\bigg]u(\plb,\slb),
\label{eq:gt32p}
\eeqy
with
\beqy
F^{T}_1&=&\left(\mlb-\mls\right) H_1-(\vq) H_2-(\vpq) H_3-\mlb H_5, \nn \\
F^{T}_2&=&\mlb H_1-\left(\mlb+\mls\right) H_2+(\vpq) H_4-\mlb H_6, \nn \\
F^{T}_3&=&\mls H_1-\left(\mlb+\mls\right) H_3-(\vq) H_4, \nn \\
F^{T}_4&=&-\left(\mlb+\mls\right) H_5+(\vq) H_6, \label{eq:ft_unnat}
\eeqy
and
\beqy
G^{T}_1&=&-\left(\mlb+\mls\right) H_1+\mls(1+\vvp)H_2+\mlb(1+\vvp)H_3+\mlb H_5+\mls H_6, \nn \\
G^{T}_2&=&\mlb H_1-\mls H_2-\mlb H_3, \nn \\
G^{T}_3&=&\mls H_1-\mls H_2-\mlb H_3-\mls H_6, \nn \\
G^{T}_4&=&-\left(\mlb-\mls\right) H_5-\mls(1-\vvp) H_6. \label{eq:gt_unnat}
\eeqy

For transitions to states with $J=3/2$, the hadronic amplitudes are 
\beqy
H_1^\mu&=&\bar{u}_\alpha(\pls,\sls)\bigg[v^\alpha\bigg(\gm^\mu\bigg(A_1+B_1\gm_5\bigg)+v^\mu\bigg(A_2+B_2\gm_5\bigg)+
v'^\mu\bigg(A_3+B_3\gm_5\bigg)\bigg)\nn\\&& +g^{\alpha\mu}\bigg(A_4+B_4\gm_5\bigg)\bigg]u(\plb,\slb), \\
H_2^\mu&=&\bar{u}_\alpha(\pls,\sls)\bigg[v^\alpha\bigg(\gm^\mu\bigg(D_1+E_1\gm_5\bigg)+v^\mu\bigg(D_2+E_2\gm_5\bigg)+
v'^\mu\bigg(D_3+E_3\gm_5\bigg)\bigg)\nn\\&& +g^{\alpha\mu}\bigg(D_4+E_4\gm_5\bigg)\bigg]u(\plb,\slb),
\eeqy
where the auxiliary form factors $A_i$, $B_i$, $D_i$, and $E_i$ are given in Eq. \ref{eq:natpar_dilep} for transitions to $\jp=3/2^-$ and Eq. \ref{eq:unnatpar_dilep} for transitions to $\jp=3/2^+$.

\subsection{$J=5/2$}

For decays to $J^P=5/2^{+}$, the matrix elements are
\begin{eqnarray}
\langle \Lambda\mid \bar{s}\gamma^{\mu}b\mid
\Lambda_{b}\rangle&=&\bar{u}_{\alpha\beta} (p_{\Lambda},s_{\Lambda})v^\alpha\bigg[v^\beta\bigg(F_{1}\gamma^{\mu}+F_{2}v^\mu +F_{3}v'^\mu\bigg) \nn \\
&& 
+F_4 g^{\beta\mu}\bigg]u(p_{\Lambda_{b}},s_{\Lambda_{b}}),
\label{eq:f52p} \\
\langle \Lambda\mid \bar{s}\gamma^{\mu}\gamma_{5}b\mid
\Lambda_{b}\rangle&=&\bar{u}_{\alpha\beta} (p_{\Lambda},s_{\Lambda})v^\alpha\bigg[v^\beta\bigg(G_{1}\gamma^{\mu}+G_{2}v^\mu +G_{3}v'^\mu\bigg) \nn \\
&& 
+G_4 g^{\beta\mu}\bigg]\gamma_{5}u(p_{\Lambda_{b}},s_{\Lambda_{b}}),
\label{eq:g52p} \\
\langle \Lambda\mid \bar{s}i\sigma^{\mu\nu}b\mid
\Lambda_{b}\rangle&=&\bar{u}_{\alpha\beta} (p_{\Lambda},s_{\Lambda}){\cal T}^{\alpha\beta\mu\nu}u(p_{\Lambda_{b}},s_{\Lambda_{b}}),
\label{eq:h52p}
\end{eqnarray}
where
\begin{eqnarray}
{\cal T}^{\alpha\beta\mu\nu}&=&v^\alpha\bigg[v^\beta\bigg(H_{1}i\sigma^{\mu\nu} +H_{2}(v^\mu\gamma^\nu-v^\nu\gamma^\mu) +
H_{3}(v'^\mu\gamma^\nu-v'^\nu\gamma^\mu) \nonumber \\ &&
+H_{4}(v^\mu v'^\nu-v^\nu v'^\mu)\bigg)+H_5 (g^{\beta\mu}\gamma^\nu-g^{\beta\nu}\gamma^\mu)+ \nonumber \\ && 
H_6 (g^{\beta\mu}v^\nu-g^{\beta\nu}v^\mu)\bigg].
\end{eqnarray}
The spinor $\bar{u}_{\alpha\beta}$ is symmetric in the indices $\alpha$ and $\beta$, and satisfies
\begin{eqnarray}
p_{\Lambda}^\alpha\bar{u}_{\alpha\beta}(p_{\Lambda},s)&=&p_{\Lambda}^\beta\bar{u}_{\alpha\beta}(p_{\Lambda},s)=0, \nonumber \\
\bar{u}_{\alpha\beta}(p_{\Lambda},s)\gamma^\alpha&=&\bar{u}_{\alpha\beta}(p_{\Lambda},s)\gamma^\beta=0, \nonumber \\ 
\bar{u}_{\alpha\beta}(p_{\Lambda},s)\slash{p}_{\Lambda}&=&m_{\Lambda}\bar{u}_{\alpha\beta}(p_{\Lambda},s) \nonumber \\
\bar{u}_{\alpha\beta}(p_{\Lambda},s)g^{\alpha\beta}&=&0.
\end{eqnarray}

The matrix elements of the tensor and axial-tensor currents in terms of their effective form factors are
\beqy
\<\Lambda(\pls)\mid\bar{s}i\sigma^{\mu\nu}q_\nu b\mid\Lambda_b(\plb)\> &=& \bar{u}_{\alpha\beta}(\pls,\sls)v^\alpha\bigg[v^\beta\bigg(F^{T}_1\gm^\mu+
F^{T}_2v^\mu+F^{T}_3v'^\mu\bigg) \nn \\ && +F^{T}_4g^{\beta\mu}\bigg]u(\plb,\slb),
\label{eq:ft52p} \\
\<\Lambda(\pls)\mid\bar{s}i\sigma^{\mu\nu}\gm_5q_\nu b\mid\Lambda_b(\plb)\> &=&
\bar{u}_{\alpha\beta}(\pls,\sls)v^\alpha\bigg[v^\beta\bigg(G^{T}_1\gm^\mu+
G^{T}_2v^\mu+G^{T}_3v'^\mu\bigg) \nn \\ && +G^{T}_4g^{\beta\mu}\bigg]\gm_5u(\plb,\slb),
\label{eq:gt52p}
\eeqy
where the effective tensor and axial-tensor form factors are given by Eqs. \ref{eq:ft_nat} and \ref{eq:gt_nat}, respectively.

The hadronic amplitudes can now be written as
\beqy
H_1^\mu&=&\bar{u}_{\alpha\beta}(\pls,\sls)v^\alpha\bigg[v^\beta\bigg(\gm^\mu\bigg(A_1+B_1\gm_5\bigg)+v^\mu\bigg(A_2+B_2\gm_5\bigg)+
v'^\mu\bigg(A_3+B_3\gm_5\bigg)\bigg)\nn\\&& +g^{\beta\mu}\bigg(A_4+B_4\gm_5\bigg)\bigg]u(\plb,\slb), \\
H_2^\mu&=&\bar{u}_{\alpha\beta}(\pls,\sls)v^\alpha\bigg[v^\beta\bigg(\gm^\mu\bigg(D_1+E_1\gm_5\bigg)+v^\mu\bigg(D_2+E_2\gm_5\bigg)+
v'^\mu\bigg(D_3+E_3\gm_5\bigg)\bigg)\nn\\&& +g^{\beta\mu}\bigg(D_4+E_4\gm_5\bigg)\bigg]u(\plb,\slb),
\eeqy
where the auxiliary form factors $A_i$, $B_i$, $D_i$, and $E_i$ are given in Eq. \ref{eq:natpar_dilep}.

\section*{Acknowledgment} We gratefully acknowledge the support of the Department of Physics, the College of Arts and Sciences, and the Office of Research at Florida State University. This research was sponsored in part by the Office of Nuclear Physics, U.S. Department of Energy.

\end{document}